\documentclass[aps,prl,notitlepage,twocolumn,superscriptaddress]{revtex4-2}

\makeatletter
\def\@hangfrom@section#1#2#3{\normalsize\@hangfrom{#1#2}#3}
\def\@hangfroms@section#1#2{\normalsize#1#2}
\makeatother

\RequirePackage[usenames, dvipsnames, x11names]{xcolor}
\usepackage{amsmath,amsfonts,amssymb,bm}
\usepackage{graphicx,color,soul}
\usepackage{dsfont}
\usepackage{enumerate}
\usepackage[colorlinks=true, allcolors=blue]{hyperref}

\usepackage{multirow}

\hypersetup{
    colorlinks=true,
    citecolor=Green,
    linkcolor=Red,
    urlcolor=NavyBlue
}

\newcommand{\nc}{\newcommand}
\nc{\ket}[1]{|#1\rangle}
\nc{\bra}[1]{\langle#1|}
\nc{\ketbra}[2]{|#1\rangle\!\langle#2|}
\nc{\braket}[2]{\langle#1|#2\rangle}
\nc{\braoprket}[3]{\langle#1|#2|#3\rangle}
\nc{\opr}[1]{\operatorname{#1}}
\nc{\avg}[1]{\langle#1\rangle}
\nc{\ketbrasame}[1]{|#1\rangle\!\langle#1|}
\nc{\tr}{\opr{tr}}
\nc{\E}{\mathbb{E}}
\nc{\var}{\opr{Var}}
\nc{\up}{\uparrow}
\nc{\dn}{\downarrow}
\nc{\bd}[1]{\boldsymbol{#1}}
\nc{\zjx}[1]{\textcolor{violet}{\textbf{#1}}}
\nc{\jx}[1]{\textcolor{violet}{\textbf{[JX: #1]}}}
\nc{\lb}[1]{\textcolor{brown}{\textbf{[LB: #1]}}}
\nc{\ww}[1]{\textcolor{LimeGreen}{\textbf{[WW: #1]}}}
\nc{\ls}[1]{\textcolor{blue}{\textbf{[LS: #1]}}}

\usepackage[normalem]{ulem}

\newcommand{\figref}[1]{Fig.\,\ref{#1}}

\newcommand{\secref}[1]{Sec.\,\ref{#1}}
\newcommand{\eqnref}[1]{Eq.\,\eqref{#1}}
\newcommand{\eqnsref}[1]{Eqs.\,\eqref{#1}}
\newcommand{\appref}[1]{Appendix~\ref{#1}}

\begin{document}
\graphicspath{{figure/}}

\title{Identifying Instabilities with Quantum Geometry in Flat Band Systems}

\author{Jia-Xin Zhang}
\thanks{These two authors contributed equally to this work.}
\affiliation{French American Center for Theoretical Science, CNRS, KITP, Santa Barbara, California 93106-4030, USA
}
\affiliation{Kavli Institute for Theoretical Physics, University of California, Santa Barbara, California 93106-4030, USA}

\author{Wen O. Wang}
\thanks{These two authors contributed equally to this work.}
\affiliation{Kavli Institute for Theoretical Physics, University of California, Santa Barbara, California 93106-4030, USA}

\author{Leon Balents}
\affiliation{Kavli Institute for Theoretical Physics, University of California, Santa Barbara, California 93106-4030, USA}
\affiliation{Canadian Institute for Advanced Research, Toronto, Ontario, Canada}
\affiliation{French American Center for Theoretical Science, CNRS, KITP, Santa Barbara, California 93106-4030, USA
}

\author{Lucile Savary}
\affiliation{French American Center for Theoretical Science, CNRS, KITP, Santa Barbara, California 93106-4030, USA
}
\affiliation{Kavli Institute for Theoretical Physics, University of California, Santa Barbara, California 93106-4030, USA}

\date{\today}
\begin{abstract}
  The absence of a well-defined Fermi surface in flat-band systems challenges the conventional understanding of instabilities toward Landau order based on nesting. We investigate the existence of an intrinsic nesting structure encoded in the band geometry (i.e. the wavefunctions of the flat band(s)), which leads to a maximal susceptibility at the mean-field level and thus determines the instability towards ordered phases. More generally, we show that for a given band structure and observable, we can define two vector fields: one which corresponds to the Bloch vector of the projection operator onto the manifold of flat bands, and another which is ``dressed'' by the observable. The overlap between the two vector fields, possibly shifted by a momentum vector $\boldsymbol{Q}$, fully determines the mean field susceptibility of the corresponding order parameter. When the overlap is maximized, so is the susceptibility, and this geometrically corresponds to ``perfect nesting'' of the band {\em structure}. In that case, we show that the correlation length of this order parameter, even for $\bm{Q}\neq \bm{0}$, is entirely characterized by a generalized quantum metric in an intuitive manner, and is therefore lower-bounded in topologically non-trivial bands. As an example, we demonstrate hidden nesting for staggered antiferromagnetic spin order in an exactly flat-band model, which is notably different from the general intuition that flat bands are closely associated with ferromagnetism. We check the actual emergence of this long-range order using the determinantal quantum Monte Carlo algorithm. Additionally, we demonstrate that a Fulde-Ferrell-Larkin-Ovchinnikov-like state (pairing with non-zero center of mass momentum) can arise in flat bands upon breaking time-reversal symmetry, even if Zeeman splitting is absent.
\end{abstract}
\maketitle

{\color{Blue4}\textit{Introduction.---}}
The Fermi surface (FS) is essential for understanding conventional Fermi-liquid theory with nonzero dispersion, as only the states near the Fermi surface are relevant to low-lying excitations~\cite{AGDbook, RevModPhys.66.129}. In this scenario, potential Fermi-surface instabilities toward specific Landau orders are encoded in the band dispersion. Indeed, when the Fermi surface exhibits a nesting structure --- i.e.\ 
when portions of FS coincide upon translation by a wave vector $\boldsymbol{Q}$ within the Brillouin zone (BZ)
--- the system exhibits ``algebraic long-range crystal order'' at $\boldsymbol{Q}$ (i.e.\ a divergent susceptibility at $\boldsymbol{Q}$)~\cite{PhysRev.128.1437, 10.1063/1.1729354, peierls1955quantum, RevModPhys.60.1129, doi:10.1073/pnas.1424791112}. True long-range order can then emerge when interactions in the corresponding channels enhance this tendency~\cite{ RevModPhys.66.129}.

However, in flat-band systems~\cite{flat,doi:10.1142/S0217979215300078, PhysRevLett.62.1201, Mielke_1992, PhysRevLett.69.1608, PhysRevB.100.035448, PhysRevLett.134.076402, 2020NatCo..11.4004K, 2020NatPh..16..725B, Berg.Hofmann.2022, yang2024fractionalquantumanomaloushall,2020NatMa..19.1265A,2018NatPh..14.1125L, 2020NatMa..19..163K, Liu2020OrbitalselectiveDF, PhysRevLett.132.036001, kitamura2025quantumgeometricferromagnetismsingular, 2025SCPMA..6897211W, PhysRevLett.131.016002, Sun.Law.2024}, 
geometric frustration and destructive hopping interference lead to kinetic energy quenching.
The absence of a well-defined FS precludes conventional nesting. 
While one might expect the largest interactions to solely determine the favorable symmetry and wavevector for ordering, diverse ordered phases emerge under identical interaction conditions across different flat-band systems.
For example, in Lieb's~\cite{PhysRevLett.62.1201}, Mielke's~\cite{Mielke_1992}, and Tasaki's~\cite{PhysRevLett.69.1608} lattices, a flat band with repulsive Hubbard interactions favors ferromagnetism, as the Stoner criterion~\cite{Stoner} is naturally satisfied due to the large density of states. On the other hand, the same interactions in certain other flat bands~\cite{PhysRevB.100.155145, 2025NatCo}, including the model discussed in our work, instead stabilize antiferromagnetism (AFM).
This strongly suggests that intrinsic features governing ordering tendencies in flat-band systems are inherent to the electronic band {\em structure}, rather than the interaction, at least when the momentum dependence of the latter is smooth.

Our work identifies criteria which determine the leading instability towards long-range order in systems with narrow bands. 
Namely we show that, even in the absence of a FS, there often exists a nesting structure within the eigenstates, and that an intuitive geometric quantity built from the order parameter and the projection operator onto the flat-band states are directly related to the susceptibility towards this order. 
This ``hidden'' nesting structure replaces 
FS nesting and determines the most favorable ordering in an arbitrary number of flat bands by maximizing the corresponding order susceptibility. More precisely, {\em for any order parameter, including nonzero momentum ones}, we construct two vector fields whose overlap fully determines the corresponding mean-field susceptibility.

To prove our results, we study a flat-band model first proposed in Ref.~\cite{Berg.Hofmann.2022} and demonstrate the `hidden' nesting of AFM order. 
Additionally, breaking time-reversal symmetry yields a Fulde-Ferrell-Larkin-Ovchinnikov (FFLO)-like state,
evoking the usual FFLO phase induced by Zeeman splitting in conventional metals~\cite{PhysRev.135.A550, Larkin:1964wok, Kinnunen_2018, annurev:/content/journals/10.1146/annurev-conmatphys-031119-050711}. Both cases are further supported by numerically exact determinantal quantum Monte Carlo (DQMC)~\cite{DQMC1,DQMC2} simulations.

Beyond the tendency towards a given instability, we examine the ``stability'' of the ordered phase. 
Crucially, one must ensure band flatness does not yield a momentum-independent RPA susceptibility, likely a sign of unstable phase.
Previous studies show that some zero-momentum orders, such as flat-band superconductivity~\cite{2015NatCo...6.8944P, PhysRevB.95.024515, PhysRevB.98.220511, PhysRevB.102.201112, PhysRevLett.117.045303, PhysRevB.106.014518, PhysRevLett.128.087002, PhysRevB.106.104514, PhysRevLett.132.026002, Bernevig.Törmä.2022, PhysRevB.110.L041105, wang2025sc} and ferromagnetism~\cite{PhysRevB.102.165118, kang2024, PhysRevB.103.205415}, are stabilized by a nonzero order-parameter-stiffness, arising from the quantum metric via the restoration of momentum dependence in the susceptibility. 
Our work shows that the high-temperature stiffness 
is fully determined by the quantum metric when the nesting condition is perfectly satisfied, even for finite ordering $\boldsymbol{Q}$. 
This in turn suggests a lower bound of the stiffness in topologically nontrivial bands.


{\color{Blue4}\textit{General Formalism.---}} 
For a generic flat-band system, wavefunction information is encoded in the 
transformation $c_{\alpha}(\boldsymbol{k})=U_{\alpha m}(\boldsymbol{k}) c_m(\boldsymbol{k})$ connecting orbital ($\alpha$) and band ($m$) operators.
The band eigenprojector matrix 
is defined as $\left(\mathcal{P}_m\right)_{\alpha \beta}(\boldsymbol{k})= U_{\alpha m}(\boldsymbol{k}) U_{m \beta}^{\dagger}(\boldsymbol{k})$. 
We consider an idealized setup, where the low-energy sector consists of $N_L$ nearly degenerate and flat bands,
assumed isolated from all other bands by a large gap, with interactions small compared to this gap but large relative to the flat-band bandwidth.
The eigenprojector to the 
low-energy flat-band manifold is 
$P(\boldsymbol{k})=\sum_{m \in L} \mathcal{P}_m(\boldsymbol{k})$, where ``$L$'' denotes the low-energy sector.  
A generic particle-hole order parameter 
at $\boldsymbol{Q}$ is
$\hat O^{\mathrm{ph}}_{ \boldsymbol{Q}}=\sum_{\boldsymbol{k},\alpha \beta}  \mathcal O_{\alpha \beta}(\boldsymbol{k}) c_\alpha^{\dagger}(\boldsymbol{k}+\boldsymbol{Q}) c_\beta(\boldsymbol{k})$. The corresponding susceptibility expressed in terms of the eigenprojector matrices is (see Supplemental Material (SM), Sec.~I~\footnote{See Supplemental Material for more technical details and additional supporting data.}) \cite{savary2017}:
\begin{equation}\label{chiph_main}
    \chi^{\mathrm{ph}}_{\boldsymbol{Q}}=\frac{1}{4T}\sum_{\boldsymbol{k}} \operatorname{Tr}\left[\mathcal O^{\dagger}(\boldsymbol{k})  P(\boldsymbol{k}+\boldsymbol{Q}) \mathcal O(\boldsymbol{k})  P(\boldsymbol{k})\right].
\end{equation}
Here we assume the temperature scale (in energy units, with Boltzmann's constant set to $1$) to be larger than the bare bandwidth of the low-energy sector and larger than the critical temperatures of all candidate orders, but still smaller than the gap above the low-energy sector.
Consequently, only the low-energy sector is relevant, and its small dispersion, which is not significantly renormalized at temperatures where the order has not yet been established, can be neglected. Importantly, since thermal excitations smooth out dispersion features, the total bandwidth of the low-energy sector need not perfectly vanish, but only has to be smaller than the interaction magnitude (more details in SM, Sec.~II.D~\cite{Note1}).

In the following, we employ the vectorial representation of the Hamiltonian's eigenprojectors for a more intuitive understanding of \eqnref{chiph_main}.
By analogy with a two-level system, where a $2\times 2$ eigenprojector is expressed via a unit Bloch vector on the $S^2$ Bloch sphere, the low-energy eigenprojector $P(\boldsymbol{k})$ in a generic $N$-band system ($N \geq 2$) can be mapped one-to-one to a $2(N-1)$-dimensional~\footnote{Due to the orthogonality relation 
 $\mathcal{P}_m(\boldsymbol{k}) \mathcal{P}_n(\boldsymbol{k})=\delta_{m n} \mathcal{P}_m(\boldsymbol{k})$, the target space of $\boldsymbol{b}(\boldsymbol{k})$ is not an $(N^2-2)$-sphere but rather a specific $2(N-1)$-dimensional subset thereof~\cite{PhysRevB.104.085114}.} generalized Bloch vector $\boldsymbol{b}(\boldsymbol{k})$  as follows~\cite{PhysRevB.104.085114}:
\begin{equation}\label{eq:pvec}
    P(\boldsymbol{k})=\frac{N_L}{N} \mathds{1}_N+\frac{1}{2} \boldsymbol{b}(\boldsymbol{k}) \cdot \boldsymbol{\lambda},
\end{equation}
where $\boldsymbol{\lambda}$ denotes the elementary generator matrices of the SU(N) Lie group ~\footnote{Specifically, for $N=2$ (resp.\ $N=3$) the $\boldsymbol{\lambda}$ matrices can be taken to be the Pauli matrices (resp.\ the Gell-Mann matrices~\cite{ PhysRev.125.1067}). }.

An arbitrary order operator can be similarly expressed as $\mathcal O(\boldsymbol{k})=o_0(\boldsymbol{k}) \mathds{1}_N+\boldsymbol{o}(\boldsymbol{k}) \cdot \boldsymbol{\lambda}$, which, together with 
Eq.~(\ref{eq:pvec}),  allows a full vectorial rewriting of Eq.~\eqref{chiph_main} (see Eq.~\eqref{mid}). Here we focus on cases where either $o_0(\boldsymbol{k})$ or $\boldsymbol{o}(\boldsymbol{k})$ vanishes, as in many realistic physical scenarios. In both cases, the susceptibility of $\hat{O}$ can be rewritten as (see SM, Sec.~I~\cite{Note1}):
\begin{align}\label{chiphmain}
\chi^{\mathrm{ph}}(\boldsymbol{Q})=\sum_{\boldsymbol{k}}\frac{\operatorname{Tr}[\mathcal O^\dagger(\boldsymbol{k})\mathcal O(\boldsymbol{k})]}{4NT}\left(\frac{N_L^2}{N}+\frac{1}{2} \zeta_{\boldsymbol{o}, \boldsymbol{Q}}(\boldsymbol{k})\right),
\end{align}
where the overlap $\zeta_{\boldsymbol{o}, \boldsymbol{Q}}(\boldsymbol{k})= \tilde{\boldsymbol{b}}_{\boldsymbol{o}}(\boldsymbol{k}+\boldsymbol{Q}) \cdot \boldsymbol{b}(\boldsymbol{k})$
entirely parametrizes $\chi^{\mathrm{ph}}(\boldsymbol{Q})$, and
$\tilde{\boldsymbol{b}}_{\boldsymbol{o}} (\boldsymbol{k}+\boldsymbol{Q})$ is a `dressed' Bloch vector, 
i.e., $\boldsymbol{b}$ `corrected' by the order parameter:
\begin{eqnarray}\label{QGN}
    \tilde{\boldsymbol{b}}_{\boldsymbol{o}}(\boldsymbol{k}+\boldsymbol{Q}) \equiv \boldsymbol{b}(\boldsymbol{k}+\boldsymbol{Q})-N\left[\hat{\boldsymbol{o}}(\boldsymbol{k}) \times \boldsymbol{b}(\boldsymbol{k}+\boldsymbol{Q}) \times \hat{\boldsymbol{o}}(\boldsymbol{k})\right],
\end{eqnarray}
where $\hat{\boldsymbol{o}}(\boldsymbol{k})$ is the unit vector of the order parameter, i.e.,  
$\hat{\boldsymbol{o}}(\boldsymbol{k}) = \boldsymbol{o}(\boldsymbol{k})/|\boldsymbol{o}(\boldsymbol{k})|$ for $\boldsymbol{o}(\boldsymbol{k})\neq\boldsymbol{0}$ and there is no `correction' when only $o_0\neq0$ (note that this latter case corresponds to a density order parameter).
Here the cross product is defined as $(\boldsymbol{m} \times \boldsymbol{n})_i=f_{i j k} m_j n_k$, where $f_{i j k} \equiv-\frac{i}{4} \operatorname{Tr}\left(\left[\boldsymbol{\lambda}_i, \boldsymbol{\lambda}_j\right] \boldsymbol{\lambda}_k\right)$ represents the antisymmetric structure constants of the $\mathfrak{s u}(N)$ Lie algebra~\cite{10.1063/1.1705141}.  Moreover, the amplitude of the dressed Bloch vector $\tilde{\boldsymbol{b}}_{\boldsymbol{o}}(\boldsymbol{k})$ cannot exceed that of $\boldsymbol{b}(\boldsymbol{k})$, implying that the susceptibility of a given particle-hole excitation $\hat{O}$ in \eqnref{chiphmain} reaches its maximum when all Bloch vectors $\boldsymbol{b}(\boldsymbol{k})$ across the Brillouin zone remain parallel to the dressed Bloch vector $\tilde{\boldsymbol{b}}_{\boldsymbol{o}}(\boldsymbol{k}+\boldsymbol{Q})$ with the same magnitude:
\begin{eqnarray}\label{QGN}
    \tilde{\boldsymbol{b}}_{\boldsymbol{o}}(\boldsymbol{k}+\boldsymbol{Q}) \;\; \parallel\;\; \boldsymbol{b}(\boldsymbol{k}),\;\;\;\;\;\;\;\;\;\; \forall \boldsymbol{k} \in \mathrm{BZ}.
\end{eqnarray}
The condition in \eqnref{QGN} implies that any region in the BZ has a ``compatible'' partner separated by $\boldsymbol{Q}$. This is analogous to ``nesting'' in conventional Fermi liquids. 


If a given order satisfies this nesting condition, its bare susceptibility can reach its theoretical maximum. 
With interactions in the same channel, the instability is further enhanced at the RPA level, leading to the corresponding susceptibility diverging at a finite temperature $T_c$ (see SM, Sec. II ~\cite{Note1}). Thus, among competing orders in flat-band systems, the one fulfilling the nesting condition attains the highest $T_c$.

Similarly, a generic particle-particle order with $\boldsymbol{Q}$ is $\hat O^{\mathrm{pp}}_{\boldsymbol{Q}}=\sum_{\boldsymbol{k}}  \mathcal O_{\alpha \beta}(\boldsymbol{k}) c_\alpha(\boldsymbol{k}+\boldsymbol{Q}) c_\beta(\boldsymbol{k})$.
Compared to the particle-hole excitation case, an additional definition is required here: the complex conjugate of the projection operator, given by  $P^*(\boldsymbol{k})=\frac{N_L}{N} \mathds{1}_N+\frac{1}{2} \boldsymbol{b}^R(\boldsymbol{k}) \cdot \boldsymbol{\lambda}$ where $\boldsymbol{b}^R(\boldsymbol{k})$ is defined by  
 $(\boldsymbol{b}(\boldsymbol{k}) \cdot \boldsymbol \lambda)^*=\boldsymbol{b}^R(\boldsymbol{k}) \cdot \boldsymbol \lambda$. Following this procedure, it becomes apparent that the condition for maximizing particle-particle fluctuations, i.e., the nesting condition, is~\cite{Note1}:
\begin{eqnarray}\label{nest_pp}
\tilde{\boldsymbol{b}}^R_{\boldsymbol{o}}(\boldsymbol{k}+\boldsymbol{Q}) \;\; \parallel\;\; \boldsymbol{b}(-\boldsymbol{k}),\;\;\;\;\;\;\;\;\;\; \forall \boldsymbol{k} \in \mathrm{BZ}.
\end{eqnarray}
The perfect nesting condition derived in \eqnsref{QGN} and (\ref{nest_pp}) can be shown to be consistent with the quantum geometric nesting formulation presented in Ref.~\cite{PhysRevX.14.041004}.

{\color{Blue4}\textit{Generic Quantum Metric and Correlation Length.---}} We now examine the coherence length for $T > T_c$, which corresponds to the `high-temperature order-parameter stiffness' and is associated with order stability.
Expanding the susceptibility of a given particle-hole order, \eqnref{chiphmain}, around the ordering $\boldsymbol{Q}$ by a small momentum $\boldsymbol{q}$ gives $\chi^{\mathrm{ph}}_{\boldsymbol{Q}+\boldsymbol{q}}\sim\chi^{\mathrm{ph}}_{\boldsymbol{Q}}-\frac{1}{T}\sum_{\boldsymbol{k}} g_{\mu \nu}(\boldsymbol{Q},\boldsymbol{k})\boldsymbol{q}_\mu \boldsymbol{q}_\nu$, where spatial fluctuations are controlled by:
\begin{equation}\label{sti}
    g_{\mu \nu}(\boldsymbol{Q},\boldsymbol{k})=\frac{1}{4}\partial_\mu \boldsymbol{b}(\boldsymbol{k})\cdot \partial_\nu \tilde{\boldsymbol{b}}_{\boldsymbol{o}}(\boldsymbol{k}+\boldsymbol{Q}).
\end{equation}

When the ordering vector is $\boldsymbol{Q} = \boldsymbol{0}$, and the order parameter takes the simple form $\mathcal{O}(\boldsymbol{k}) = o_0(\boldsymbol{k}) \mathds{1}_N$, i.e.\ a simple zero-momentum density, \eqnref{sti} simplifies to 
\begin{equation}\label{metric}
    g_{\mu \nu}(\boldsymbol{k})=\frac{1}{4}\partial_\mu \boldsymbol{b}(\boldsymbol{k})\cdot \partial_\nu \boldsymbol{b}(\boldsymbol{k}),
\end{equation}
which equals the quantum metric~\cite{metric}.
Based on Ginzburg-Landau theory, the correlation length~\cite{Coleman_2015}, i.e., the characteristic length scale beyond which the internal structure of the order parameter becomes irrelevant, can be extracted from the spatial fluctuation term in $\chi_{\boldsymbol{Q}+\boldsymbol{q}}$, so that (see SM, Sec. II~\cite{Note1})
\begin{equation}\label{col} \xi=\frac{(\operatorname{det} \bar{g}_{\mu\nu})^{1 / 4}}{\sqrt{ N_L}} f(T/T_c),
\end{equation}
where $\bar{g}_{\mu \nu} \equiv \sum_k g_{\mu \nu}(\boldsymbol{k}) / \mathcal{N}$ is the quantum metric averaged over the BZ, with $\mathcal{N}$ denoting the total number of sites. At mean-field level, $f(T/T_c)=\left|1-T/T_c\right|^{-1/2}$. This is consistent with recent reports~\cite{PhysRevLett.132.026002, 10.1038/s42005-024-01930-0} of the coherence length in flat-band superconductors~\footnote{Given that $\boldsymbol{b}_{\uparrow}^R(-\boldsymbol{k}) = \boldsymbol{b}_{\downarrow}(\boldsymbol{k})$ in systems with time-reversal symmetry, the expansion of the uniform pairing susceptibility $\chi^\Delta(\boldsymbol{q})$ with respect to small $\boldsymbol{q}$ reduces to \eqnref{metric}.}.

\begin{figure}[t]
    \centering
    \includegraphics[width=0.75\linewidth]{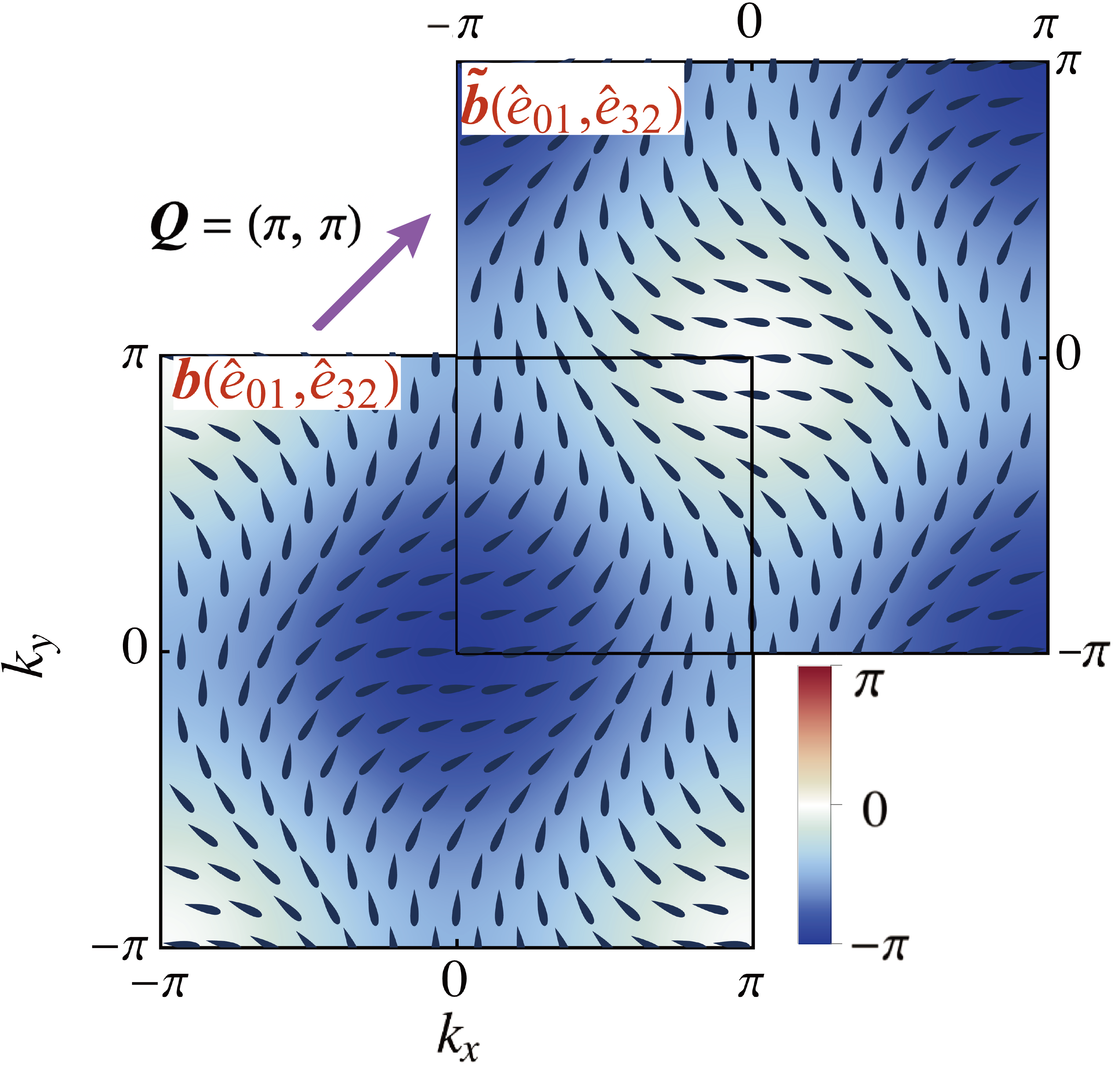}
    \caption{The Bloch vector $\boldsymbol{b}$ in left-bottom and the dressed Bloch vector $\tilde{\boldsymbol{b}}_{\boldsymbol{N}}$ in right-top for the staggered spin order $\boldsymbol{N}_i^x$. The drop arrows indicate the direction of the (dressed) Bloch vector, while the background color represents the angle of the (dressed) Bloch vector, i.e., $\mathrm{Arg}(\boldsymbol{b})=\mathrm{arctan}(\frac{\boldsymbol{b}\cdot \hat{e}_{01}}{\boldsymbol{b}\cdot \hat{e}_{32}})$. A perfect nesting exists between $\boldsymbol{b}$ and $\tilde{\boldsymbol{b}}_{\boldsymbol{N}}$ at $\boldsymbol{Q} = (\pi, \pi)$. With parameter: $\eta=0.75$.}
    \label{fig:spin}
\end{figure}

However, for a generic finite-$\boldsymbol{Q}$ order, interpreting \eqnref{sti} as a quantum metric is no longer valid because the reference axes $\boldsymbol{b}(\boldsymbol{k})$ and $\tilde{\boldsymbol{b}}_{\boldsymbol{o}}(\boldsymbol{k}+\boldsymbol{Q})$ can differ significantly. Consequently, 
$g_{\mu \nu}(\boldsymbol{Q},\boldsymbol{k})$ in \eqnref{sti} is no longer local and not necessarily non-negative, undermining its interpretation as a metric.
Nonetheless, 
when the nesting condition \eqnref{QGN} is satisfied, the reference axes $\tilde{\boldsymbol{b}}_{\boldsymbol{o}}(\boldsymbol{k}+\boldsymbol{Q})$ and $\boldsymbol{b}(\boldsymbol{k})$ are exactly aligned, leading to the identification  
$\tilde{\boldsymbol{b}}_{\boldsymbol{o}}(\boldsymbol{k}+\boldsymbol{Q}) \sim \boldsymbol{b}(\boldsymbol{k})$. This alignment implies that the stiffness in \eqnref{sti} retains the same interpretation as in \eqnref{metric}. It may therefore be considered a generalized quantum metric and is directly related to the correlation length, \eqnref{col}, of the order parameter. Moreover, given the bounds established in Refs.~\cite{2015NatCo...6.8944P,PhysRevB.104.045103, PhysRevB.90.165139, PhysRevB.95.024515}, the correlation length in \eqnref{col} will always remain finite when the low-energy sectors are topologically nontrivial, i.e., it is lower-bounded by the total Chern number~\cite{Note1, PhysRevB.109.214518}, $\xi\geq \sqrt{|\sum_{m \in L }\mathcal{C}_m|/4\pi N_L}f(T/T_c)$, where $\mathcal{C}_m$ is the Chern number of band $m$. 
Note that the lower bound is determined by the total rather than the summed absolute Chern number (see SM, Sec.~II.C \cite{Note1}),
due to interband components of the quantum metric in composite bands with \( N_L \neq 1 \) ~\cite{PhysRevB.109.214518}.

{\color{Blue4}\textit{Staggered Spin Order in the Flat Band.---}} In the following, we present several examples to illustrate our method to determine leading instabilities and the geometric nesting hidden in flat-band systems. We begin with a two-orbital spinful electronic  model (with orbitals labeled $A$ and $B$)~\cite{Berg.Hofmann.2022,PhysRevLett.130.226001}, with the non-interacting Hamiltonian $H_0= \sum_k \psi_k^{\dagger}h_{\boldsymbol{k}}\psi_k$, where the basis is defined as  
 $\psi_k=\left[c_{A\uparrow}(\boldsymbol{k})\;\;c_{B\uparrow}(\boldsymbol{k})\;\;c_{A\downarrow}(\boldsymbol{k})\;\;c_{B\downarrow}(\boldsymbol{k})\right]^T$ and the Hamiltonian matrix is:
\begin{eqnarray}\label{hk}
h_{\boldsymbol{k}}=t\left(\begin{array}{cccc}
-\mu & -i e^{i\alpha_k^{\uparrow}} & 0 & 0 \\
i e^{-i \alpha_k^{\uparrow}} & -\mu & 0 & 0 \\
0 & 0 & -\mu & ie^{-i \alpha_k^{\downarrow }} \\
0 & 0 & -ie^{i \alpha_k^{\downarrow }} & -\mu
\end{array}\right)
\end{eqnarray}
where $\alpha_k^\sigma=\eta_{\sigma}\left(\cos k_x+\cos k_y\right)$, and $\eta_{\sigma}$ controls the locality of the Wannier wave function for spin $\sigma$ \cite{souza2000,Vanderbilt_2018}. Time-reversal symmetry is explicitly broken when $\eta_\sigma$ takes different values for opposite spins.
This free model has two pairs of perfectly flat bands (with $N=4$) at energies $t(\pm1-\mu)$, independent of the parameters $\eta_{\sigma}$. For filling number $\nu < 2$, all electrons remain in the lower band sector (with $N_L=2$), and the energy gap between the two sectors is $2t$.

Considering a local repulsive interaction given by  $H_V=U\sum_{i}\left(n_{iA}^2+n_{iB}^2\right)$, where $U>0$, and $n_{iA/B}$ denotes the total electron number at orbital $A/B$, various interaction channels can arise. Two important order parameters are the total spin order   $\boldsymbol{M}_i=\boldsymbol{S}_{iA}+\boldsymbol{S}_{i B}$ and the staggered spin order  
 $\boldsymbol{N}_i=\boldsymbol{S}_{iA}-\boldsymbol{S}_{i B}$, where $\boldsymbol{S}_{iA/B}$ represents the spin operator at orbital $A/B$. 
 
When 
$\alpha_k^\uparrow = \alpha_k^\downarrow=\alpha_k$, it is straightforward to check that $\boldsymbol{M}_i$ does not satisfy the nesting condition~\cite{Note1}, whereas the AFM $\boldsymbol{N}_i$ in the $x$-$y$ plane does. More specifically, in the basis $\boldsymbol{\lambda}_{\alpha\beta}=\boldsymbol{\sigma}_\alpha \otimes \boldsymbol{\tau}_\beta /\sqrt{2} $, where $\sigma$ and $\tau$ are Pauli matrices for spin and orbital degrees of freedom, the generalized Bloch vector of \eqnref{hk} is $\boldsymbol{b}(\boldsymbol{k})=-\sqrt{2}\sin \alpha_k \hat{e}_{01}-\sqrt{2}\cos \alpha_k \hat{e}_{32}$, where $\hat e_{\alpha\beta}$ is the $N$-dimensional unit vector associated to $\boldsymbol{\lambda}_{\alpha\beta}$. Since the unit vector of the staggered spin order along the $x$-direction is $\boldsymbol{N}_i^x=\hat e_{13}$, the corresponding dressed Bloch vector is  
$\tilde{\boldsymbol{b}}_{\boldsymbol{N}}(\boldsymbol{k})=\sqrt{2}\sin \alpha_k \hat{e}_{01}-\sqrt{2}\cos \alpha_k \hat{e}_{32}$ (see SM, Sec. III~\cite{Note1}). The patterns of  $\boldsymbol{b}(\boldsymbol{k})$ and $\tilde{\boldsymbol{b}}_{\boldsymbol{N}}(\boldsymbol{k})$ are shown in \figref{fig:spin}, demonstrating perfect nesting with $\boldsymbol{Q}=(\pi,\pi)$, i.e., $\boldsymbol{b}(\boldsymbol{k})$ is completely `compatible' with $\tilde{\boldsymbol{b}}_{\boldsymbol{N}}(\boldsymbol{k})$ upon shifting by a momentum $(\pi,\pi)$.

\begin{figure}[t]
    \centering
    \includegraphics[width=1\linewidth]{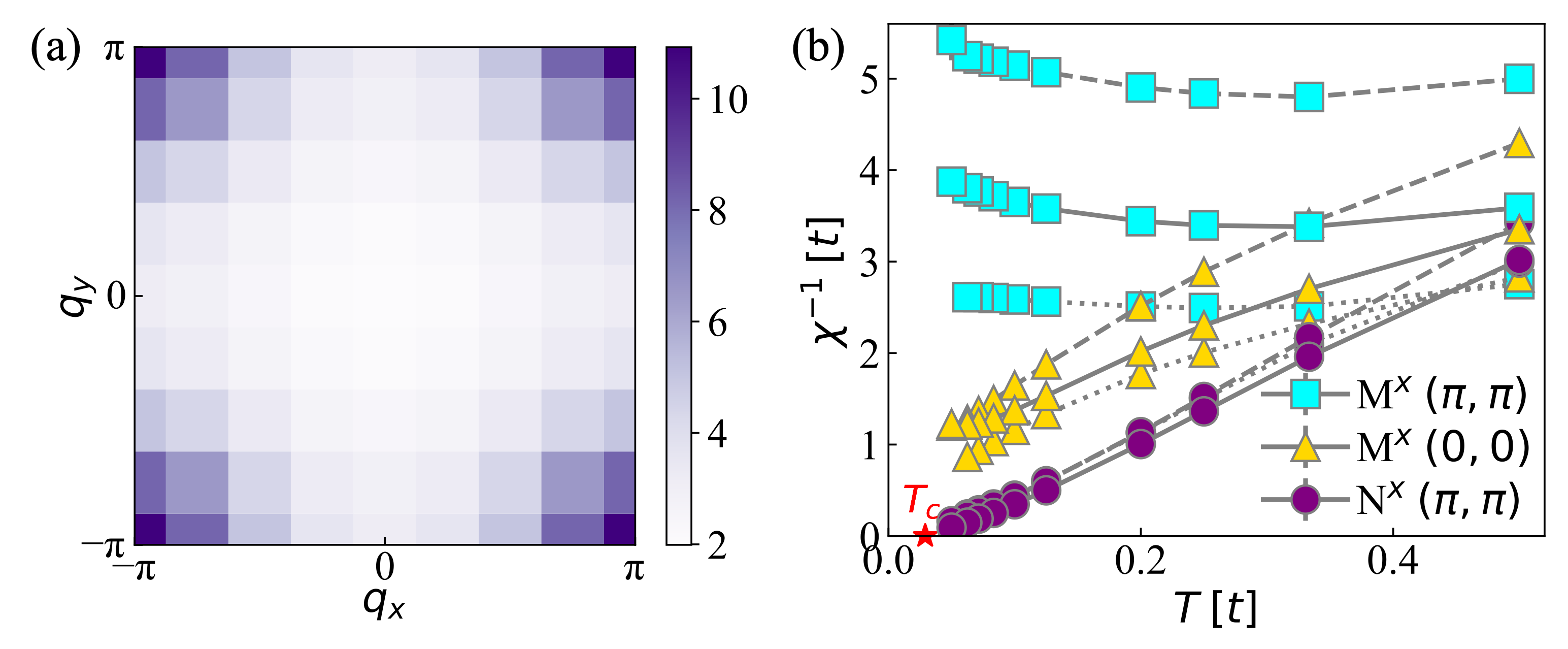}
    \caption{
    Momentum distribution of the $\boldsymbol{N}^x$ spin susceptibility from DQMC with $\eta_\sigma = 0.75$ for $\nu=1$ at $T/t=0.05$.
    (b) Temperature dependence of the inverse $\boldsymbol{M}^x$ spin susceptibility at $\mathbf{Q}=(0,0)$ and $(\pi,\pi)$, and the inverse $\boldsymbol{N}^x$ susceptibility at $(\pi,\pi)$.
    Results are shown for three filling numbers: $\nu=0.75$ (dashed), $\nu=1$ (solid), and $\nu=1.25$ (dotted). The star marks the extrapolated critical temperature $T_c$ for $\nu=1$.}
    \label{fig:QMCspin}
\end{figure}

To provide further evidence, we perform DQMC simulations 
on an $8\times 8$ square lattice with periodic boundary conditions, fixing $|U|/t=1$ for all considered cases (see SM~\cite{Note1} Sec.~IV). 
In \figref{fig:QMCspin}~(a),
the susceptibility for the $\boldsymbol{N}^x$ order peaks at $\mathbf{Q}=(\pi,\pi)$.
We demonstrate 
a divergent susceptibility for the AFM in-plane spin order at a finite 
$T_c$ through 
temperature scaling 
in \figref{fig:QMCspin}~(b),
which confirms the AFM forms as an actual long-range order across 
fillings. These results are consistent with our aforementioned theoretical predictions, 
and are markedly different from the ferromagnetism often observed in flat-band systems~\cite {PhysRevLett.62.1201,Mielke_1992,PhysRevLett.69.1608, 10.1143/PTP.99.489, PhysRevB.93.144418, 2018NatPh..14.1125L}.

Furthermore, the average quantum metric in this case is $\bar{g}_{\mu \nu}=\mathds{1}_2\eta^2/4$. Consequently, according to \eqnref{col}, the correlation length for the $\boldsymbol{N}^x$ order is given by $\frac{|\eta|}{2\sqrt{2}}\left|1-\frac{T}{T_c}\right|^{-1/2}$. The linear dependence of the correlation length on $\eta$ is also verified numerically (see \figref{fig:corr_len} in SM, Sec. IV \cite{Note1}).

{\color{Blue4}\textit{FFLO in a Time-reversal Breaking Model.---}} Next, we examine particle-particle excitations in the presence of a local attractive interaction, i.e., $U<0$ in $H_V$. Given the spin $U(1)$ rotational symmetry, the projection matrix can be decomposed into distinct spin sectors as   $P(\boldsymbol{k})=P_\uparrow(\boldsymbol{k})\oplus P_\downarrow(\boldsymbol{k})$, where $P_{\sigma}(\boldsymbol{k})=\frac{N_L}{N} \mathds{1}_{N/2}+\frac{1}{2} \boldsymbol{b}_{\sigma}(\boldsymbol{k}) \cdot \boldsymbol{\lambda}$ involves only the states with spin $\sigma$. The intra-orbital singlet pairing order parameter at momentum $\boldsymbol{Q}$ is given by~\cite{Note1}  $\Delta_{\boldsymbol{Q}}=\sum_{\boldsymbol{k}} \left[c_{A \uparrow}(\boldsymbol{k}+\boldsymbol{Q}) c_{A \downarrow}(-\boldsymbol{k})+c_{B \uparrow}(\boldsymbol{k}+\boldsymbol{Q}) c_{B \downarrow}(-\boldsymbol{k})\right]$, with the corresponding susceptibility $\chi^\Delta (\boldsymbol{Q})\sim \sum_{\boldsymbol{k}}\boldsymbol{b}_{\uparrow}^R(\boldsymbol{k}+\boldsymbol{Q}) \cdot \boldsymbol{b}_{\downarrow}(-\boldsymbol{k})$, indicating that the pairing susceptibility is enhanced when the Bloch vector $\boldsymbol{b}_{\uparrow}^R$ at $\boldsymbol{k}+\boldsymbol{Q}$ is compatible (parallel) with $\boldsymbol{b}_{\downarrow}$ at $-\boldsymbol{k}$.

\begin{figure}[t]
    \centering
    \includegraphics[width=1\linewidth]{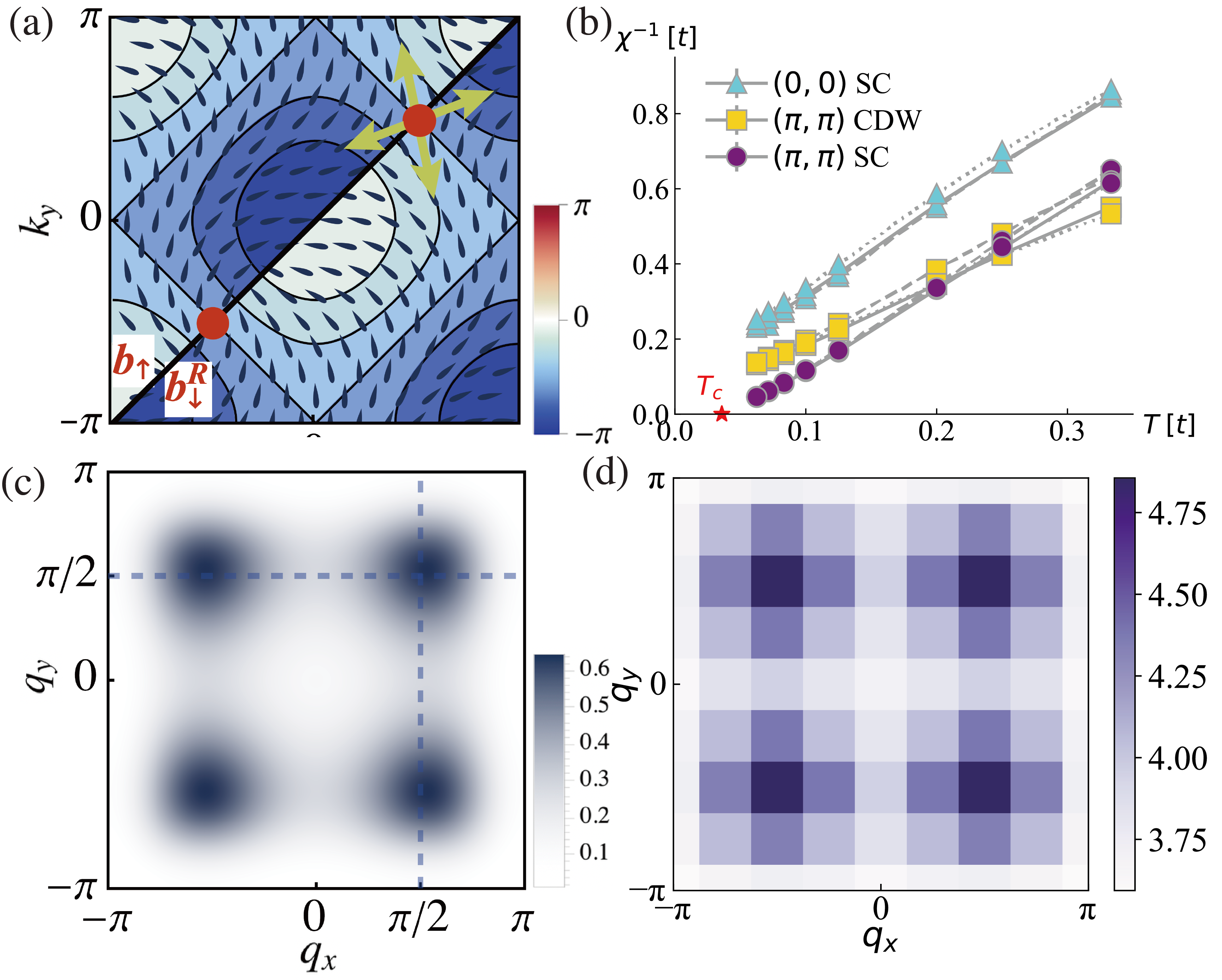}
    \caption{
    (a,b) $(\pi,\pi)$ SC with $\eta_{\sigma}=0.75 \sigma$ and (c,d) $(\pi/2,\pi/2)$ SC with $\eta_\sigma = 1.25+2.5\sigma $.
    (a) The Bloch vector $\boldsymbol{b}_{\uparrow}$ (top-left) and $\boldsymbol{b}^R_{\downarrow}$ (bottom-right) for the intra-orbital singlet pairing order. The drop arrows indicate the direction of the Bloch vector, while the background color represents their angle.
    The yellow arrow serves as a eye guide for the parallel Bloch vectors at the center of momentum $(\pm\pi/2, \pm\pi/2)$ (red points). (b)
   DQMC results for temperature dependence of the inverse SC susceptibilities at $\mathbf{Q}=(0,0)$ and $(\pi,\pi)$ and the inverse CDW susceptibility at $\mathbf{Q}=(\pi,\pi)$. Dashed, solid, and dotted lines represent $\nu=0.75$, $1$, and $1.25$, respectively, and the star marks the extrapolated $T_c$ for $\nu=1$ for $(\pi,\pi)$ SC. 
   (c) The momentum distribution of $\sum_{\boldsymbol{k}} \boldsymbol{b}_{\uparrow}^R(\boldsymbol{k}+\boldsymbol{q}) \cdot \boldsymbol{b}_{\downarrow}(-\boldsymbol{k})$ (d) DQMC result of the SC susceptibility at $\nu=1.25$ and $T/t=0.0625$. 
    }
    \label{fig:sc}
\end{figure}

Given the Hamiltonian in \eqnref{hk} with general $\eta_{\sigma}$, the Bloch vectors are~\cite{Note1} $\boldsymbol{b}_{\sigma}(\boldsymbol{k})=- \sin \alpha_k^{\sigma} \hat{e}_1-\sigma \cos \alpha_k^{\sigma} \hat{e}_2$ and $\boldsymbol{b}^R_{\sigma}(\boldsymbol{k})=- \sin \alpha_k^{\sigma} \hat{e}_1+\sigma \cos \alpha_k^{\sigma} \hat{e}_2$, where $\alpha$ in $\hat{e}_{\alpha}$ corresponds to the orbital index of the Pauli matrix $\boldsymbol{\tau}_\alpha$. The patterns of $\boldsymbol{b}$ and $\boldsymbol{b}^R$ 
in \figref{fig:sc}~(a) for the case with $\eta_{\sigma}=0.75\sigma$, which 
illustrate that for any Bloch vector $\boldsymbol{b}_{\uparrow}$ at an arbitrary momentum $\boldsymbol{Q}/2+\boldsymbol{k}$, there exists a compatible $\boldsymbol{b}_{\downarrow}^R$ at $\boldsymbol{Q}/2-\boldsymbol{k}$, with $\boldsymbol{Q}=(\pi,\pi)$, indicating a perfect nesting structure for superconductivity (SC) with a center-of-mass momentum of $(\pi,\pi)$. Using the same parameters, from DQMC, the SC susceptibility peaks at $(\pi,\pi)$ (see \figref{fig:supp_q_scaling} (a) in SM, Sec. IV~\cite{Note1}). The susceptibility temperature evolution in Fig.~\ref{fig:sc} (b) indicates that the $(\pi,\pi)$ SC exhibits significantly stronger fluctuations than both the uniform SC and the $(\pi,\pi)$ charge density wave, consistent with the fact that the latter two do not satisfy the perfect nesting condition.

Moreover, independently tuning the magnitude of $\eta_{\sigma}$ further for opposite spins can induce an independent momentum shift in the Bloch vector patterns of
$\boldsymbol{b}$ and $\boldsymbol{b}^R$ (see \figref{fig:supp} in SM \cite{Note1}). While perfect nesting for the pairing order $\Delta_{\boldsymbol{Q}}$ no longer holds for any center-of-mass momentum $\boldsymbol{Q}$, the resulting shifted nesting scenario resembles that of conventional metals under Zeeman splitting, where an energy shift between opposite-spin Fermi surfaces disrupts the original pairing condition but may favor an FFLO state~\cite{PhysRev.135.A550, Larkin:1964wok, Kinnunen_2018, annurev:/content/journals/10.1146/annurev-conmatphys-031119-050711}. 

More specifically, even though perfect nesting is absent, the degree of nesting for different momentum shifts $\boldsymbol{Q}$ can still be quantified using $\sum_{\boldsymbol{k}}\zeta_{\Delta, \boldsymbol{Q}}(\boldsymbol{k})=\sum_{\boldsymbol{k}}\boldsymbol{b}_{\uparrow}^R(\boldsymbol{k}+\boldsymbol{Q}) \cdot \boldsymbol{b}_{\downarrow}(-\boldsymbol{k})$. With parameters $\eta_{\uparrow}=3.75$ and $\eta_{\downarrow}=-1.25$, this `nesting extent' for pairings with different center-of-mass momenta $\boldsymbol{Q}$ is shown in \figref{fig:sc}~(c), and exhibits a peak near $(\pi/2, \pi/2)$. The consequences of this result are further corroborated by DQMC simulations, which reveal an enhanced pairing susceptibility at $(\pi/2, \pi/2)$, shown in \figref{fig:sc}~(d). Although additional numerical results (see \figref{fig:supp_q_scaling}~(b) in SM, Sec. IV~\cite{Note1}) cannot definitively determine the true ground-state order due to the presence of competing orders, the exotic FFLO state at $(\pi/2,\pi/2)$  consistently exhibits greater coherence than uniform SC, making it a strong low-lying energy candidate. Some previous works have also shown that band geometry can drive finite-momentum pairing~\cite{Chen2022PairDW, 2025SCPMA..6897211W, PhysRevLett.131.016002, Sun.Law.2024}. Within our generalized Bloch-vector framework, we provide a general and intuitive geometric picture of how such finite-$Q$ pairing tendencies naturally emerge in flat bands from the underlying Bloch-state structure.

{\color{Blue4}\textit{Discussion.---}}Our work demonstrates that, within the Bloch vector representation, an intrinsic geometric structure is encoded in the band structure of flat-band systems. 
This underlying structure fundamentally influences the favored order in flat-band systems, even though such systems lack a well-defined Fermi surface. Our results also provide a roadmap for the search for exotic topological phases, such as fractional Chern insulators (FCI)~\cite{PhysRevX.1.021014, PhysRevB.85.075116, PhysRevLett.106.236804, PhysRevLett.106.236802, PhysRevLett.106.236803, PhysRevLett.111.126802}, by identifying parameter regimes that circumvent strong instabilities toward competing ordered states --- such as charge density waves --- which are usually proximate to the FCI~\cite{ yang2024fractionalquantumanomaloushall,PhysRevB.103.125406,PhysRevB.109.085143, PhysRevB.109.115116}.

Importantly, our nesting-based formalism, grounded in band geometry at high temperatures before any order develops, identifies the instability channel with the highest potential $T_c$, and naturally interfaces with finite-temperature numerical methods such as DQMC, which sensitively capture thermal fluctuations and dynamics. It thus provides a finite-$T$ counterpart to the conventional Hartree–Fock and DMRG framework that start from ground states of flat-band systems. 


Recent studies have linked the local quantum metric~\cite{metric} to experimentally relevant zero-momentum quantities, 
such as the stiffness for SC~\cite{2015NatCo...6.8944P, PhysRevB.95.024515, PhysRevB.98.220511, PhysRevB.102.201112, PhysRevLett.117.045303, PhysRevB.106.014518, PhysRevLett.128.087002, PhysRevB.106.104514, PhysRevLett.132.026002, Bernevig.Törmä.2022} / FM~\cite{PhysRevB.102.165118, kang2024, PhysRevB.103.205415}, and electromagnetic responses~\cite{PhysRevB.105.085154, verma2024quantummetricstepresponse, PhysRevLett.96.137601, Onishi_2024, PhysRevB.77.054438, PhysRevB.87.245103, PhysRev.109.741}.  Here, we show that a generalized quantum metric, which reduces to the usual quantum metric in the case of perfect nesting, plays this role for any order, including nonzero momentum ones, as it is directly proportional to the correlation length. This provides a framework to relate geometric quantities to physical observables at nonzero momentum.

\begin{acknowledgements}
{\color{Blue4}\textit{Acknowledgements.---}}J.-X.Z.\ and W.O.W.\ acknowledge stimulating discussions with Junkai Dong, Edwin W.\ Huang, Zhaoyu Han, Xuepeng Wang, Zhouquan Wan, Chuyi Tuo, Hiroaki Ishizuka,
and Shuai A.\ Chen. 
J.-X.Z.\ and L.S.\ were funded by the European Research Council (ERC) under the European Union’s Horizon 2020 research and innovation program (Grant Agreement No.\ 853116, acronym TRANSPORT). 
W.O.W.\ acknowledges support from the Gordon and Betty Moore Foundation through Grant GBMF8690 to the University of California, Santa Barbara, to the Kavli Institute for Theoretical Physics (KITP).
This research was supported in part by grant NSF PHY-2309135 to the Kavli Institute for Theoretical Physics (KITP). 
L.B.\ acknowledges support by the DOE through grant No. DE-SC0020305 for theoretical analysis of superconductivity, and by the Simons Collaboration on Ultra-Quantum Matter, which is a grant from the Simons Foundation (Grant No. 651440) for theory of magnetism and guidance of numerics.
Use was made of computational facilities purchased with funds from the National Science Foundation (CNS-1725797) and administered by the Center for Scientific Computing (CSC). The C.S.C. is supported by the California NanoSystems Institute and the Materials Research Science and Engineering Center (MRSEC; NSF DMR 2308708) at UC Santa Barbara.
\\
{\color{Blue4}\textit{Code and Data Availability.}} The DQMC code and data needed to reproduce the figures can be found at \url{https://doi.org/10.5281/zenodo.15128407}.
\\
{\color{Blue4}\textit{Author Contributions.}} J.-X.Z.\ and L.S.\ conceived the project. L.S., L.B., and J.-X.Z.\ guided the theoretical analysis, and J.-X.Z.\ performed the bulk of the theoretical calculations.  W.O.W.\ carried out the numerical simulations. All authors contributed to the discussion and writing of the manuscript.
\end{acknowledgements}
\bibliographystyle{apsreve}
\bibliography{main}

\begin{thebibliography}{89}%
\makeatletter
\providecommand \@ifxundefined [1]{%
 \@ifx{#1\undefined}
}%
\providecommand \@ifnum [1]{%
 \ifnum #1\expandafter \@firstoftwo
 \else \expandafter \@secondoftwo
 \fi
}%
\providecommand \@ifx [1]{%
 \ifx #1\expandafter \@firstoftwo
 \else \expandafter \@secondoftwo
 \fi
}%
\providecommand \natexlab [1]{#1}%
\providecommand \bibnamefont  [1]{#1}%
\providecommand \bibfnamefont [1]{#1}%
\providecommand \citenamefont [1]{#1}%
\providecommand \href@noop [0]{\@secondoftwo}%
\providecommand \href [0]{\begingroup \@sanitize@url \@href}%
\providecommand \@href[1]{\@@startlink{#1}\@@href}%
\providecommand \@@href[1]{\endgroup#1\@@endlink}%
\providecommand \@sanitize@url [0]{\catcode `\\12\catcode `\$12\catcode `\&12\catcode `\#12\catcode `\^12\catcode `\_12\catcode `\%12\relax}%
\providecommand \@@startlink[1]{}%
\providecommand \@@endlink[0]{}%
\providecommand \url  [0]{\begingroup\@sanitize@url \@url }%
\providecommand \@url [1]{\endgroup\@href {#1}{\urlprefix }}%
\providecommand \urlprefix  [0]{URL }%
\providecommand \Eprint [0]{\href }%
\providecommand \doibase [0]{http://dx.doi.org/}%
\providecommand \selectlanguage [0]{\@gobble}%
\providecommand \bibinfo  [0]{\@secondoftwo}%
\providecommand \bibfield  [0]{\@secondoftwo}%
\providecommand \translation [1]{[#1]}%
\providecommand \BibitemOpen [0]{}%
\providecommand \bibitemStop [0]{}%
\providecommand \bibitemNoStop [0]{.\EOS\space}%
\providecommand \EOS [0]{\spacefactor3000\relax}%
\providecommand \BibitemShut  [1]{\csname bibitem#1\endcsname}%
\let\auto@bib@innerbib\@empty
\bibitem [{\citenamefont {Abrikosov}\ \emph {et~al.}(2012)\citenamefont {Abrikosov}, \citenamefont {Gorkov},\ and\ \citenamefont {Dzyaloshinski}}]{AGDbook}%
  \BibitemOpen
  \bibfield  {author} {\bibinfo {author} {\bibfnamefont {A.~A.}\ \bibnamefont {Abrikosov}}, \bibinfo {author} {\bibfnamefont {L.~P.}\ \bibnamefont {Gorkov}}, \ and\ \bibinfo {author} {\bibfnamefont {I.~E.}\ \bibnamefont {Dzyaloshinski}},\ }Methods of quantum field theory in statistical physics\ (\bibinfo  {publisher} {Dover, New York},\ \bibinfo {year} {2012})\BibitemShut {NoStop}%
\bibitem [{\citenamefont {Shankar}(1994)}]{RevModPhys.66.129}%
  \BibitemOpen
  \bibfield  {author} {\bibinfo {author} {\bibfnamefont {R.}~\bibnamefont {Shankar}},\ }\bibfield  {title} {Renormalization-group approach to interacting fermions,\ }\href {\doibase 10.1103/RevModPhys.66.129} {\bibfield  {journal} {\bibinfo  {journal} {Rev. Mod. Phys.}\ }\textbf {\bibinfo {volume} {66}},\ \bibinfo {pages} {129} (\bibinfo {year} {1994})}\BibitemShut {NoStop}%
\bibitem [{\citenamefont {Overhauser}(1962)}]{PhysRev.128.1437}%
  \BibitemOpen
  \bibfield  {author} {\bibinfo {author} {\bibfnamefont {A.~W.}\ \bibnamefont {Overhauser}},\ }\bibfield  {title} {Spin density waves in an electron gas,\ }\href {\doibase 10.1103/PhysRev.128.1437} {\bibfield  {journal} {\bibinfo  {journal} {Phys. Rev.}\ }\textbf {\bibinfo {volume} {128}},\ \bibinfo {pages} {1437} (\bibinfo {year} {1962})}\BibitemShut {NoStop}%
\bibitem [{\citenamefont {Overhauser}(1963)}]{10.1063/1.1729354}%
  \BibitemOpen
  \bibfield  {author} {\bibinfo {author} {\bibfnamefont {A.~W.}\ \bibnamefont {Overhauser}},\ }\bibfield  {title} {Spin‐density‐wave mechanisms of antiferromagnetism,\ }\href {\doibase 10.1063/1.1729354} {\bibfield  {journal} {\bibinfo  {journal} {Journal of Applied Physics}\ }\textbf {\bibinfo {volume} {34}},\ \bibinfo {pages} {1019} (\bibinfo {year} {1963})}\BibitemShut {NoStop}%
\bibitem [{\citenamefont {Peierls}(1955)}]{peierls1955quantum}%
  \BibitemOpen
  \bibfield  {author} {\bibinfo {author} {\bibfnamefont {R.}~\bibnamefont {Peierls}},\ }Quantum theory of solids,\ International series of monographs on physics\ (\bibinfo  {publisher} {Clarendon Press},\ \bibinfo {year} {1955})\BibitemShut {NoStop}%
\bibitem [{\citenamefont {Gr\"uner}(1988)}]{RevModPhys.60.1129}%
  \BibitemOpen
  \bibfield  {author} {\bibinfo {author} {\bibfnamefont {G.}~\bibnamefont {Gr\"uner}},\ }\bibfield  {title} {The dynamics of charge-density waves,\ }\href {\doibase 10.1103/RevModPhys.60.1129} {\bibfield  {journal} {\bibinfo  {journal} {Rev. Mod. Phys.}\ }\textbf {\bibinfo {volume} {60}},\ \bibinfo {pages} {1129} (\bibinfo {year} {1988})}\BibitemShut {NoStop}%
\bibitem [{\citenamefont {Zhu}\ \emph {et~al.}(2015)\citenamefont {Zhu}, \citenamefont {Cao}, \citenamefont {Zhang}, \citenamefont {Plummer},\ and\ \citenamefont {Guo}}]{doi:10.1073/pnas.1424791112}%
  \BibitemOpen
  \bibfield  {author} {\bibinfo {author} {\bibfnamefont {X.}~\bibnamefont {Zhu}}, \bibinfo {author} {\bibfnamefont {Y.}~\bibnamefont {Cao}}, \bibinfo {author} {\bibfnamefont {J.}~\bibnamefont {Zhang}}, \bibinfo {author} {\bibfnamefont {E.~W.}\ \bibnamefont {Plummer}}, \ and\ \bibinfo {author} {\bibfnamefont {J.}~\bibnamefont {Guo}},\ }\bibfield  {title} {Classification of charge density waves based on their nature,\ }\href {\doibase 10.1073/pnas.1424791112} {\bibfield  {journal} {\bibinfo  {journal} {Proceedings of the National Academy of Sciences}\ }\textbf {\bibinfo {volume} {112}},\ \bibinfo {pages} {2367} (\bibinfo {year} {2015})}\BibitemShut {NoStop}%
\bibitem [{\citenamefont {{Leykam}}\ \emph {et~al.}(2018)\citenamefont {{Leykam}}, \citenamefont {{Andreanov}},\ and\ \citenamefont {{Flach}}}]{flat}%
  \BibitemOpen
  \bibfield  {author} {\bibinfo {author} {\bibfnamefont {D.}~\bibnamefont {{Leykam}}}, \bibinfo {author} {\bibfnamefont {A.}~\bibnamefont {{Andreanov}}}, \ and\ \bibinfo {author} {\bibfnamefont {S.}~\bibnamefont {{Flach}}},\ }\bibfield  {title} {{Artificial flat band systems: from lattice models to experiments},\ }\href {\doibase 10.1080/23746149.2018.1473052} {\bibfield  {journal} {\bibinfo  {journal} {Advances in Physics X}\ }\textbf {\bibinfo {volume} {3}},\ \bibinfo {pages} {1473052} (\bibinfo {year} {2018})}\BibitemShut {NoStop}%
\bibitem [{\citenamefont {Derzhko}\ \emph {et~al.}(2015)\citenamefont {Derzhko}, \citenamefont {Richter},\ and\ \citenamefont {Maksymenko}}]{doi:10.1142/S0217979215300078}%
  \BibitemOpen
  \bibfield  {author} {\bibinfo {author} {\bibfnamefont {O.}~\bibnamefont {Derzhko}}, \bibinfo {author} {\bibfnamefont {J.}~\bibnamefont {Richter}}, \ and\ \bibinfo {author} {\bibfnamefont {M.}~\bibnamefont {Maksymenko}},\ }\bibfield  {title} {Strongly correlated flat-band systems: The route from {Heisenberg} spins to {Hubbard} electrons,\ }\href {\doibase 10.1142/S0217979215300078} {\bibfield  {journal} {\bibinfo  {journal} {International Journal of Modern Physics B}\ }\textbf {\bibinfo {volume} {29}},\ \bibinfo {pages} {1530007} (\bibinfo {year} {2015})}\BibitemShut {NoStop}%
\bibitem [{\citenamefont {Lieb}(1989)}]{PhysRevLett.62.1201}%
  \BibitemOpen
  \bibfield  {author} {\bibinfo {author} {\bibfnamefont {E.~H.}\ \bibnamefont {Lieb}},\ }\bibfield  {title} {Two theorems on the {Hubbard} model,\ }\href {\doibase 10.1103/PhysRevLett.62.1201} {\bibfield  {journal} {\bibinfo  {journal} {Phys. Rev. Lett.}\ }\textbf {\bibinfo {volume} {62}},\ \bibinfo {pages} {1201} (\bibinfo {year} {1989})}\BibitemShut {NoStop}%
\bibitem [{\citenamefont {Mielke}(1992)}]{Mielke_1992}%
  \BibitemOpen
  \bibfield  {author} {\bibinfo {author} {\bibfnamefont {A.}~\bibnamefont {Mielke}},\ }\bibfield  {title} {Exact ground states for the {Hubbard} model on the kagome lattice,\ }\href {\doibase 10.1088/0305-4470/25/16/011} {\bibfield  {journal} {\bibinfo  {journal} {Journal of Physics A: Mathematical and General}\ }\textbf {\bibinfo {volume} {25}},\ \bibinfo {pages} {4335} (\bibinfo {year} {1992})}\BibitemShut {NoStop}%
\bibitem [{\citenamefont {Tasaki}(1992)}]{PhysRevLett.69.1608}%
  \BibitemOpen
  \bibfield  {author} {\bibinfo {author} {\bibfnamefont {H.}~\bibnamefont {Tasaki}},\ }\bibfield  {title} {Ferromagnetism in the {Hubbard} models with degenerate single-electron ground states,\ }\href {\doibase 10.1103/PhysRevLett.69.1608} {\bibfield  {journal} {\bibinfo  {journal} {Phys. Rev. Lett.}\ }\textbf {\bibinfo {volume} {69}},\ \bibinfo {pages} {1608} (\bibinfo {year} {1992})}\BibitemShut {NoStop}%
\bibitem [{\citenamefont {Bi}\ \emph {et~al.}(2019)\citenamefont {Bi}, \citenamefont {Yuan},\ and\ \citenamefont {Fu}}]{PhysRevB.100.035448}%
  \BibitemOpen
  \bibfield  {author} {\bibinfo {author} {\bibfnamefont {Z.}~\bibnamefont {Bi}}, \bibinfo {author} {\bibfnamefont {N.~F.~Q.}\ \bibnamefont {Yuan}}, \ and\ \bibinfo {author} {\bibfnamefont {L.}~\bibnamefont {Fu}},\ }\bibfield  {title} {Designing flat bands by strain,\ }\href {\doibase 10.1103/PhysRevB.100.035448} {\bibfield  {journal} {\bibinfo  {journal} {Phys. Rev. B}\ }\textbf {\bibinfo {volume} {100}},\ \bibinfo {pages} {035448} (\bibinfo {year} {2019})}\BibitemShut {NoStop}%
\bibitem [{\citenamefont {Li}\ \emph {et~al.}(2025)\citenamefont {Li}, \citenamefont {Ji}, \citenamefont {Yan}, \citenamefont {Fan}, \citenamefont {Zhang}, \citenamefont {Sun}, \citenamefont {Miao}, \citenamefont {Chen}, \citenamefont {Wan},\ and\ \citenamefont {Ding}}]{PhysRevLett.134.076402}%
  \BibitemOpen
  \bibfield  {author} {\bibinfo {author} {\bibfnamefont {H.~T.}\ \bibnamefont {Li}}, \bibinfo {author} {\bibfnamefont {T.~Z.}\ \bibnamefont {Ji}}, \bibinfo {author} {\bibfnamefont {R.~G.}\ \bibnamefont {Yan}}, \bibinfo {author} {\bibfnamefont {W.~L.}\ \bibnamefont {Fan}}, \bibinfo {author} {\bibfnamefont {Z.~X.}\ \bibnamefont {Zhang}}, \bibinfo {author} {\bibfnamefont {L.}~\bibnamefont {Sun}}, \bibinfo {author} {\bibfnamefont {B.~F.}\ \bibnamefont {Miao}}, \bibinfo {author} {\bibfnamefont {G.}~\bibnamefont {Chen}}, \bibinfo {author} {\bibfnamefont {X.~G.}\ \bibnamefont {Wan}}, \ and\ \bibinfo {author} {\bibfnamefont {H.~F.}\ \bibnamefont {Ding}},\ }\bibfield  {title} {General method to construct flat bands in two-dimensional lattices,\ }\href {\doibase 10.1103/PhysRevLett.134.076402} {\bibfield  {journal} {\bibinfo  {journal} {Phys. Rev. Lett.}\ }\textbf {\bibinfo {volume} {134}},\ \bibinfo {pages} {076402} (\bibinfo {year} {2025})}\BibitemShut {NoStop}%
\bibitem [{\citenamefont {{Kang}}\ \emph {et~al.}(2020{\natexlab{a}})\citenamefont {{Kang}}, \citenamefont {{Fang}}, \citenamefont {{Ye}}, \citenamefont {{Po}}, \citenamefont {{Denlinger}}, \citenamefont {{Jozwiak}}, \citenamefont {{Bostwick}}, \citenamefont {{Rotenberg}}, \citenamefont {{Kaxiras}}, \citenamefont {{Checkelsky}},\ and\ \citenamefont {{Comin}}}]{2020NatCo..11.4004K}%
  \BibitemOpen
  \bibfield  {author} {\bibinfo {author} {\bibfnamefont {M.}~\bibnamefont {{Kang}}}, \bibinfo {author} {\bibfnamefont {S.}~\bibnamefont {{Fang}}}, \bibinfo {author} {\bibfnamefont {L.}~\bibnamefont {{Ye}}}, \bibinfo {author} {\bibfnamefont {H.~C.}\ \bibnamefont {{Po}}}, \bibinfo {author} {\bibfnamefont {J.}~\bibnamefont {{Denlinger}}}, \bibinfo {author} {\bibfnamefont {C.}~\bibnamefont {{Jozwiak}}}, \bibinfo {author} {\bibfnamefont {A.}~\bibnamefont {{Bostwick}}}, \bibinfo {author} {\bibfnamefont {E.}~\bibnamefont {{Rotenberg}}}, \bibinfo {author} {\bibfnamefont {E.}~\bibnamefont {{Kaxiras}}}, \bibinfo {author} {\bibfnamefont {J.~G.}\ \bibnamefont {{Checkelsky}}}, \ and\ \bibinfo {author} {\bibfnamefont {R.}~\bibnamefont {{Comin}}},\ }\bibfield  {title} {{Topological flat bands in frustrated kagome lattice {CoSn}},\ }\href {\doibase 10.1038/s41467-020-17465-1} {\bibfield  {journal} {\bibinfo  {journal} {Nature Communications}\ }\textbf {\bibinfo {volume} {11}},\ \bibinfo {eid} {4004} (\bibinfo {year}
  {2020}{\natexlab{a}})}\BibitemShut {NoStop}%
\bibitem [{\citenamefont {{Balents}}\ \emph {et~al.}(2020)\citenamefont {{Balents}}, \citenamefont {{Dean}}, \citenamefont {{Efetov}},\ and\ \citenamefont {{Young}}}]{2020NatPh..16..725B}%
  \BibitemOpen
  \bibfield  {author} {\bibinfo {author} {\bibfnamefont {L.}~\bibnamefont {{Balents}}}, \bibinfo {author} {\bibfnamefont {C.~R.}\ \bibnamefont {{Dean}}}, \bibinfo {author} {\bibfnamefont {D.~K.}\ \bibnamefont {{Efetov}}}, \ and\ \bibinfo {author} {\bibfnamefont {A.~F.}\ \bibnamefont {{Young}}},\ }\bibfield  {title} {{Superconductivity and strong correlations in moir{\'e} flat bands},\ }\href {\doibase 10.1038/s41567-020-0906-9} {\bibfield  {journal} {\bibinfo  {journal} {Nature Physics}\ }\textbf {\bibinfo {volume} {16}},\ \bibinfo {pages} {725} (\bibinfo {year} {2020})}\BibitemShut {NoStop}%
\bibitem [{\citenamefont {Hofmann}\ \emph {et~al.}(2022)\citenamefont {Hofmann}, \citenamefont {Chowdhury}, \citenamefont {Kivelson},\ and\ \citenamefont {Berg}}]{Berg.Hofmann.2022}%
  \BibitemOpen
  \bibfield  {author} {\bibinfo {author} {\bibfnamefont {J.~S.}\ \bibnamefont {Hofmann}}, \bibinfo {author} {\bibfnamefont {D.}~\bibnamefont {Chowdhury}}, \bibinfo {author} {\bibfnamefont {S.~A.}\ \bibnamefont {Kivelson}}, \ and\ \bibinfo {author} {\bibfnamefont {E.}~\bibnamefont {Berg}},\ }\bibfield  {title} {{Heuristic bounds on superconductivity and how to exceed them},\ }\href {\doibase 10.1038/s41535-022-00491-1} {\bibfield  {journal} {\bibinfo  {journal} {npj Quantum Materials}\ }\textbf {\bibinfo {volume} {7}},\ \bibinfo {pages} {83} (\bibinfo {year} {2022})}\BibitemShut {NoStop}%
\bibitem [{\citenamefont {Yang}\ \emph {et~al.}(2024)\citenamefont {Yang}, \citenamefont {Zhai}, \citenamefont {Tan}, \citenamefont {Fan}, \citenamefont {Lin},\ and\ \citenamefont {Yao}}]{yang2024fractionalquantumanomaloushall}%
  \BibitemOpen
  \bibfield  {author} {\bibinfo {author} {\bibfnamefont {W.}~\bibnamefont {Yang}}, \bibinfo {author} {\bibfnamefont {D.}~\bibnamefont {Zhai}}, \bibinfo {author} {\bibfnamefont {T.}~\bibnamefont {Tan}}, \bibinfo {author} {\bibfnamefont {F.-R.}\ \bibnamefont {Fan}}, \bibinfo {author} {\bibfnamefont {Z.}~\bibnamefont {Lin}}, \ and\ \bibinfo {author} {\bibfnamefont {W.}~\bibnamefont {Yao}},\ }\href {https://arxiv.org/abs/2405.01829} {Fractional quantum anomalous {Hall} effect in a singular flat band} (\bibinfo {year} {2024}),\ \Eprint {http://arxiv.org/abs/2405.01829} {arXiv:2405.01829 [cond-mat.mes-hall]} \BibitemShut {NoStop}%
\bibitem [{\citenamefont {{Andrei}}\ and\ \citenamefont {{MacDonald}}(2020)}]{2020NatMa..19.1265A}%
  \BibitemOpen
  \bibfield  {author} {\bibinfo {author} {\bibfnamefont {E.~Y.}\ \bibnamefont {{Andrei}}}\ and\ \bibinfo {author} {\bibfnamefont {A.~H.}\ \bibnamefont {{MacDonald}}},\ }\bibfield  {title} {{Graphene bilayers with a twist},\ }\href {\doibase 10.1038/s41563-020-00840-0} {\bibfield  {journal} {\bibinfo  {journal} {Nature Materials}\ }\textbf {\bibinfo {volume} {19}},\ \bibinfo {pages} {1265} (\bibinfo {year} {2020})}\BibitemShut {NoStop}%
\bibitem [{\citenamefont {{Liu}}\ \emph {et~al.}(2018)\citenamefont {{Liu}}, \citenamefont {{Sun}}, \citenamefont {{Kumar}}, \citenamefont {{Muechler}}, \citenamefont {{Sun}}, \citenamefont {{Jiao}}, \citenamefont {{Yang}}, \citenamefont {{Liu}}, \citenamefont {{Liang}}, \citenamefont {{Xu}}, \citenamefont {{Kroder}}, \citenamefont {{S{\"u}{\ss}}}, \citenamefont {{Borrmann}}, \citenamefont {{Shekhar}}, \citenamefont {{Wang}}, \citenamefont {{Xi}}, \citenamefont {{Wang}}, \citenamefont {{Schnelle}}, \citenamefont {{Wirth}}, \citenamefont {{Chen}}, \citenamefont {{Goennenwein}},\ and\ \citenamefont {{Felser}}}]{2018NatPh..14.1125L}%
  \BibitemOpen
  \bibfield  {author} {\bibinfo {author} {\bibfnamefont {E.}~\bibnamefont {{Liu}}}, \bibinfo {author} {\bibfnamefont {Y.}~\bibnamefont {{Sun}}}, \bibinfo {author} {\bibfnamefont {N.}~\bibnamefont {{Kumar}}}, \bibinfo {author} {\bibfnamefont {L.}~\bibnamefont {{Muechler}}}, \bibinfo {author} {\bibfnamefont {A.}~\bibnamefont {{Sun}}}, \bibinfo {author} {\bibfnamefont {L.}~\bibnamefont {{Jiao}}}, \bibinfo {author} {\bibfnamefont {S.-Y.}\ \bibnamefont {{Yang}}}, \bibinfo {author} {\bibfnamefont {D.}~\bibnamefont {{Liu}}}, \bibinfo {author} {\bibfnamefont {A.}~\bibnamefont {{Liang}}}, \bibinfo {author} {\bibfnamefont {Q.}~\bibnamefont {{Xu}}}, \bibinfo {author} {\bibfnamefont {J.}~\bibnamefont {{Kroder}}}, \bibinfo {author} {\bibfnamefont {V.}~\bibnamefont {{S{\"u}{\ss}}}}, \bibinfo {author} {\bibfnamefont {H.}~\bibnamefont {{Borrmann}}}, \bibinfo {author} {\bibfnamefont {C.}~\bibnamefont {{Shekhar}}}, \bibinfo {author} {\bibfnamefont {Z.}~\bibnamefont {{Wang}}}, \bibinfo {author} {\bibfnamefont {C.}~\bibnamefont
  {{Xi}}}, \bibinfo {author} {\bibfnamefont {W.}~\bibnamefont {{Wang}}}, \bibinfo {author} {\bibfnamefont {W.}~\bibnamefont {{Schnelle}}}, \bibinfo {author} {\bibfnamefont {S.}~\bibnamefont {{Wirth}}}, \bibinfo {author} {\bibfnamefont {Y.}~\bibnamefont {{Chen}}}, \bibinfo {author} {\bibfnamefont {S.~T.~B.}\ \bibnamefont {{Goennenwein}}}, \ and\ \bibinfo {author} {\bibfnamefont {C.}~\bibnamefont {{Felser}}},\ }\bibfield  {title} {{Giant anomalous {Hall} effect in a ferromagnetic kagome-lattice semimetal},\ }\href {\doibase 10.1038/s41567-018-0234-5} {\bibfield  {journal} {\bibinfo  {journal} {Nature Physics}\ }\textbf {\bibinfo {volume} {14}},\ \bibinfo {pages} {1125} (\bibinfo {year} {2018})}\BibitemShut {NoStop}%
\bibitem [{\citenamefont {{Kang}}\ \emph {et~al.}(2020{\natexlab{b}})\citenamefont {{Kang}}, \citenamefont {{Ye}}, \citenamefont {{Fang}}, \citenamefont {{You}}, \citenamefont {{Levitan}}, \citenamefont {{Han}}, \citenamefont {{Facio}}, \citenamefont {{Jozwiak}}, \citenamefont {{Bostwick}}, \citenamefont {{Rotenberg}}, \citenamefont {{Chan}}, \citenamefont {{McDonald}}, \citenamefont {{Graf}}, \citenamefont {{Kaznatcheev}}, \citenamefont {{Vescovo}}, \citenamefont {{Bell}}, \citenamefont {{Kaxiras}}, \citenamefont {{van den Brink}}, \citenamefont {{Richter}}, \citenamefont {{Prasad Ghimire}}, \citenamefont {{Checkelsky}},\ and\ \citenamefont {{Comin}}}]{2020NatMa..19..163K}%
  \BibitemOpen
  \bibfield  {author} {\bibinfo {author} {\bibfnamefont {M.}~\bibnamefont {{Kang}}}, \bibinfo {author} {\bibfnamefont {L.}~\bibnamefont {{Ye}}}, \bibinfo {author} {\bibfnamefont {S.}~\bibnamefont {{Fang}}}, \bibinfo {author} {\bibfnamefont {J.-S.}\ \bibnamefont {{You}}}, \bibinfo {author} {\bibfnamefont {A.}~\bibnamefont {{Levitan}}}, \bibinfo {author} {\bibfnamefont {M.}~\bibnamefont {{Han}}}, \bibinfo {author} {\bibfnamefont {J.~I.}\ \bibnamefont {{Facio}}}, \bibinfo {author} {\bibfnamefont {C.}~\bibnamefont {{Jozwiak}}}, \bibinfo {author} {\bibfnamefont {A.}~\bibnamefont {{Bostwick}}}, \bibinfo {author} {\bibfnamefont {E.}~\bibnamefont {{Rotenberg}}}, \bibinfo {author} {\bibfnamefont {M.~K.}\ \bibnamefont {{Chan}}}, \bibinfo {author} {\bibfnamefont {R.~D.}\ \bibnamefont {{McDonald}}}, \bibinfo {author} {\bibfnamefont {D.}~\bibnamefont {{Graf}}}, \bibinfo {author} {\bibfnamefont {K.}~\bibnamefont {{Kaznatcheev}}}, \bibinfo {author} {\bibfnamefont {E.}~\bibnamefont {{Vescovo}}}, \bibinfo {author}
  {\bibfnamefont {D.~C.}\ \bibnamefont {{Bell}}}, \bibinfo {author} {\bibfnamefont {E.}~\bibnamefont {{Kaxiras}}}, \bibinfo {author} {\bibfnamefont {J.}~\bibnamefont {{van den Brink}}}, \bibinfo {author} {\bibfnamefont {M.}~\bibnamefont {{Richter}}}, \bibinfo {author} {\bibfnamefont {M.}~\bibnamefont {{Prasad Ghimire}}}, \bibinfo {author} {\bibfnamefont {J.~G.}\ \bibnamefont {{Checkelsky}}}, \ and\ \bibinfo {author} {\bibfnamefont {R.}~\bibnamefont {{Comin}}},\ }\bibfield  {title} {{Dirac fermions and flat bands in the ideal kagome metal {FeSn}},\ }\href {\doibase 10.1038/s41563-019-0531-0} {\bibfield  {journal} {\bibinfo  {journal} {Nature Materials}\ }\textbf {\bibinfo {volume} {19}},\ \bibinfo {pages} {163} (\bibinfo {year} {2020}{\natexlab{b}})}\BibitemShut {NoStop}%
\bibitem [{\citenamefont {Liu}\ \emph {et~al.}(2020)\citenamefont {Liu}, \citenamefont {Li}, \citenamefont {Wang}, \citenamefont {Wang}, \citenamefont {Wen}, \citenamefont {Jiang}, \citenamefont {Lu}, \citenamefont {Yan}, \citenamefont {Huang}, \citenamefont {Shen}, \citenamefont {Yin}, \citenamefont {Wang}, \citenamefont {Yin}, \citenamefont {Lei},\ and\ \citenamefont {Wang}}]{Liu2020OrbitalselectiveDF}%
  \BibitemOpen
  \bibfield  {author} {\bibinfo {author} {\bibfnamefont {Z.}~\bibnamefont {Liu}}, \bibinfo {author} {\bibfnamefont {M.}~\bibnamefont {Li}}, \bibinfo {author} {\bibfnamefont {Q.}~\bibnamefont {Wang}}, \bibinfo {author} {\bibfnamefont {G.}~\bibnamefont {Wang}}, \bibinfo {author} {\bibfnamefont {C.}~\bibnamefont {Wen}}, \bibinfo {author} {\bibfnamefont {K.}~\bibnamefont {Jiang}}, \bibinfo {author} {\bibfnamefont {X.}~\bibnamefont {Lu}}, \bibinfo {author} {\bibfnamefont {S.}~\bibnamefont {Yan}}, \bibinfo {author} {\bibfnamefont {Y.}~\bibnamefont {Huang}}, \bibinfo {author} {\bibfnamefont {D.}~\bibnamefont {Shen}}, \bibinfo {author} {\bibfnamefont {J.-X.}\ \bibnamefont {Yin}}, \bibinfo {author} {\bibfnamefont {Z.}~\bibnamefont {Wang}}, \bibinfo {author} {\bibfnamefont {Z.}~\bibnamefont {Yin}}, \bibinfo {author} {\bibfnamefont {H.}~\bibnamefont {Lei}}, \ and\ \bibinfo {author} {\bibfnamefont {S.}~\bibnamefont {Wang}},\ }\bibfield  {title} {Orbital-selective {Dirac} fermions and extremely flat bands in frustrated
  kagome-lattice metal {CoSn},\ }\href {https://api.semanticscholar.org/CorpusID:221084388} {\bibfield  {journal} {\bibinfo  {journal} {Nature Communications}\ }\textbf {\bibinfo {volume} {11}} (\bibinfo {year} {2020})}\BibitemShut {NoStop}%
\bibitem [{\citenamefont {Kitamura}\ \emph {et~al.}(2024)\citenamefont {Kitamura}, \citenamefont {Daido},\ and\ \citenamefont {Yanase}}]{PhysRevLett.132.036001}%
  \BibitemOpen
  \bibfield  {author} {\bibinfo {author} {\bibfnamefont {T.}~\bibnamefont {Kitamura}}, \bibinfo {author} {\bibfnamefont {A.}~\bibnamefont {Daido}}, \ and\ \bibinfo {author} {\bibfnamefont {Y.}~\bibnamefont {Yanase}},\ }\bibfield  {title} {Spin-triplet superconductivity from quantum-geometry-induced ferromagnetic fluctuation,\ }\href {\doibase 10.1103/PhysRevLett.132.036001} {\bibfield  {journal} {\bibinfo  {journal} {Phys. Rev. Lett.}\ }\textbf {\bibinfo {volume} {132}},\ \bibinfo {pages} {036001} (\bibinfo {year} {2024})}\BibitemShut {NoStop}%
\bibitem [{\citenamefont {Kitamura}\ \emph {et~al.}(2025)\citenamefont {Kitamura}, \citenamefont {Nakai}, \citenamefont {Daido},\ and\ \citenamefont {Yanase}}]{kitamura2025quantumgeometricferromagnetismsingular}%
  \BibitemOpen
  \bibfield  {author} {\bibinfo {author} {\bibfnamefont {T.}~\bibnamefont {Kitamura}}, \bibinfo {author} {\bibfnamefont {H.}~\bibnamefont {Nakai}}, \bibinfo {author} {\bibfnamefont {A.}~\bibnamefont {Daido}}, \ and\ \bibinfo {author} {\bibfnamefont {Y.}~\bibnamefont {Yanase}},\ }\href {https://arxiv.org/abs/2505.01089} {Quantum geometric ferromagnetism by singular saddle point} (\bibinfo {year} {2025}),\ \Eprint {http://arxiv.org/abs/2505.01089} {arXiv:2505.01089 [cond-mat.str-el]} \BibitemShut {NoStop}%
\bibitem [{\citenamefont {{Wang}}\ and\ \citenamefont {{Huang}}(2025)}]{2025SCPMA..6897211W}%
  \BibitemOpen
  \bibfield  {author} {\bibinfo {author} {\bibfnamefont {H.-X.}\ \bibnamefont {{Wang}}}\ and\ \bibinfo {author} {\bibfnamefont {W.}~\bibnamefont {{Huang}}},\ }\bibfield  {title} {{Density matrix renormalization group study of the quantum-geometry-facilitated pair density wave order},\ }\href {\doibase 10.1007/s11433-025-2701-1} {\bibfield  {journal} {\bibinfo  {journal} {Science China Physics, Mechanics, and Astronomy}\ }\textbf {\bibinfo {volume} {68}},\ \bibinfo {eid} {297211} (\bibinfo {year} {2025})}\BibitemShut {NoStop}%
\bibitem [{\citenamefont {Jiang}\ and\ \citenamefont {Barlas}(2023)}]{PhysRevLett.131.016002}%
  \BibitemOpen
  \bibfield  {author} {\bibinfo {author} {\bibfnamefont {G.}~\bibnamefont {Jiang}}\ and\ \bibinfo {author} {\bibfnamefont {Y.}~\bibnamefont {Barlas}},\ }\bibfield  {title} {Pair density waves from local band geometry,\ }\href {\doibase 10.1103/PhysRevLett.131.016002} {\bibfield  {journal} {\bibinfo  {journal} {Phys. Rev. Lett.}\ }\textbf {\bibinfo {volume} {131}},\ \bibinfo {pages} {016002} (\bibinfo {year} {2023})}\BibitemShut {NoStop}%
\bibitem [{\citenamefont {Sun}\ \emph {et~al.}(2024)\citenamefont {Sun}, \citenamefont {Yu}, \citenamefont {Chen}, \citenamefont {Hu},\ and\ \citenamefont {Law}}]{Sun.Law.2024}%
  \BibitemOpen
  \bibfield  {author} {\bibinfo {author} {\bibfnamefont {Z.-T.}\ \bibnamefont {Sun}}, \bibinfo {author} {\bibfnamefont {R.-P.}\ \bibnamefont {Yu}}, \bibinfo {author} {\bibfnamefont {S.~A.}\ \bibnamefont {Chen}}, \bibinfo {author} {\bibfnamefont {J.-X.}\ \bibnamefont {Hu}}, \ and\ \bibinfo {author} {\bibfnamefont {K.}~\bibnamefont {Law}},\ }\href@noop {} {Flat-band {Fulde-Ferrell-Larkin-Ovchinnikov} state from quantum geometry discrepancy} (\bibinfo {year} {2024}),\ \Eprint {http://arxiv.org/abs/2408.00548} {arXiv:2408.00548 [cond-mat.str-el]} \BibitemShut {NoStop}%
\bibitem [{\citenamefont {{Stoner}}(1938)}]{Stoner}%
  \BibitemOpen
  \bibfield  {author} {\bibinfo {author} {\bibfnamefont {E.~C.}\ \bibnamefont {{Stoner}}},\ }\bibfield  {title} {{Collective Electron Ferromagnetism},\ }\href {\doibase 10.1098/rspa.1938.0066} {\bibfield  {journal} {\bibinfo  {journal} {Proceedings of the Royal Society of London Series A}\ }\textbf {\bibinfo {volume} {165}},\ \bibinfo {pages} {372} (\bibinfo {year} {1938})}\BibitemShut {NoStop}%
\bibitem [{\citenamefont {Klebl}\ and\ \citenamefont {Honerkamp}(2019)}]{PhysRevB.100.155145}%
  \BibitemOpen
  \bibfield  {author} {\bibinfo {author} {\bibfnamefont {L.}~\bibnamefont {Klebl}}\ and\ \bibinfo {author} {\bibfnamefont {C.}~\bibnamefont {Honerkamp}},\ }\bibfield  {title} {Inherited and flatband-induced ordering in twisted graphene bilayers,\ }\href {\doibase 10.1103/PhysRevB.100.155145} {\bibfield  {journal} {\bibinfo  {journal} {Phys. Rev. B}\ }\textbf {\bibinfo {volume} {100}},\ \bibinfo {pages} {155145} (\bibinfo {year} {2019})}\BibitemShut {NoStop}%
\bibitem [{\citenamefont {{Wu}}\ \emph {et~al.}(2025)\citenamefont {{Wu}}, \citenamefont {{Xu}}, \citenamefont {{Wang}}, \citenamefont {{Lin}}, \citenamefont {{Cao}},\ and\ \citenamefont {{Cao}}}]{2025NatCo}%
  \BibitemOpen
  \bibfield  {author} {\bibinfo {author} {\bibfnamefont {S.}~\bibnamefont {{Wu}}}, \bibinfo {author} {\bibfnamefont {C.}~\bibnamefont {{Xu}}}, \bibinfo {author} {\bibfnamefont {X.}~\bibnamefont {{Wang}}}, \bibinfo {author} {\bibfnamefont {H.-Q.}\ \bibnamefont {{Lin}}}, \bibinfo {author} {\bibfnamefont {C.}~\bibnamefont {{Cao}}}, \ and\ \bibinfo {author} {\bibfnamefont {G.-H.}\ \bibnamefont {{Cao}}},\ }\bibfield  {title} {{Flat-band enhanced antiferromagnetic fluctuations and superconductivity in pressurized {CsCr}$_{3}${Sb}$_{5}$},\ }\href {\doibase 10.1038/s41467-025-56582-7} {\bibfield  {journal} {\bibinfo  {journal} {Nature Communications}\ }\textbf {\bibinfo {volume} {16}},\ \bibinfo {eid} {1375} (\bibinfo {year} {2025})}\BibitemShut {NoStop}%
\bibitem [{\citenamefont {Fulde}\ and\ \citenamefont {Ferrell}(1964)}]{PhysRev.135.A550}%
  \BibitemOpen
  \bibfield  {author} {\bibinfo {author} {\bibfnamefont {P.}~\bibnamefont {Fulde}}\ and\ \bibinfo {author} {\bibfnamefont {R.~A.}\ \bibnamefont {Ferrell}},\ }\bibfield  {title} {Superconductivity in a strong spin-exchange field,\ }\href {\doibase 10.1103/PhysRev.135.A550} {\bibfield  {journal} {\bibinfo  {journal} {Phys. Rev.}\ }\textbf {\bibinfo {volume} {135}},\ \bibinfo {pages} {A550} (\bibinfo {year} {1964})}\BibitemShut {NoStop}%
\bibitem [{\citenamefont {Larkin}\ and\ \citenamefont {Ovchinnikov}(1964)}]{Larkin:1964wok}%
  \BibitemOpen
  \bibfield  {author} {\bibinfo {author} {\bibfnamefont {A.~I.}\ \bibnamefont {Larkin}}\ and\ \bibinfo {author} {\bibfnamefont {Y.~N.}\ \bibnamefont {Ovchinnikov}},\ }\bibfield  {title} {{Nonuniform state of superconductors},\ }\href@noop {} {\bibfield  {journal} {\bibinfo  {journal} {Zh. Eksp. Teor. Fiz.}\ }\textbf {\bibinfo {volume} {47}},\ \bibinfo {pages} {1136} (\bibinfo {year} {1964})}\BibitemShut {NoStop}%
\bibitem [{\citenamefont {Kinnunen}\ \emph {et~al.}(2018)\citenamefont {Kinnunen}, \citenamefont {Baarsma}, \citenamefont {Martikainen},\ and\ \citenamefont {Törmä}}]{Kinnunen_2018}%
  \BibitemOpen
  \bibfield  {author} {\bibinfo {author} {\bibfnamefont {J.~J.}\ \bibnamefont {Kinnunen}}, \bibinfo {author} {\bibfnamefont {J.~E.}\ \bibnamefont {Baarsma}}, \bibinfo {author} {\bibfnamefont {J.-P.}\ \bibnamefont {Martikainen}}, \ and\ \bibinfo {author} {\bibfnamefont {P.}~\bibnamefont {Törmä}},\ }\bibfield  {title} {The {Fulde–Ferrell–Larkin–Ovchinnikov} state for ultracold fermions in lattice and harmonic potentials: a review,\ }\href {\doibase 10.1088/1361-6633/aaa4ad} {\bibfield  {journal} {\bibinfo  {journal} {Reports on Progress in Physics}\ }\textbf {\bibinfo {volume} {81}},\ \bibinfo {pages} {046401} (\bibinfo {year} {2018})}\BibitemShut {NoStop}%
\bibitem [{\citenamefont {Agterberg}\ \emph {et~al.}(2020)\citenamefont {Agterberg}, \citenamefont {Davis}, \citenamefont {Edkins}, \citenamefont {Fradkin}, \citenamefont {Van~Harlingen}, \citenamefont {Kivelson}, \citenamefont {Lee}, \citenamefont {Radzihovsky}, \citenamefont {Tranquada},\ and\ \citenamefont {Wang}}]{annurev:/content/journals/10.1146/annurev-conmatphys-031119-050711}%
  \BibitemOpen
  \bibfield  {author} {\bibinfo {author} {\bibfnamefont {D.~F.}\ \bibnamefont {Agterberg}}, \bibinfo {author} {\bibfnamefont {J.~S.}\ \bibnamefont {Davis}}, \bibinfo {author} {\bibfnamefont {S.~D.}\ \bibnamefont {Edkins}}, \bibinfo {author} {\bibfnamefont {E.}~\bibnamefont {Fradkin}}, \bibinfo {author} {\bibfnamefont {D.~J.}\ \bibnamefont {Van~Harlingen}}, \bibinfo {author} {\bibfnamefont {S.~A.}\ \bibnamefont {Kivelson}}, \bibinfo {author} {\bibfnamefont {P.~A.}\ \bibnamefont {Lee}}, \bibinfo {author} {\bibfnamefont {L.}~\bibnamefont {Radzihovsky}}, \bibinfo {author} {\bibfnamefont {J.~M.}\ \bibnamefont {Tranquada}}, \ and\ \bibinfo {author} {\bibfnamefont {Y.}~\bibnamefont {Wang}},\ }\bibfield  {title} {The physics of pair-density waves: Cuprate superconductors and beyond,\ }\href {\doibase https://doi.org/10.1146/annurev-conmatphys-031119-050711} {\bibfield  {journal} {\bibinfo  {journal} {Annual Review of Condensed Matter Physics}\ }\textbf {\bibinfo {volume} {11}},\ \bibinfo {pages} {231} (\bibinfo
  {year} {2020})}\BibitemShut {NoStop}%
\bibitem [{\citenamefont {Blankenbecler}\ \emph {et~al.}(1981)\citenamefont {Blankenbecler}, \citenamefont {Scalapino},\ and\ \citenamefont {Sugar}}]{DQMC1}%
  \BibitemOpen
  \bibfield  {author} {\bibinfo {author} {\bibfnamefont {R.}~\bibnamefont {Blankenbecler}}, \bibinfo {author} {\bibfnamefont {D.~J.}\ \bibnamefont {Scalapino}}, \ and\ \bibinfo {author} {\bibfnamefont {R.~L.}\ \bibnamefont {Sugar}},\ }\bibfield  {title} {\uppercase{M}onte \uppercase{C}arlo calculations of coupled boson-fermion systems. {I},\ }\href {\doibase 10.1103/PhysRevD.24.2278} {\bibfield  {journal} {\bibinfo  {journal} {Phys. Rev. D}\ }\textbf {\bibinfo {volume} {24}},\ \bibinfo {pages} {2278} (\bibinfo {year} {1981})}\BibitemShut {NoStop}%
\bibitem [{\citenamefont {White}\ \emph {et~al.}(1989)\citenamefont {White}, \citenamefont {Scalapino}, \citenamefont {Sugar}, \citenamefont {Loh}, \citenamefont {Gubernatis},\ and\ \citenamefont {Scalettar}}]{DQMC2}%
  \BibitemOpen
  \bibfield  {author} {\bibinfo {author} {\bibfnamefont {S.~R.}\ \bibnamefont {White}}, \bibinfo {author} {\bibfnamefont {D.~J.}\ \bibnamefont {Scalapino}}, \bibinfo {author} {\bibfnamefont {R.~L.}\ \bibnamefont {Sugar}}, \bibinfo {author} {\bibfnamefont {E.~Y.}\ \bibnamefont {Loh}}, \bibinfo {author} {\bibfnamefont {J.~E.}\ \bibnamefont {Gubernatis}}, \ and\ \bibinfo {author} {\bibfnamefont {R.~T.}\ \bibnamefont {Scalettar}},\ }\bibfield  {title} {Numerical study of the two-dimensional \uppercase{H}ubbard model,\ }\href {\doibase 10.1103/PhysRevB.40.506} {\bibfield  {journal} {\bibinfo  {journal} {Phys. Rev. B}\ }\textbf {\bibinfo {volume} {40}},\ \bibinfo {pages} {506} (\bibinfo {year} {1989})}\BibitemShut {NoStop}%
\bibitem [{\citenamefont {{Peotta}}\ and\ \citenamefont {{T{\"o}rm{\"a}}}(2015)}]{2015NatCo...6.8944P}%
  \BibitemOpen
  \bibfield  {author} {\bibinfo {author} {\bibfnamefont {S.}~\bibnamefont {{Peotta}}}\ and\ \bibinfo {author} {\bibfnamefont {P.}~\bibnamefont {{T{\"o}rm{\"a}}}},\ }\bibfield  {title} {{Superfluidity in topologically nontrivial flat bands},\ }\href {\doibase 10.1038/ncomms9944} {\bibfield  {journal} {\bibinfo  {journal} {Nature Communications}\ }\textbf {\bibinfo {volume} {6}},\ \bibinfo {eid} {8944} (\bibinfo {year} {2015})}\BibitemShut {NoStop}%
\bibitem [{\citenamefont {Liang}\ \emph {et~al.}(2017)\citenamefont {Liang}, \citenamefont {Vanhala}, \citenamefont {Peotta}, \citenamefont {Siro}, \citenamefont {Harju},\ and\ \citenamefont {T\"orm\"a}}]{PhysRevB.95.024515}%
  \BibitemOpen
  \bibfield  {author} {\bibinfo {author} {\bibfnamefont {L.}~\bibnamefont {Liang}}, \bibinfo {author} {\bibfnamefont {T.~I.}\ \bibnamefont {Vanhala}}, \bibinfo {author} {\bibfnamefont {S.}~\bibnamefont {Peotta}}, \bibinfo {author} {\bibfnamefont {T.}~\bibnamefont {Siro}}, \bibinfo {author} {\bibfnamefont {A.}~\bibnamefont {Harju}}, \ and\ \bibinfo {author} {\bibfnamefont {P.}~\bibnamefont {T\"orm\"a}},\ }\bibfield  {title} {Band geometry, {Berry} curvature, and superfluid weight,\ }\href {\doibase 10.1103/PhysRevB.95.024515} {\bibfield  {journal} {\bibinfo  {journal} {Phys. Rev. B}\ }\textbf {\bibinfo {volume} {95}},\ \bibinfo {pages} {024515} (\bibinfo {year} {2017})}\BibitemShut {NoStop}%
\bibitem [{\citenamefont {T\"orm\"a}\ \emph {et~al.}(2018)\citenamefont {T\"orm\"a}, \citenamefont {Liang},\ and\ \citenamefont {Peotta}}]{PhysRevB.98.220511}%
  \BibitemOpen
  \bibfield  {author} {\bibinfo {author} {\bibfnamefont {P.}~\bibnamefont {T\"orm\"a}}, \bibinfo {author} {\bibfnamefont {L.}~\bibnamefont {Liang}}, \ and\ \bibinfo {author} {\bibfnamefont {S.}~\bibnamefont {Peotta}},\ }\bibfield  {title} {Quantum metric and effective mass of a two-body bound state in a flat band,\ }\href {\doibase 10.1103/PhysRevB.98.220511} {\bibfield  {journal} {\bibinfo  {journal} {Phys. Rev. B}\ }\textbf {\bibinfo {volume} {98}},\ \bibinfo {pages} {220511} (\bibinfo {year} {2018})}\BibitemShut {NoStop}%
\bibitem [{\citenamefont {Hofmann}\ \emph {et~al.}(2020)\citenamefont {Hofmann}, \citenamefont {Berg},\ and\ \citenamefont {Chowdhury}}]{PhysRevB.102.201112}%
  \BibitemOpen
  \bibfield  {author} {\bibinfo {author} {\bibfnamefont {J.~S.}\ \bibnamefont {Hofmann}}, \bibinfo {author} {\bibfnamefont {E.}~\bibnamefont {Berg}}, \ and\ \bibinfo {author} {\bibfnamefont {D.}~\bibnamefont {Chowdhury}},\ }\bibfield  {title} {Superconductivity, pseudogap, and phase separation in topological flat bands,\ }\href {\doibase 10.1103/PhysRevB.102.201112} {\bibfield  {journal} {\bibinfo  {journal} {Phys. Rev. B}\ }\textbf {\bibinfo {volume} {102}},\ \bibinfo {pages} {201112} (\bibinfo {year} {2020})}\BibitemShut {NoStop}%
\bibitem [{\citenamefont {Julku}\ \emph {et~al.}(2016)\citenamefont {Julku}, \citenamefont {Peotta}, \citenamefont {Vanhala}, \citenamefont {Kim},\ and\ \citenamefont {T\"orm\"a}}]{PhysRevLett.117.045303}%
  \BibitemOpen
  \bibfield  {author} {\bibinfo {author} {\bibfnamefont {A.}~\bibnamefont {Julku}}, \bibinfo {author} {\bibfnamefont {S.}~\bibnamefont {Peotta}}, \bibinfo {author} {\bibfnamefont {T.~I.}\ \bibnamefont {Vanhala}}, \bibinfo {author} {\bibfnamefont {D.-H.}\ \bibnamefont {Kim}}, \ and\ \bibinfo {author} {\bibfnamefont {P.}~\bibnamefont {T\"orm\"a}},\ }\bibfield  {title} {Geometric origin of superfluidity in the {Lieb}-lattice flat band,\ }\href {\doibase 10.1103/PhysRevLett.117.045303} {\bibfield  {journal} {\bibinfo  {journal} {Phys. Rev. Lett.}\ }\textbf {\bibinfo {volume} {117}},\ \bibinfo {pages} {045303} (\bibinfo {year} {2016})}\BibitemShut {NoStop}%
\bibitem [{\citenamefont {Huhtinen}\ \emph {et~al.}(2022)\citenamefont {Huhtinen}, \citenamefont {Herzog-Arbeitman}, \citenamefont {Chew}, \citenamefont {Bernevig},\ and\ \citenamefont {T\"orm\"a}}]{PhysRevB.106.014518}%
  \BibitemOpen
  \bibfield  {author} {\bibinfo {author} {\bibfnamefont {K.-E.}\ \bibnamefont {Huhtinen}}, \bibinfo {author} {\bibfnamefont {J.}~\bibnamefont {Herzog-Arbeitman}}, \bibinfo {author} {\bibfnamefont {A.}~\bibnamefont {Chew}}, \bibinfo {author} {\bibfnamefont {B.~A.}\ \bibnamefont {Bernevig}}, \ and\ \bibinfo {author} {\bibfnamefont {P.}~\bibnamefont {T\"orm\"a}},\ }\bibfield  {title} {Revisiting flat band superconductivity: Dependence on minimal quantum metric and band touchings,\ }\href {\doibase 10.1103/PhysRevB.106.014518} {\bibfield  {journal} {\bibinfo  {journal} {Phys. Rev. B}\ }\textbf {\bibinfo {volume} {106}},\ \bibinfo {pages} {014518} (\bibinfo {year} {2022})}\BibitemShut {NoStop}%
\bibitem [{\citenamefont {Herzog-Arbeitman}\ \emph {et~al.}(2022)\citenamefont {Herzog-Arbeitman}, \citenamefont {Peri}, \citenamefont {Schindler}, \citenamefont {Huber},\ and\ \citenamefont {Bernevig}}]{PhysRevLett.128.087002}%
  \BibitemOpen
  \bibfield  {author} {\bibinfo {author} {\bibfnamefont {J.}~\bibnamefont {Herzog-Arbeitman}}, \bibinfo {author} {\bibfnamefont {V.}~\bibnamefont {Peri}}, \bibinfo {author} {\bibfnamefont {F.}~\bibnamefont {Schindler}}, \bibinfo {author} {\bibfnamefont {S.~D.}\ \bibnamefont {Huber}}, \ and\ \bibinfo {author} {\bibfnamefont {B.~A.}\ \bibnamefont {Bernevig}},\ }\bibfield  {title} {Superfluid weight bounds from symmetry and quantum geometry in flat bands,\ }\href {\doibase 10.1103/PhysRevLett.128.087002} {\bibfield  {journal} {\bibinfo  {journal} {Phys. Rev. Lett.}\ }\textbf {\bibinfo {volume} {128}},\ \bibinfo {pages} {087002} (\bibinfo {year} {2022})}\BibitemShut {NoStop}%
\bibitem [{\citenamefont {Chan}\ \emph {et~al.}(2022)\citenamefont {Chan}, \citenamefont {Gr\'emaud},\ and\ \citenamefont {Batrouni}}]{PhysRevB.106.104514}%
  \BibitemOpen
  \bibfield  {author} {\bibinfo {author} {\bibfnamefont {S.~M.}\ \bibnamefont {Chan}}, \bibinfo {author} {\bibfnamefont {B.}~\bibnamefont {Gr\'emaud}}, \ and\ \bibinfo {author} {\bibfnamefont {G.~G.}\ \bibnamefont {Batrouni}},\ }\bibfield  {title} {Designer flat bands: Topology and enhancement of superconductivity,\ }\href {\doibase 10.1103/PhysRevB.106.104514} {\bibfield  {journal} {\bibinfo  {journal} {Phys. Rev. B}\ }\textbf {\bibinfo {volume} {106}},\ \bibinfo {pages} {104514} (\bibinfo {year} {2022})}\BibitemShut {NoStop}%
\bibitem [{\citenamefont {Chen}\ and\ \citenamefont {Law}(2024)}]{PhysRevLett.132.026002}%
  \BibitemOpen
  \bibfield  {author} {\bibinfo {author} {\bibfnamefont {S.~A.}\ \bibnamefont {Chen}}\ and\ \bibinfo {author} {\bibfnamefont {K.~T.}\ \bibnamefont {Law}},\ }\bibfield  {title} {Ginzburg-{Landau} theory of flat-band superconductors with quantum metric,\ }\href {\doibase 10.1103/PhysRevLett.132.026002} {\bibfield  {journal} {\bibinfo  {journal} {Phys. Rev. Lett.}\ }\textbf {\bibinfo {volume} {132}},\ \bibinfo {pages} {026002} (\bibinfo {year} {2024})}\BibitemShut {NoStop}%
\bibitem [{\citenamefont {Törmä}\ \emph {et~al.}(2022)\citenamefont {Törmä}, \citenamefont {Peotta},\ and\ \citenamefont {Bernevig}}]{Bernevig.Törmä.2022}%
  \BibitemOpen
  \bibfield  {author} {\bibinfo {author} {\bibfnamefont {P.}~\bibnamefont {Törmä}}, \bibinfo {author} {\bibfnamefont {S.}~\bibnamefont {Peotta}}, \ and\ \bibinfo {author} {\bibfnamefont {B.~A.}\ \bibnamefont {Bernevig}},\ }\bibfield  {title} {{Superconductivity, superfluidity and quantum geometry in twisted multilayer systems},\ }\href {\doibase 10.1038/s42254-022-00466-y} {\bibfield  {journal} {\bibinfo  {journal} {Nature Reviews Physics}\ }\textbf {\bibinfo {volume} {4}},\ \bibinfo {pages} {528} (\bibinfo {year} {2022})}\BibitemShut {NoStop}%
\bibitem [{\citenamefont {Wang}\ \emph {et~al.}(2024)\citenamefont {Wang}, \citenamefont {Mendez-Valderrama}, \citenamefont {Hofmann},\ and\ \citenamefont {Chowdhury}}]{PhysRevB.110.L041105}%
  \BibitemOpen
  \bibfield  {author} {\bibinfo {author} {\bibfnamefont {X.}~\bibnamefont {Wang}}, \bibinfo {author} {\bibfnamefont {J.~F.}\ \bibnamefont {Mendez-Valderrama}}, \bibinfo {author} {\bibfnamefont {J.~S.}\ \bibnamefont {Hofmann}}, \ and\ \bibinfo {author} {\bibfnamefont {D.}~\bibnamefont {Chowdhury}},\ }\bibfield  {title} {Intertwined magnetism and superconductivity in isolated correlated flat bands,\ }\href {\doibase 10.1103/PhysRevB.110.L041105} {\bibfield  {journal} {\bibinfo  {journal} {Phys. Rev. B}\ }\textbf {\bibinfo {volume} {110}},\ \bibinfo {pages} {L041105} (\bibinfo {year} {2024})}\BibitemShut {NoStop}%
\bibitem [{\citenamefont {Wang}\ \emph {et~al.}(2025)\citenamefont {Wang}, \citenamefont {Mendez-Valderrama}, \citenamefont {Hofmann},\ and\ \citenamefont {Chowdhury}}]{wang2025sc}%
  \BibitemOpen
  \bibfield  {author} {\bibinfo {author} {\bibfnamefont {X.}~\bibnamefont {Wang}}, \bibinfo {author} {\bibfnamefont {J.~F.}\ \bibnamefont {Mendez-Valderrama}}, \bibinfo {author} {\bibfnamefont {J.~S.}\ \bibnamefont {Hofmann}}, \ and\ \bibinfo {author} {\bibfnamefont {D.}~\bibnamefont {Chowdhury}},\ }\href {https://arxiv.org/abs/2507.22971} {Spin-polaron mediated superconductivity in doped chern antiferromagnets} (\bibinfo {year} {2025}),\ \Eprint {http://arxiv.org/abs/2507.22971} {arXiv:2507.22971 [cond-mat.str-el]} \BibitemShut {NoStop}%
\bibitem [{\citenamefont {Wu}\ and\ \citenamefont {Das~Sarma}(2020)}]{PhysRevB.102.165118}%
  \BibitemOpen
  \bibfield  {author} {\bibinfo {author} {\bibfnamefont {F.}~\bibnamefont {Wu}}\ and\ \bibinfo {author} {\bibfnamefont {S.}~\bibnamefont {Das~Sarma}},\ }\bibfield  {title} {Quantum geometry and stability of moir\'e flatband ferromagnetism,\ }\href {\doibase 10.1103/PhysRevB.102.165118} {\bibfield  {journal} {\bibinfo  {journal} {Phys. Rev. B}\ }\textbf {\bibinfo {volume} {102}},\ \bibinfo {pages} {165118} (\bibinfo {year} {2020})}\BibitemShut {NoStop}%
\bibitem [{\citenamefont {Kang}\ \emph {et~al.}(2024)\citenamefont {Kang}, \citenamefont {Oh}, \citenamefont {Lee},\ and\ \citenamefont {Yang}}]{kang2024}%
  \BibitemOpen
  \bibfield  {author} {\bibinfo {author} {\bibfnamefont {J.}~\bibnamefont {Kang}}, \bibinfo {author} {\bibfnamefont {T.}~\bibnamefont {Oh}}, \bibinfo {author} {\bibfnamefont {J.}~\bibnamefont {Lee}}, \ and\ \bibinfo {author} {\bibfnamefont {B.-J.}\ \bibnamefont {Yang}},\ }\href {https://arxiv.org/abs/2402.07171} {Quantum geometric bound for saturated ferromagnetism} (\bibinfo {year} {2024}),\ \Eprint {http://arxiv.org/abs/2402.07171} {arXiv:2402.07171 [cond-mat.str-el]} \BibitemShut {NoStop}%
\bibitem [{\citenamefont {Bernevig}\ \emph {et~al.}(2021)\citenamefont {Bernevig}, \citenamefont {Lian}, \citenamefont {Cowsik}, \citenamefont {Xie}, \citenamefont {Regnault},\ and\ \citenamefont {Song}}]{PhysRevB.103.205415}%
  \BibitemOpen
  \bibfield  {author} {\bibinfo {author} {\bibfnamefont {B.~A.}\ \bibnamefont {Bernevig}}, \bibinfo {author} {\bibfnamefont {B.}~\bibnamefont {Lian}}, \bibinfo {author} {\bibfnamefont {A.}~\bibnamefont {Cowsik}}, \bibinfo {author} {\bibfnamefont {F.}~\bibnamefont {Xie}}, \bibinfo {author} {\bibfnamefont {N.}~\bibnamefont {Regnault}}, \ and\ \bibinfo {author} {\bibfnamefont {Z.-D.}\ \bibnamefont {Song}},\ }\bibfield  {title} {Twisted bilayer graphene. {V.} exact analytic many-body excitations in coulomb hamiltonians: Charge gap, {Goldstone} modes, and absence of {Cooper} pairing,\ }\href {\doibase 10.1103/PhysRevB.103.205415} {\bibfield  {journal} {\bibinfo  {journal} {Phys. Rev. B}\ }\textbf {\bibinfo {volume} {103}},\ \bibinfo {pages} {205415} (\bibinfo {year} {2021})}\BibitemShut {NoStop}%
\bibitem [{Note1()}]{Note1}%
  \BibitemOpen
  \bibinfo {note} {See Supplemental Material for more technical details and additional supporting data.}\BibitemShut {Stop}%
\bibitem [{\citenamefont {Savary}\ \emph {et~al.}(2017)\citenamefont {Savary}, \citenamefont {Ruhman}, \citenamefont {Venderbos}, \citenamefont {Fu},\ and\ \citenamefont {Lee}}]{savary2017}%
  \BibitemOpen
  \bibfield  {author} {\bibinfo {author} {\bibfnamefont {L.}~\bibnamefont {Savary}}, \bibinfo {author} {\bibfnamefont {J.}~\bibnamefont {Ruhman}}, \bibinfo {author} {\bibfnamefont {J.~W.~F.}\ \bibnamefont {Venderbos}}, \bibinfo {author} {\bibfnamefont {L.}~\bibnamefont {Fu}}, \ and\ \bibinfo {author} {\bibfnamefont {P.~A.}\ \bibnamefont {Lee}},\ }\bibfield  {title} {Superconductivity in three-dimensional spin-orbit coupled semimetals,\ }\href {\doibase 10.1103/PhysRevB.96.214514} {\bibfield  {journal} {\bibinfo  {journal} {Phys. Rev. B}\ }\textbf {\bibinfo {volume} {96}},\ \bibinfo {pages} {214514} (\bibinfo {year} {2017})}\BibitemShut {NoStop}%
\bibitem [{Note2()}]{Note2}%
  \BibitemOpen
  \bibinfo {note} {Due to the orthogonality relation $\protect \mathcal {P}_m(\protect \bm {k}) \protect \mathcal {P}_n(\protect \bm {k})=\delta _{m n} \protect \mathcal {P}_m(\protect \bm {k})$, the target space of $\protect \bm {b}(\protect \bm {k})$ is not an $(N^2-2)$-sphere but rather a specific $2(N-1)$-dimensional subset thereof~\cite {PhysRevB.104.085114}.}\BibitemShut {Stop}%
\bibitem [{\citenamefont {Graf}\ and\ \citenamefont {Pi\'echon}(2021)}]{PhysRevB.104.085114}%
  \BibitemOpen
  \bibfield  {author} {\bibinfo {author} {\bibfnamefont {A.}~\bibnamefont {Graf}}\ and\ \bibinfo {author} {\bibfnamefont {F.}~\bibnamefont {Pi\'echon}},\ }\bibfield  {title} {Berry curvature and quantum metric in ${N}$-band systems: An eigenprojector approach,\ }\href {\doibase 10.1103/PhysRevB.104.085114} {\bibfield  {journal} {\bibinfo  {journal} {Phys. Rev. B}\ }\textbf {\bibinfo {volume} {104}},\ \bibinfo {pages} {085114} (\bibinfo {year} {2021})}\BibitemShut {NoStop}%
\bibitem [{Note3()}]{Note3}%
  \BibitemOpen
  \bibinfo {note} {Specifically, for $N=2$ (resp.\ $N=3$) the $\protect \bm {\lambda }$ matrices can be taken to be the Pauli matrices (resp.\ the Gell-Mann matrices~\cite { PhysRev.125.1067}).}\BibitemShut {Stop}%
\bibitem [{\citenamefont {Kaplan}\ and\ \citenamefont {Resnikoff}(1967)}]{10.1063/1.1705141}%
  \BibitemOpen
  \bibfield  {author} {\bibinfo {author} {\bibfnamefont {L.~M.}\ \bibnamefont {Kaplan}}\ and\ \bibinfo {author} {\bibfnamefont {M.}~\bibnamefont {Resnikoff}},\ }\bibfield  {title} {Matrix products and the explicit 3, 6, 9, and 12‐j coefficients of the regular representation of {SU(}n),\ }\href {\doibase 10.1063/1.1705141} {\bibfield  {journal} {\bibinfo  {journal} {Journal of Mathematical Physics}\ }\textbf {\bibinfo {volume} {8}},\ \bibinfo {pages} {2194} (\bibinfo {year} {1967})}\BibitemShut {NoStop}%
\bibitem [{\citenamefont {Han}\ \emph {et~al.}(2024)\citenamefont {Han}, \citenamefont {Herzog-Arbeitman}, \citenamefont {Bernevig},\ and\ \citenamefont {Kivelson}}]{PhysRevX.14.041004}%
  \BibitemOpen
  \bibfield  {author} {\bibinfo {author} {\bibfnamefont {Z.}~\bibnamefont {Han}}, \bibinfo {author} {\bibfnamefont {J.}~\bibnamefont {Herzog-Arbeitman}}, \bibinfo {author} {\bibfnamefont {B.~A.}\ \bibnamefont {Bernevig}}, \ and\ \bibinfo {author} {\bibfnamefont {S.~A.}\ \bibnamefont {Kivelson}},\ }\bibfield  {title} {``quantum geometric nesting'' and solvable model flat-band systems,\ }\href {\doibase 10.1103/PhysRevX.14.041004} {\bibfield  {journal} {\bibinfo  {journal} {Phys. Rev. X}\ }\textbf {\bibinfo {volume} {14}},\ \bibinfo {pages} {041004} (\bibinfo {year} {2024})}\BibitemShut {NoStop}%
\bibitem [{\citenamefont {{Provost}}\ and\ \citenamefont {{Vallee}}(1980)}]{metric}%
  \BibitemOpen
  \bibfield  {author} {\bibinfo {author} {\bibfnamefont {J.~P.}\ \bibnamefont {{Provost}}}\ and\ \bibinfo {author} {\bibfnamefont {G.}~\bibnamefont {{Vallee}}},\ }\bibfield  {title} {{Riemannian structure on manifolds of quantum states},\ }\href {\doibase 10.1007/BF02193559} {\bibfield  {journal} {\bibinfo  {journal} {Communications in Mathematical Physics}\ }\textbf {\bibinfo {volume} {76}},\ \bibinfo {pages} {289} (\bibinfo {year} {1980})}\BibitemShut {NoStop}%
\bibitem [{\citenamefont {Coleman}(2015)}]{Coleman_2015}%
  \BibitemOpen
  \bibfield  {author} {\bibinfo {author} {\bibfnamefont {P.}~\bibnamefont {Coleman}},\ }Introduction to many-body physics\ (\bibinfo  {publisher} {Cambridge University Press},\ \bibinfo {year} {2015})\BibitemShut {NoStop}%
\bibitem [{\citenamefont {Hu}\ \emph {et~al.}(2025)\citenamefont {Hu}, \citenamefont {Chen},\ and\ \citenamefont {Law}}]{10.1038/s42005-024-01930-0}%
  \BibitemOpen
  \bibfield  {author} {\bibinfo {author} {\bibfnamefont {J.-X.}\ \bibnamefont {Hu}}, \bibinfo {author} {\bibfnamefont {S.~A.}\ \bibnamefont {Chen}}, \ and\ \bibinfo {author} {\bibfnamefont {K.~T.}\ \bibnamefont {Law}},\ }\bibfield  {title} {{Anomalous coherence length in superconductors with quantum metric},\ }\href {\doibase 10.1038/s42005-024-01930-0} {\bibfield  {journal} {\bibinfo  {journal} {Communications Physics}\ }\textbf {\bibinfo {volume} {8}},\ \bibinfo {pages} {20} (\bibinfo {year} {2025})}\BibitemShut {NoStop}%
\bibitem [{Note4()}]{Note4}%
  \BibitemOpen
  \bibinfo {note} {Given that $\protect \bm {b}_{\uparrow }^R(-\protect \bm {k}) = \protect \bm {b}_{\downarrow }(\protect \bm {k})$ in systems with time-reversal symmetry, the expansion of the uniform pairing susceptibility $\chi ^\Delta (\protect \bm {q})$ with respect to small $\protect \bm {q}$ reduces to Eq.\protect \,\protect \eqref {metric}.}\BibitemShut {Stop}%
\bibitem [{\citenamefont {Ozawa}\ and\ \citenamefont {Mera}(2021)}]{PhysRevB.104.045103}%
  \BibitemOpen
  \bibfield  {author} {\bibinfo {author} {\bibfnamefont {T.}~\bibnamefont {Ozawa}}\ and\ \bibinfo {author} {\bibfnamefont {B.}~\bibnamefont {Mera}},\ }\bibfield  {title} {Relations between topology and the quantum metric for {Chern} insulators,\ }\href {\doibase 10.1103/PhysRevB.104.045103} {\bibfield  {journal} {\bibinfo  {journal} {Phys. Rev. B}\ }\textbf {\bibinfo {volume} {104}},\ \bibinfo {pages} {045103} (\bibinfo {year} {2021})}\BibitemShut {NoStop}%
\bibitem [{\citenamefont {Roy}(2014)}]{PhysRevB.90.165139}%
  \BibitemOpen
  \bibfield  {author} {\bibinfo {author} {\bibfnamefont {R.}~\bibnamefont {Roy}},\ }\bibfield  {title} {Band geometry of fractional topological insulators,\ }\href {\doibase 10.1103/PhysRevB.90.165139} {\bibfield  {journal} {\bibinfo  {journal} {Phys. Rev. B}\ }\textbf {\bibinfo {volume} {90}},\ \bibinfo {pages} {165139} (\bibinfo {year} {2014})}\BibitemShut {NoStop}%
\bibitem [{\citenamefont {Jiang}\ and\ \citenamefont {Barlas}(2024)}]{PhysRevB.109.214518}%
  \BibitemOpen
  \bibfield  {author} {\bibinfo {author} {\bibfnamefont {G.}~\bibnamefont {Jiang}}\ and\ \bibinfo {author} {\bibfnamefont {Y.}~\bibnamefont {Barlas}},\ }\bibfield  {title} {Geometric superfluid weight of composite bands in multiorbital superconductors,\ }\href {\doibase 10.1103/PhysRevB.109.214518} {\bibfield  {journal} {\bibinfo  {journal} {Phys. Rev. B}\ }\textbf {\bibinfo {volume} {109}},\ \bibinfo {pages} {214518} (\bibinfo {year} {2024})}\BibitemShut {NoStop}%
\bibitem [{\citenamefont {Hofmann}\ \emph {et~al.}(2023)\citenamefont {Hofmann}, \citenamefont {Berg},\ and\ \citenamefont {Chowdhury}}]{PhysRevLett.130.226001}%
  \BibitemOpen
  \bibfield  {author} {\bibinfo {author} {\bibfnamefont {J.~S.}\ \bibnamefont {Hofmann}}, \bibinfo {author} {\bibfnamefont {E.}~\bibnamefont {Berg}}, \ and\ \bibinfo {author} {\bibfnamefont {D.}~\bibnamefont {Chowdhury}},\ }\bibfield  {title} {Superconductivity, charge density wave, and supersolidity in flat bands with a tunable quantum metric,\ }\href {\doibase 10.1103/PhysRevLett.130.226001} {\bibfield  {journal} {\bibinfo  {journal} {Phys. Rev. Lett.}\ }\textbf {\bibinfo {volume} {130}},\ \bibinfo {pages} {226001} (\bibinfo {year} {2023})}\BibitemShut {NoStop}%
\bibitem [{\citenamefont {Souza}\ \emph {et~al.}(2000)\citenamefont {Souza}, \citenamefont {Wilkens},\ and\ \citenamefont {Martin}}]{souza2000}%
  \BibitemOpen
  \bibfield  {author} {\bibinfo {author} {\bibfnamefont {I.}~\bibnamefont {Souza}}, \bibinfo {author} {\bibfnamefont {T.}~\bibnamefont {Wilkens}}, \ and\ \bibinfo {author} {\bibfnamefont {R.~M.}\ \bibnamefont {Martin}},\ }\bibfield  {title} {Polarization and localization in insulators: Generating function approach,\ }\href {\doibase 10.1103/PhysRevB.62.1666} {\bibfield  {journal} {\bibinfo  {journal} {Phys. Rev. B}\ }\textbf {\bibinfo {volume} {62}},\ \bibinfo {pages} {1666} (\bibinfo {year} {2000})}\BibitemShut {NoStop}%
\bibitem [{\citenamefont {Vanderbilt}(2018)}]{Vanderbilt_2018}%
  \BibitemOpen
  \bibfield  {author} {\bibinfo {author} {\bibfnamefont {D.}~\bibnamefont {Vanderbilt}},\ }Berry phases in electronic structure theory: Electric polarization, orbital magnetization and topological insulators\ (\bibinfo  {publisher} {Cambridge University Press},\ \bibinfo {year} {2018})\BibitemShut {NoStop}%
\bibitem [{\citenamefont {Tasaki}(1998)}]{10.1143/PTP.99.489}%
  \BibitemOpen
  \bibfield  {author} {\bibinfo {author} {\bibfnamefont {H.}~\bibnamefont {Tasaki}},\ }\bibfield  {title} {From {Nagaoka}'s ferromagnetism to flat-band ferromagnetism and beyond: An introduction to ferromagnetism in the {Hubbard} model,\ }\href {\doibase 10.1143/PTP.99.489} {\bibfield  {journal} {\bibinfo  {journal} {Progress of Theoretical Physics}\ }\textbf {\bibinfo {volume} {99}},\ \bibinfo {pages} {489} (\bibinfo {year} {1998})}\BibitemShut {NoStop}%
\bibitem [{\citenamefont {M\"uller}\ \emph {et~al.}(2016)\citenamefont {M\"uller}, \citenamefont {Richter},\ and\ \citenamefont {Derzhko}}]{PhysRevB.93.144418}%
  \BibitemOpen
  \bibfield  {author} {\bibinfo {author} {\bibfnamefont {P.}~\bibnamefont {M\"uller}}, \bibinfo {author} {\bibfnamefont {J.}~\bibnamefont {Richter}}, \ and\ \bibinfo {author} {\bibfnamefont {O.}~\bibnamefont {Derzhko}},\ }\bibfield  {title} {Hubbard models with nearly flat bands: Ground-state ferromagnetism driven by kinetic energy,\ }\href {\doibase 10.1103/PhysRevB.93.144418} {\bibfield  {journal} {\bibinfo  {journal} {Phys. Rev. B}\ }\textbf {\bibinfo {volume} {93}},\ \bibinfo {pages} {144418} (\bibinfo {year} {2016})}\BibitemShut {NoStop}%
\bibitem [{\citenamefont {Chen}\ and\ \citenamefont {Huang}(2022)}]{Chen2022PairDW}%
  \BibitemOpen
  \bibfield  {author} {\bibinfo {author} {\bibfnamefont {W.}~\bibnamefont {Chen}}\ and\ \bibinfo {author} {\bibfnamefont {W.}~\bibnamefont {Huang}},\ }\bibfield  {title} {Pair density wave facilitated by bloch quantum geometry in nearly flat band multiorbital superconductors,\ }\href {https://api.semanticscholar.org/CorpusID:251320574} {\bibfield  {journal} {\bibinfo  {journal} {Science China Physics, Mechanics \& Astronomy}\ }\textbf {\bibinfo {volume} {66}},\ \bibinfo {pages} {1} (\bibinfo {year} {2022})}\BibitemShut {NoStop}%
\bibitem [{\citenamefont {Regnault}\ and\ \citenamefont {Bernevig}(2011)}]{PhysRevX.1.021014}%
  \BibitemOpen
  \bibfield  {author} {\bibinfo {author} {\bibfnamefont {N.}~\bibnamefont {Regnault}}\ and\ \bibinfo {author} {\bibfnamefont {B.~A.}\ \bibnamefont {Bernevig}},\ }\bibfield  {title} {Fractional {Chern} insulator,\ }\href {\doibase 10.1103/PhysRevX.1.021014} {\bibfield  {journal} {\bibinfo  {journal} {Phys. Rev. X}\ }\textbf {\bibinfo {volume} {1}},\ \bibinfo {pages} {021014} (\bibinfo {year} {2011})}\BibitemShut {NoStop}%
\bibitem [{\citenamefont {Wu}\ \emph {et~al.}(2012)\citenamefont {Wu}, \citenamefont {Bernevig},\ and\ \citenamefont {Regnault}}]{PhysRevB.85.075116}%
  \BibitemOpen
  \bibfield  {author} {\bibinfo {author} {\bibfnamefont {Y.-L.}\ \bibnamefont {Wu}}, \bibinfo {author} {\bibfnamefont {B.~A.}\ \bibnamefont {Bernevig}}, \ and\ \bibinfo {author} {\bibfnamefont {N.}~\bibnamefont {Regnault}},\ }\bibfield  {title} {Zoology of fractional {Chern} insulators,\ }\href {\doibase 10.1103/PhysRevB.85.075116} {\bibfield  {journal} {\bibinfo  {journal} {Phys. Rev. B}\ }\textbf {\bibinfo {volume} {85}},\ \bibinfo {pages} {075116} (\bibinfo {year} {2012})}\BibitemShut {NoStop}%
\bibitem [{\citenamefont {Neupert}\ \emph {et~al.}(2011)\citenamefont {Neupert}, \citenamefont {Santos}, \citenamefont {Chamon},\ and\ \citenamefont {Mudry}}]{PhysRevLett.106.236804}%
  \BibitemOpen
  \bibfield  {author} {\bibinfo {author} {\bibfnamefont {T.}~\bibnamefont {Neupert}}, \bibinfo {author} {\bibfnamefont {L.}~\bibnamefont {Santos}}, \bibinfo {author} {\bibfnamefont {C.}~\bibnamefont {Chamon}}, \ and\ \bibinfo {author} {\bibfnamefont {C.}~\bibnamefont {Mudry}},\ }\bibfield  {title} {Fractional quantum {Hall} states at zero magnetic field,\ }\href {\doibase 10.1103/PhysRevLett.106.236804} {\bibfield  {journal} {\bibinfo  {journal} {Phys. Rev. Lett.}\ }\textbf {\bibinfo {volume} {106}},\ \bibinfo {pages} {236804} (\bibinfo {year} {2011})}\BibitemShut {NoStop}%
\bibitem [{\citenamefont {Tang}\ \emph {et~al.}(2011)\citenamefont {Tang}, \citenamefont {Mei},\ and\ \citenamefont {Wen}}]{PhysRevLett.106.236802}%
  \BibitemOpen
  \bibfield  {author} {\bibinfo {author} {\bibfnamefont {E.}~\bibnamefont {Tang}}, \bibinfo {author} {\bibfnamefont {J.-W.}\ \bibnamefont {Mei}}, \ and\ \bibinfo {author} {\bibfnamefont {X.-G.}\ \bibnamefont {Wen}},\ }\bibfield  {title} {High-temperature fractional quantum {Hall} states,\ }\href {\doibase 10.1103/PhysRevLett.106.236802} {\bibfield  {journal} {\bibinfo  {journal} {Phys. Rev. Lett.}\ }\textbf {\bibinfo {volume} {106}},\ \bibinfo {pages} {236802} (\bibinfo {year} {2011})}\BibitemShut {NoStop}%
\bibitem [{\citenamefont {Sun}\ \emph {et~al.}(2011)\citenamefont {Sun}, \citenamefont {Gu}, \citenamefont {Katsura},\ and\ \citenamefont {Das~Sarma}}]{PhysRevLett.106.236803}%
  \BibitemOpen
  \bibfield  {author} {\bibinfo {author} {\bibfnamefont {K.}~\bibnamefont {Sun}}, \bibinfo {author} {\bibfnamefont {Z.}~\bibnamefont {Gu}}, \bibinfo {author} {\bibfnamefont {H.}~\bibnamefont {Katsura}}, \ and\ \bibinfo {author} {\bibfnamefont {S.}~\bibnamefont {Das~Sarma}},\ }\bibfield  {title} {Nearly flatbands with nontrivial topology,\ }\href {\doibase 10.1103/PhysRevLett.106.236803} {\bibfield  {journal} {\bibinfo  {journal} {Phys. Rev. Lett.}\ }\textbf {\bibinfo {volume} {106}},\ \bibinfo {pages} {236803} (\bibinfo {year} {2011})}\BibitemShut {NoStop}%
\bibitem [{\citenamefont {L\"auchli}\ \emph {et~al.}(2013)\citenamefont {L\"auchli}, \citenamefont {Liu}, \citenamefont {Bergholtz},\ and\ \citenamefont {Moessner}}]{PhysRevLett.111.126802}%
  \BibitemOpen
  \bibfield  {author} {\bibinfo {author} {\bibfnamefont {A.~M.}\ \bibnamefont {L\"auchli}}, \bibinfo {author} {\bibfnamefont {Z.}~\bibnamefont {Liu}}, \bibinfo {author} {\bibfnamefont {E.~J.}\ \bibnamefont {Bergholtz}}, \ and\ \bibinfo {author} {\bibfnamefont {R.}~\bibnamefont {Moessner}},\ }\bibfield  {title} {Hierarchy of fractional {Chern} insulators and competing compressible states,\ }\href {\doibase 10.1103/PhysRevLett.111.126802} {\bibfield  {journal} {\bibinfo  {journal} {Phys. Rev. Lett.}\ }\textbf {\bibinfo {volume} {111}},\ \bibinfo {pages} {126802} (\bibinfo {year} {2013})}\BibitemShut {NoStop}%
\bibitem [{\citenamefont {Wilhelm}\ \emph {et~al.}(2021)\citenamefont {Wilhelm}, \citenamefont {Lang},\ and\ \citenamefont {L\"auchli}}]{PhysRevB.103.125406}%
  \BibitemOpen
  \bibfield  {author} {\bibinfo {author} {\bibfnamefont {P.}~\bibnamefont {Wilhelm}}, \bibinfo {author} {\bibfnamefont {T.~C.}\ \bibnamefont {Lang}}, \ and\ \bibinfo {author} {\bibfnamefont {A.~M.}\ \bibnamefont {L\"auchli}},\ }\bibfield  {title} {Interplay of fractional {Chern} insulator and charge density wave phases in twisted bilayer graphene,\ }\href {\doibase 10.1103/PhysRevB.103.125406} {\bibfield  {journal} {\bibinfo  {journal} {Phys. Rev. B}\ }\textbf {\bibinfo {volume} {103}},\ \bibinfo {pages} {125406} (\bibinfo {year} {2021})}\BibitemShut {NoStop}%
\bibitem [{\citenamefont {Song}\ \emph {et~al.}(2024{\natexlab{a}})\citenamefont {Song}, \citenamefont {Zhang},\ and\ \citenamefont {Senthil}}]{PhysRevB.109.085143}%
  \BibitemOpen
  \bibfield  {author} {\bibinfo {author} {\bibfnamefont {X.-Y.}\ \bibnamefont {Song}}, \bibinfo {author} {\bibfnamefont {Y.-H.}\ \bibnamefont {Zhang}}, \ and\ \bibinfo {author} {\bibfnamefont {T.}~\bibnamefont {Senthil}},\ }\bibfield  {title} {Phase transitions out of quantum {Hall} states in moir\'e materials,\ }\href {\doibase 10.1103/PhysRevB.109.085143} {\bibfield  {journal} {\bibinfo  {journal} {Phys. Rev. B}\ }\textbf {\bibinfo {volume} {109}},\ \bibinfo {pages} {085143} (\bibinfo {year} {2024}{\natexlab{a}})}\BibitemShut {NoStop}%
\bibitem [{\citenamefont {Song}\ \emph {et~al.}(2024{\natexlab{b}})\citenamefont {Song}, \citenamefont {Jian}, \citenamefont {Fu},\ and\ \citenamefont {Xu}}]{PhysRevB.109.115116}%
  \BibitemOpen
  \bibfield  {author} {\bibinfo {author} {\bibfnamefont {X.-Y.}\ \bibnamefont {Song}}, \bibinfo {author} {\bibfnamefont {C.-M.}\ \bibnamefont {Jian}}, \bibinfo {author} {\bibfnamefont {L.}~\bibnamefont {Fu}}, \ and\ \bibinfo {author} {\bibfnamefont {C.}~\bibnamefont {Xu}},\ }\bibfield  {title} {Intertwined fractional quantum anomalous {Hall} states and charge density waves,\ }\href {\doibase 10.1103/PhysRevB.109.115116} {\bibfield  {journal} {\bibinfo  {journal} {Phys. Rev. B}\ }\textbf {\bibinfo {volume} {109}},\ \bibinfo {pages} {115116} (\bibinfo {year} {2024}{\natexlab{b}})}\BibitemShut {NoStop}%
\bibitem [{\citenamefont {Mitscherling}\ and\ \citenamefont {Holder}(2022)}]{PhysRevB.105.085154}%
  \BibitemOpen
  \bibfield  {author} {\bibinfo {author} {\bibfnamefont {J.}~\bibnamefont {Mitscherling}}\ and\ \bibinfo {author} {\bibfnamefont {T.}~\bibnamefont {Holder}},\ }\bibfield  {title} {Bound on resistivity in flat-band materials due to the quantum metric,\ }\href {\doibase 10.1103/PhysRevB.105.085154} {\bibfield  {journal} {\bibinfo  {journal} {Phys. Rev. B}\ }\textbf {\bibinfo {volume} {105}},\ \bibinfo {pages} {085154} (\bibinfo {year} {2022})}\BibitemShut {NoStop}%
\bibitem [{\citenamefont {Verma}\ and\ \citenamefont {Queiroz}(2024)}]{verma2024quantummetricstepresponse}%
  \BibitemOpen
  \bibfield  {author} {\bibinfo {author} {\bibfnamefont {N.}~\bibnamefont {Verma}}\ and\ \bibinfo {author} {\bibfnamefont {R.}~\bibnamefont {Queiroz}},\ }\href {https://arxiv.org/abs/2406.17845} {Quantum metric in step response} (\bibinfo {year} {2024}),\ \Eprint {http://arxiv.org/abs/2406.17845} {arXiv:2406.17845 [cond-mat.mes-hall]} \BibitemShut {NoStop}%
\bibitem [{\citenamefont {Resta}(2006)}]{PhysRevLett.96.137601}%
  \BibitemOpen
  \bibfield  {author} {\bibinfo {author} {\bibfnamefont {R.}~\bibnamefont {Resta}},\ }\bibfield  {title} {Polarization fluctuations in insulators and metals: New and old theories merge,\ }\href {\doibase 10.1103/PhysRevLett.96.137601} {\bibfield  {journal} {\bibinfo  {journal} {Phys. Rev. Lett.}\ }\textbf {\bibinfo {volume} {96}},\ \bibinfo {pages} {137601} (\bibinfo {year} {2006})}\BibitemShut {NoStop}%
\bibitem [{\citenamefont {Onishi}\ and\ \citenamefont {Fu}(2024)}]{Onishi_2024}%
  \BibitemOpen
  \bibfield  {author} {\bibinfo {author} {\bibfnamefont {Y.}~\bibnamefont {Onishi}}\ and\ \bibinfo {author} {\bibfnamefont {L.}~\bibnamefont {Fu}},\ }\bibfield  {title} {Topological bound on the structure factor,\ }\href {\doibase 10.1103/physrevlett.133.206602} {\bibfield  {journal} {\bibinfo  {journal} {Physical Review Letters}\ }\textbf {\bibinfo {volume} {133}} (\bibinfo {year} {2024}),\ 10.1103/physrevlett.133.206602}\BibitemShut {NoStop}%
\bibitem [{\citenamefont {Souza}\ and\ \citenamefont {Vanderbilt}(2008)}]{PhysRevB.77.054438}%
  \BibitemOpen
  \bibfield  {author} {\bibinfo {author} {\bibfnamefont {I.}~\bibnamefont {Souza}}\ and\ \bibinfo {author} {\bibfnamefont {D.}~\bibnamefont {Vanderbilt}},\ }\bibfield  {title} {Dichroic $f$-sum rule and the orbital magnetization of crystals,\ }\href {\doibase 10.1103/PhysRevB.77.054438} {\bibfield  {journal} {\bibinfo  {journal} {Phys. Rev. B}\ }\textbf {\bibinfo {volume} {77}},\ \bibinfo {pages} {054438} (\bibinfo {year} {2008})}\BibitemShut {NoStop}%
\bibitem [{\citenamefont {Neupert}\ \emph {et~al.}(2013)\citenamefont {Neupert}, \citenamefont {Chamon},\ and\ \citenamefont {Mudry}}]{PhysRevB.87.245103}%
  \BibitemOpen
  \bibfield  {author} {\bibinfo {author} {\bibfnamefont {T.}~\bibnamefont {Neupert}}, \bibinfo {author} {\bibfnamefont {C.}~\bibnamefont {Chamon}}, \ and\ \bibinfo {author} {\bibfnamefont {C.}~\bibnamefont {Mudry}},\ }\bibfield  {title} {Measuring the quantum geometry of {Bloch} bands with current noise,\ }\href {\doibase 10.1103/PhysRevB.87.245103} {\bibfield  {journal} {\bibinfo  {journal} {Phys. Rev. B}\ }\textbf {\bibinfo {volume} {87}},\ \bibinfo {pages} {245103} (\bibinfo {year} {2013})}\BibitemShut {NoStop}%
\bibitem [{\citenamefont {Nozi\`eres}\ and\ \citenamefont {Pines}(1958)}]{PhysRev.109.741}%
  \BibitemOpen
  \bibfield  {author} {\bibinfo {author} {\bibfnamefont {P.}~\bibnamefont {Nozi\`eres}}\ and\ \bibinfo {author} {\bibfnamefont {D.}~\bibnamefont {Pines}},\ }\bibfield  {title} {Electron interaction in solids. general formulation,\ }\href {\doibase 10.1103/PhysRev.109.741} {\bibfield  {journal} {\bibinfo  {journal} {Phys. Rev.}\ }\textbf {\bibinfo {volume} {109}},\ \bibinfo {pages} {741} (\bibinfo {year} {1958})}\BibitemShut {NoStop}%
\bibitem [{\citenamefont {Gell-Mann}(1962)}]{PhysRev.125.1067}%
  \BibitemOpen
  \bibfield  {author} {\bibinfo {author} {\bibfnamefont {M.}~\bibnamefont {Gell-Mann}},\ }\bibfield  {title} {Symmetries of baryons and mesons,\ }\href {\doibase 10.1103/PhysRev.125.1067} {\bibfield  {journal} {\bibinfo  {journal} {Phys. Rev.}\ }\textbf {\bibinfo {volume} {125}},\ \bibinfo {pages} {1067} (\bibinfo {year} {1962})}\BibitemShut {NoStop}%
\bibitem [{\citenamefont {Deng}\ \emph {et~al.}(2024)\citenamefont {Deng}, \citenamefont {Li}, \citenamefont {Wu}, \citenamefont {Sun},\ and\ \citenamefont {Chen}}]{DENG2024169639}%
  \BibitemOpen
  \bibfield  {author} {\bibinfo {author} {\bibfnamefont {H.}~\bibnamefont {Deng}}, \bibinfo {author} {\bibfnamefont {C.}~\bibnamefont {Li}}, \bibinfo {author} {\bibfnamefont {Y.}~\bibnamefont {Wu}}, \bibinfo {author} {\bibfnamefont {L.}~\bibnamefont {Sun}}, \ and\ \bibinfo {author} {\bibfnamefont {Q.}~\bibnamefont {Chen}},\ }\bibfield  {title} {Flat band effects on the ground-state bcs-bec crossover in atomic fermi gases in a quasi-two-dimensional lieb lattice,\ }\href {\doibase https://doi.org/10.1016/j.aop.2024.169639} {\bibfield  {journal} {\bibinfo  {journal} {Annals of Physics}\ }\textbf {\bibinfo {volume} {463}},\ \bibinfo {pages} {169639} (\bibinfo {year} {2024})}\BibitemShut {NoStop}%
\end{thebibliography}%

\clearpage
\newpage
\widetext

\begin{center}
\textbf{\large Supplemental Material for \\``Identifying Instabilities with Quantum Geometry in Flat Band Systems''}
\end{center}
\addtocontents{toc}{\protect\setcounter{tocdepth}{0}}
{
\tableofcontents
}

\renewcommand{\thefigure}{S\arabic{figure}}
\setcounter{figure}{0}
\renewcommand{\theequation}{S\arabic{equation}}
\setcounter{equation}{0}
\renewcommand{\thesection}{\Roman{section}}
\setcounter{section}{0}
\setcounter{secnumdepth}{4}

\section{Derivation Details of the Nesting Condition}

In this section, we provide a more detailed derivation of Eq.~(6) and Eq.~(7) in the main text.

\subsection{Particle-Hole Excitation}\label{IA}

For a generic multi-orbital (multi-band) system, the transformation between the orbital and band representations is given by  
\begin{equation}\label{Uc}
    c_\alpha(\boldsymbol{k})=U_{\alpha m}(\boldsymbol{k}) c_m(\boldsymbol{k}),
\end{equation}
where $c_m(\boldsymbol{k})$ and $c_\alpha(\boldsymbol{k})$ are the electronic operators in the band and orbital (spin) representations, respectively. The eigenprojector matrix onto a given band $m$ is given by  
\begin{equation}\label{P0}
    \left(\mathcal{P}_m\right)_{\alpha \beta}(\boldsymbol{k})= U_{\alpha m}(\boldsymbol{k}) U_{m \beta}^{\dagger}(\boldsymbol{k}).
\end{equation}
When the low-energy sector consists of $N_L$ nearly degenerate bands, whose bandwidths are much smaller than the energy gap separating them from the high-energy sector, the eigenprojector onto the low-energy sector is \cite{savary2017}
\begin{eqnarray}\label{PL}
    P(\boldsymbol{k})=\sum_{m \in L}\mathcal{P}_m(\boldsymbol{k}),
\end{eqnarray}
where ``$L$'' denotes the low-energy sector.
The general particle-hole order parameter with ordering vector $\boldsymbol{Q}$, which is defined in the physical orbital representation, is given by  
\begin{equation}\label{Oph}
    \hat O^{\mathrm{ph}}_{\boldsymbol{Q}}=\sum_{\boldsymbol{k}}  \mathcal O_{\alpha \beta}(\boldsymbol{k}) c_\alpha^{\dagger}(\boldsymbol{k}+\boldsymbol{Q}) c_\beta(\boldsymbol{k})=\sum_{m, n \in L} \sum_{\boldsymbol{k}} c_m^{\dagger}(\boldsymbol{k}+\boldsymbol{Q})  U_{m \alpha}^{\dagger}(\boldsymbol{k}+\boldsymbol{Q}) \mathcal O_{\alpha \beta}(\boldsymbol{k})  U_{\beta n}(\boldsymbol{k}) c_n(\boldsymbol{k}),
\end{equation}
where the repeated indices $\alpha$ and $\beta$ imply summation. The susceptibility corresponding to \eqnref{Oph} at temperature $T=1/\beta$ is given by:
\begin{eqnarray}\label{chiph}
    \begin{aligned}
    \chi^{\mathrm{ph}}_{\boldsymbol{Q}}=&\int_0^\beta d \tau\left(\left\langle \hat O^{\mathrm{ph},\dagger}_{\boldsymbol{Q}} (\tau) \hat O^{\mathrm{ph}}_{\boldsymbol{Q}} (0)\right\rangle-\left\langle \hat O^{\mathrm{ph},\dagger}_{\boldsymbol{Q}} (\tau)\right\rangle\left\langle \hat O^{\mathrm{ph}}_{\boldsymbol{Q}} (0)\right\rangle\right)\\
    =& \sum_{m, m' \in L} \sum_{\boldsymbol{k}} U_{m' \alpha^{\prime}}^{\dagger}(\boldsymbol{k}) \mathcal O_{\alpha^{\prime} \beta^{\prime}}^{\dagger}(\boldsymbol{k})  U_{\beta^{\prime} m}(\boldsymbol{k}+\boldsymbol{Q}) U_{m \alpha}^{\dagger}(\boldsymbol{k}+\boldsymbol{Q}) \mathcal O_{\alpha \beta}(\boldsymbol{k})  U_{\beta m'}(\boldsymbol{k})\\
    &\times\sum _{\omega_n}\left\langle c_m(\boldsymbol{k}+\boldsymbol{Q}, i\omega_n) c_m^{\dagger}(\boldsymbol{k}+\boldsymbol{Q}, i\omega_n)\right\rangle\left\langle c_{m'}^{\dagger}(\boldsymbol{k}, i\omega_n) c_{m'}(\boldsymbol{k}, i\omega_n)\right\rangle\\
    =&\sum_{m, m' \in L} \sum_{\boldsymbol{k}} U_{m' \alpha^{\prime}}^{\dagger}(\boldsymbol{k}) \mathcal O_{\alpha^{\prime} \beta^{\prime}}^{\dagger}(\boldsymbol{k})  U_{\beta^{\prime} m}(\boldsymbol{k}+\boldsymbol{Q}) U_{m \alpha}^{\dagger}(\boldsymbol{k}+\boldsymbol{Q}) \mathcal O_{\alpha \beta}(\boldsymbol{k})  U_{\beta m'}(\boldsymbol{k})\frac{n_F\left[\epsilon_m(\boldsymbol{k}+\boldsymbol{Q})\right]-n_F\left[\epsilon_{m'}(\boldsymbol{k})\right]}{\epsilon_{m'}(\boldsymbol{k})-\epsilon_m(\boldsymbol{k}+\boldsymbol{Q})}
    \\
    \approx&\frac{1}{4T}\sum_{\boldsymbol{k}} \operatorname{Tr}\left[\mathcal O^{\dagger}(\boldsymbol{k})  P(\boldsymbol{k}+\boldsymbol{Q}) \mathcal O(\boldsymbol{k})  P(\boldsymbol{k})\right],
    \end{aligned}
\end{eqnarray}
where $n_F(\omega)=1/(e^{\beta \omega}+1)$ is the Fermi-Dirac distribution function, and $\epsilon_m(\boldsymbol{k})$ denotes the dispersion of band $m$ \cite{savary2017}.
In the last line of \eqnref{chiph}, the approximation assumes that the temperature is larger than both the low-energy sector bandwidth and the small energy gap between bands in the low-energy sector.  

In the following, we use the Bloch vector representation to derive a compact expression for the susceptibility in \eqnref{chiph}. First, the projection operator for a given band in \eqnref{P0} can be expressed in terms of the Bloch vector as:
\begin{equation}
    \mathcal{P}_m(\boldsymbol{k})=\frac{1}{N} \mathds{1}_N+\frac{1}{2} \boldsymbol{v}_m(\boldsymbol{k}) \cdot \boldsymbol{\lambda},
\end{equation}
where $\boldsymbol{\lambda}$ are the elementary generator matrices of the SU(N) Lie group.
Note that, due to the orthogonality relation $\mathcal{P}_m(\boldsymbol{k}) \mathcal{P}_n(\boldsymbol{k})=\delta_{m n} \mathcal{P}_m(\boldsymbol{k})$, the target space of $\boldsymbol{v}_m(\boldsymbol{k})$ is not an $(N^2-2)$-sphere but rather a specific $2(N-1)$-dimensional subset.
Similarly, the projection operator for the low-energy sector in \eqnref{PL} can be expressed as:
\begin{equation}
    P(\boldsymbol{k})=\sum_{m \in L} \mathcal P_m(\boldsymbol{k})=n \mathds{1}_N+\frac{1}{2} \boldsymbol{b}(\boldsymbol{k}) \cdot \boldsymbol{\lambda},
\end{equation}
where $n=\frac{N_{L}}{N}$, with $N_{L}$ denoting the number of bands in the low-energy sector. Similarly, the order parameter in \eqnref{Oph} can be expressed as:
\begin{equation}
    \mathcal O(\boldsymbol{k})=o_0(\boldsymbol{k}) \mathds{1}_N+\boldsymbol{o}(\boldsymbol{k}) \cdot \boldsymbol{\lambda}.
\end{equation}
Assume that $\boldsymbol{o}(\boldsymbol{k})$ is either a purely real or purely imaginary vector [when $\boldsymbol{o}(\boldsymbol{k})$ is a purely imaginary vector, then let $\boldsymbol{o}(\boldsymbol{k}) \rightarrow-i \boldsymbol{o}(\boldsymbol{k})$, so $\boldsymbol{o}(\boldsymbol{k})$ can be a purely real vector]. Under this assumption, the susceptibility for a particle-hole excitation in \eqnref{chiph} goes like:
\begin{eqnarray}\label{mid}
    \begin{aligned}
        &\operatorname{Tr}\left[O^{\dagger}(\boldsymbol{k}) P(\boldsymbol{k}+\boldsymbol{Q}) O(\boldsymbol{k}) P(\boldsymbol{k})\right]\\
        =&N  n^2 o_0^2(\boldsymbol{k}) + 2n o_0(\boldsymbol{k}) \boldsymbol{o}(\boldsymbol{k}) \cdot \boldsymbol{b}(\boldsymbol{k}+\boldsymbol{Q})+2n^2  |\boldsymbol{o}(\boldsymbol{k})|^2 +2n o_0(\boldsymbol{k}) \boldsymbol{o}(\boldsymbol{k}) \cdot \boldsymbol{b}(\boldsymbol{k})\\
        &+\frac{o_0^2(\boldsymbol{k})}{2} \boldsymbol{b}(\boldsymbol{k}+\boldsymbol{Q}) \cdot \boldsymbol{b}(\boldsymbol{k})+\frac{o_0(\boldsymbol{k})}{2}\boldsymbol{o}(\boldsymbol{k}) \star \boldsymbol{b}(\boldsymbol{k}+\boldsymbol{Q}) \cdot \boldsymbol{b}(\boldsymbol{k})\\
        &+\frac{o_0(\boldsymbol{k})}{2}\boldsymbol{b}(\boldsymbol{k}+\boldsymbol{Q}) \star \boldsymbol{o}(\boldsymbol{k}) \cdot \boldsymbol{b}(\boldsymbol{k})+\frac{1}{N} \boldsymbol{o}(\boldsymbol{k}) \cdot \boldsymbol{b}(\boldsymbol{k}+\boldsymbol{Q}) \boldsymbol{o}(\boldsymbol{k}) \cdot \boldsymbol{b}(\boldsymbol{k})\\
        &+\frac{1}{2}(\boldsymbol{o}(\boldsymbol{k}) \star \boldsymbol{b}(\boldsymbol{k}+\boldsymbol{Q})+i \boldsymbol{o}(\boldsymbol{k}) \times \boldsymbol{b}(\boldsymbol{k}+\boldsymbol{Q})) \cdot(\boldsymbol{o}(\boldsymbol{k}) \star \boldsymbol{b}(\boldsymbol{k})+i\boldsymbol{o}(\boldsymbol{k}) \times \boldsymbol{b}(\boldsymbol{k})),
    \end{aligned}
\end{eqnarray}
where we used the relations:
\begin{eqnarray}
    (\boldsymbol{m} \cdot \boldsymbol{\lambda})(\boldsymbol{n} \cdot \boldsymbol{\lambda})&=&\frac{2}{N} \boldsymbol{m} \cdot \boldsymbol{n} \mathds{1}_N+(\boldsymbol{m} \star \boldsymbol{n}+i\boldsymbol{m} \times \boldsymbol{n}) \cdot \boldsymbol{\lambda},\\
    \operatorname{Tr}[(\boldsymbol{m} \cdot \boldsymbol{\lambda})(\boldsymbol{n} \cdot \boldsymbol{\lambda})(\boldsymbol{v} \cdot \boldsymbol{\lambda})]&=&2(\boldsymbol{m} \star \boldsymbol{n}+i\boldsymbol{m} \times \boldsymbol{n}) \cdot \boldsymbol{v},\\
    \operatorname{Tr}[(\boldsymbol{m} \cdot \boldsymbol{\lambda})(\boldsymbol{n} \cdot \boldsymbol{\lambda})(\boldsymbol{v} \cdot \boldsymbol{\lambda})(\boldsymbol{h} \cdot \boldsymbol{\lambda})]&=&\frac{4}{N} (\boldsymbol{m} \cdot \boldsymbol{n}) (\boldsymbol{v} \cdot \boldsymbol{h}) +2(\boldsymbol{m} \star \boldsymbol{n}+i \boldsymbol{m} \times \boldsymbol{n}) \cdot(\boldsymbol{v} \star \boldsymbol{h}+i \boldsymbol{v} \times \boldsymbol{h}).
\end{eqnarray}
The symbols $\cdot, \times, \star$ are defined as follows:
\begin{eqnarray}
    \begin{aligned}
& \boldsymbol{m} \cdot \boldsymbol{n}=m_i n_i, \\
& (\boldsymbol{m} \times \boldsymbol{n})_i=f_{ijk} m_j n_k, \\
& (\boldsymbol{m} \star \boldsymbol{n})_i=d_{ijk} m_j n_k,
\end{aligned}
\end{eqnarray}
where
\begin{equation}\label{li}
    f_{ijk} \equiv-\frac{i}{4} \operatorname{Tr}\left(\left[\boldsymbol{\lambda}_i, \boldsymbol{\lambda}_j\right] \boldsymbol{\lambda}_k\right), \;\;\;\;\;\;\;d_{ijk} \equiv \frac{1}{4} \operatorname{Tr}\left(\left\{\boldsymbol{\lambda}_i, \boldsymbol{\lambda}_j\right\} \boldsymbol{\lambda}_k\right).
\end{equation}
Some relations are also helpful when deriving the expression \eqnref{mid}:
\begin{eqnarray}
    \begin{aligned}
    &\boldsymbol{o}(\boldsymbol{k}) \times \boldsymbol{o}(\boldsymbol{k})=0,\\
    &(\boldsymbol m \star \boldsymbol n) \cdot \boldsymbol h= \boldsymbol m \cdot(\boldsymbol n \star \boldsymbol h).
    \end{aligned}
\end{eqnarray}
The relation $\boldsymbol{o}(\boldsymbol{k}) \star \boldsymbol{o}(\boldsymbol{k}) = 0$ can further simplify the expression in such cases.

Due to the complexity of \eqnref{mid}, we will analyze it case by case.
\begin{itemize}
    \item case 1: $o_0(\boldsymbol{k}) \neq 0$ and $\boldsymbol{o}(\boldsymbol{k})=\boldsymbol{0}$
\end{itemize}
Then, \eqnsref{chiph} and (\ref{mid}) become
\begin{eqnarray}\label{chiphre1}
    \chi^{\mathrm{ph}}_{\boldsymbol{Q}} =  \frac{1}{4T}\sum_{\boldsymbol{k}}\operatorname{Tr}\left[O^{\dagger}(\boldsymbol{k}) P(\boldsymbol{k}+\boldsymbol{Q}) O(\boldsymbol{k}) P(\boldsymbol{k})\right]=\frac{1}{4T}\sum_{\boldsymbol{k}}o_0^2(\boldsymbol{k})\left(N n^2+\frac{1}{2} \boldsymbol{b}(\boldsymbol{k}+\boldsymbol{Q}) \cdot \boldsymbol{b}(\boldsymbol{k})\right).
\end{eqnarray}
Note that the magnitude of  the Bloch vector $b(\boldsymbol{k})$ is constant and given by:
\begin{equation}\label{magni}
    |\boldsymbol{b}(\boldsymbol{k})|^2=\sum_{m, m' \in L} \boldsymbol{v}_m(\boldsymbol{k}) \cdot \boldsymbol{v}_{m'}(\boldsymbol{k})=2 N\left(n-n^2\right).
\end{equation}
where we used the relation $\boldsymbol{v}_m(\boldsymbol{k}) \cdot \boldsymbol{v}_{m'}(\boldsymbol{k})=2\left(\delta_{m m'}-\frac{1}{N}\right)$. Therefore, \eqnref{chiphre1} reaches it maximum when all Bloch vectors $\boldsymbol{b}(\boldsymbol{k})$ across the BZ are parallel to the Bloch vector  $\boldsymbol{b}(\boldsymbol{k}+\boldsymbol{Q})$:
\begin{equation}
    \boldsymbol{b}(\boldsymbol{k}+\boldsymbol{Q}) \;\; \parallel\;\; \boldsymbol{b}(\boldsymbol{k}),\;\;\;\;\;\;\;\;\;\; \forall \boldsymbol{k} \in \mathrm{BZ}.
\end{equation}
\begin{itemize}
    \item case 2: $o_0(\boldsymbol{k}) = 0$ and $\boldsymbol{o}(\boldsymbol{k})\neq \boldsymbol{0}$
\end{itemize}
Now, \eqnref{mid} reduces to
\begin{eqnarray}
    \begin{aligned}
        &\operatorname{Tr}\left[O^{\dagger}(\boldsymbol{k}) P(\boldsymbol{k}+\boldsymbol{Q}) O(\boldsymbol{k}) P(\boldsymbol{k})\right]\\
        =&2 n^2 |\boldsymbol{o}(\boldsymbol{k})|^2 +\frac{1}{N} \boldsymbol{o}(\boldsymbol{k}) \cdot \boldsymbol{b}(\boldsymbol{k}+\boldsymbol{Q}) \boldsymbol{o}(\boldsymbol{k}) \cdot \boldsymbol{b}(\boldsymbol{k})+\frac{1}{2}(\boldsymbol{o}(\boldsymbol{k}) \star \boldsymbol{b}(\boldsymbol{k}+\boldsymbol{Q})) \cdot(\boldsymbol{o}(\boldsymbol{k}) \star \boldsymbol{b}(\boldsymbol{k}))-\frac{1}{2}(\boldsymbol{o}(\boldsymbol{k}) \times \boldsymbol{b}(\boldsymbol{k}+\boldsymbol{Q})) \cdot(\boldsymbol{o}(\boldsymbol{k}) \times \boldsymbol{b}(-\boldsymbol{k})),
    \end{aligned}
\end{eqnarray}
where we used:
\begin{equation}
    (\boldsymbol{m} \times \boldsymbol n) \cdot(\boldsymbol o \star \boldsymbol p)+(\boldsymbol m \star \boldsymbol o) \cdot(\boldsymbol p \star \boldsymbol n)+(\boldsymbol m \star \boldsymbol p) \cdot(\boldsymbol n \star \boldsymbol o)=0.
\end{equation}
Furthermore, we have 
\begin{eqnarray}
    \begin{aligned}
    &(\boldsymbol o(\boldsymbol k) \star \boldsymbol b(\boldsymbol k+\boldsymbol Q)) \cdot(\boldsymbol o(\boldsymbol k) \star \boldsymbol b(\boldsymbol k))\\
    =&(\boldsymbol{o}(\boldsymbol{k}) \times \boldsymbol{b}(\boldsymbol{k})) \cdot(\boldsymbol{b}(\boldsymbol{k}+\boldsymbol{Q}) \times \boldsymbol{o}(\boldsymbol{k}))-\frac{2}{N} \boldsymbol{o}(\boldsymbol{k}) \cdot \boldsymbol{b}(\boldsymbol{k}+\boldsymbol{Q}) \boldsymbol{b}(\boldsymbol{k}) \cdot \boldsymbol{o}(\boldsymbol{k})+\frac{2}{N}|\boldsymbol{o}(\boldsymbol{k}) |^2\boldsymbol{b}(\boldsymbol{k}+\boldsymbol{Q}) \cdot \boldsymbol{b}(\boldsymbol{k}),
    \end{aligned}
\end{eqnarray}
since
\begin{eqnarray}\label{re1}
    (\boldsymbol m \times \boldsymbol n) \cdot(\boldsymbol o \times \boldsymbol p)=\frac{2}{N}[(\boldsymbol m \cdot \boldsymbol o)(\boldsymbol n \cdot \boldsymbol p)-(\boldsymbol m \cdot \boldsymbol p)(\boldsymbol n \cdot \boldsymbol o)]+(\boldsymbol m \star \boldsymbol o) \cdot(\boldsymbol n \star \boldsymbol p)-(\boldsymbol m \star \boldsymbol p) \cdot(\boldsymbol n \star \boldsymbol o).
\end{eqnarray}
In turn we can write \eqnref{chiph} as follows:
\begin{eqnarray}  \chi^{\mathrm{ph}}_{\boldsymbol{Q}}=\frac{1}{4T}\sum_{\boldsymbol{k}}|\boldsymbol{o}(\boldsymbol{k})|^2\left[2 n^2+\frac{1}{N} \tilde{\boldsymbol{b}}_{\boldsymbol{o}}(\boldsymbol{k}+\boldsymbol{Q}) \cdot \boldsymbol{b}(\boldsymbol{k})\right],
\end{eqnarray}
where  $\tilde{\boldsymbol{b}}_{\boldsymbol{o}}(\boldsymbol{k}+\boldsymbol{Q})$ is the dressed Bloch vector, corrected by the order parameter:
\begin{eqnarray} \tilde{\boldsymbol{b}}_{\boldsymbol{o}}(\boldsymbol{k}+\boldsymbol{Q}) \equiv \boldsymbol{b}(\boldsymbol{k}+\boldsymbol{Q})-N(\hat{\boldsymbol{o}}(\boldsymbol{k}) \times \boldsymbol{b}(\boldsymbol{k}+\boldsymbol{Q}) \times \hat{\boldsymbol{o}}(\boldsymbol{k})).
\end{eqnarray}
Here $\hat{\boldsymbol{o}}(\boldsymbol{k})$ is the unit vector of the order parameter, defined by:
\begin{eqnarray}
    \hat{\boldsymbol{o}}(\boldsymbol{k}) = \frac{\boldsymbol{o}(\boldsymbol{k})}{|\boldsymbol{o}(\boldsymbol{k})|}.
\end{eqnarray}
Using \eqnref{re1}, we can prove that the amplitude of the dressed Bloch vector $\tilde{\boldsymbol{b}}_{\boldsymbol{o}}(\boldsymbol{k})$ cannot exceed that of $\boldsymbol{b}(\boldsymbol{k})$:
\begin{eqnarray}   |\tilde{\boldsymbol{b}}_{\boldsymbol{o}}(\boldsymbol{k})|^2=|\boldsymbol b(\boldsymbol{k})|^2-N^2|(\hat{\boldsymbol{o}}(\boldsymbol{k}) \times \boldsymbol{b}(\boldsymbol{k})) \star \hat{\boldsymbol{o}}(\boldsymbol{k})|^2\leq|\boldsymbol b(\boldsymbol k)|^2.
\end{eqnarray}
Then, the condition for obtaining the maximum of \eqnsref{chiph} can be rewritten as:
\begin{eqnarray}\label{phnest} \tilde{\boldsymbol{b}}_{\boldsymbol{o}}(\boldsymbol{k}+\boldsymbol{Q}) \;\; \parallel\;\; \boldsymbol{b}(\boldsymbol{k}),\;\;\;\;\;\;\;\;\;\; \forall \boldsymbol k \in \mathrm{BZ}.
\end{eqnarray}
This indicates that the maximum susceptibility is achieved when all Bloch vectors $\boldsymbol{b}(\boldsymbol{k})$ across the BZ are parallel to the dressed Bloch vector $\tilde{\boldsymbol{b}}_{\boldsymbol{o}}(\boldsymbol{k}+\boldsymbol{Q})$.

\subsection{Particle-Particle Excitation}\label{IB}
Similarly, a general particle-particle order parameter with wave vector $\boldsymbol{Q}$ is given by:
\begin{equation}\label{Opp}
    \hat O^{\mathrm{pp}}_{\boldsymbol{Q}}=\sum_{\boldsymbol{k}}  \mathcal O_{\alpha \beta}(\boldsymbol{k}) c_\alpha(\boldsymbol{k}+\boldsymbol{Q}) c_\beta(\boldsymbol{k})=\sum_{m, n \in L} \sum_{\boldsymbol{k}} c_m(\boldsymbol{k}+\boldsymbol{Q})  U_{\alpha m}(\boldsymbol{k}+\boldsymbol{Q}) \mathcal O_{\alpha \beta}(\boldsymbol{k})  U_{\beta n}(-\boldsymbol{k}) c_n(-\boldsymbol{k}),
\end{equation}
so that the corresponding susceptibility is given by:
\begin{eqnarray}\label{chipp}
    \begin{aligned}
    \chi^{\mathrm{pp}}_{\boldsymbol{Q}}=&\int_0^\beta d \tau\left(\left\langle \hat O^{\mathrm{pp},\dagger}_{\boldsymbol{Q}} (\tau) \hat O^{\mathrm{pp}}_{\boldsymbol{Q}} (0)\right\rangle-\left\langle \hat O^{\mathrm{pp},\dagger}_{\boldsymbol{Q}} (\tau)\right\rangle\left\langle \hat O^{\mathrm{pp}}_{\boldsymbol{Q}} (0)\right\rangle\right)\\
    \approx&\frac{1}{4T} \frac{|\nu|}{\mathrm{arctanh}|\nu|}\sum_{\boldsymbol{k}} \operatorname{Tr}\left[\mathcal O^{\dagger}(\boldsymbol{k})  P^*(\boldsymbol{k}+\boldsymbol{Q}) \mathcal O(\boldsymbol{k})  P(\boldsymbol{-k})\right] ,
    \end{aligned}
\end{eqnarray}
where $\nu \equiv \tanh \frac{\mu}{2 T} \in(-1,1)$ denotes the filling fraction of the flat bands, and $\mu$ is the chemical potential measured from charge neutrality. 
Compared to particle-hole excitations, we additionally need to define the complex conjugate of the projection operator, i.e.:
\begin{equation}
    P^*(\boldsymbol{k})=n \mathds{1}_N+\frac{1}{2} \boldsymbol{b}^R(\boldsymbol{k}) \cdot \boldsymbol{\lambda},
\end{equation}
where $\boldsymbol{b}^R(\boldsymbol{k})$ is:
\begin{equation}
    (\boldsymbol{b}(\boldsymbol{k}) \cdot \boldsymbol \lambda)^*=\boldsymbol{b}^R(\boldsymbol{k}) \cdot \boldsymbol \lambda.
\end{equation}
Then, to obtain $\chi^{\mathrm{pp}}_{\boldsymbol{Q}}$ (\eqnref{chipp}) from $\chi^{\mathrm{ph}}_{\boldsymbol{Q}}$ (\eqnref{chiph}), we only need to make the following replacements:
\begin{equation}
    \begin{aligned}
        \boldsymbol{b}(\boldsymbol{k}) &\rightarrow \boldsymbol{b}(-\boldsymbol{k}),\\\boldsymbol{b}(\boldsymbol{k}+\boldsymbol{Q}) &\rightarrow \boldsymbol{b}^R(\boldsymbol{k}+\boldsymbol{Q}).
        \end{aligned}
\end{equation}
The conclusions for different cases are:
\begin{itemize}
    \item case 1: $o_0(\boldsymbol{k}) \neq 0$ and $\boldsymbol{o}(\boldsymbol{k})=\boldsymbol{0}$
\end{itemize}
The susceptibility in \eqnref{chipp} is then:
\begin{eqnarray}\label{chippfinal}
    \chi^{\mathrm{pp}}_{\boldsymbol{Q}}=\frac{1}{4T} \frac{|\nu|}{\mathrm{arctanh}|\nu|}\sum_{\boldsymbol{k}} o_0^2(\boldsymbol{k})\left(N n^2+\frac{1}{2} \boldsymbol{b}^R(\boldsymbol{k}+\boldsymbol{Q}) \cdot \boldsymbol{b}(-\boldsymbol{k})\right).
\end{eqnarray}
Therefore, the condition for obtaining the maximum of \eqnref{chippfinal} is
\begin{equation}
    \boldsymbol{b}^R(\boldsymbol{k}+\boldsymbol{Q}) \;\; \parallel\;\; \boldsymbol{b}(-\boldsymbol{k}),\;\;\;\;\;\;\;\;\;\; \forall \boldsymbol{k} \in \mathrm{BZ},
\end{equation}
which indicates that the maximum susceptibility is achieved when all Bloch vectors $\boldsymbol{b}(-\boldsymbol{k})$ across the BZ are parallel to the Bloch vector $\boldsymbol{b}^R(\boldsymbol{k}+\boldsymbol{Q})$.
\begin{itemize}
    \item case 2: $\boldsymbol{o}_0(\boldsymbol{k}) = 0$ and $\boldsymbol{o}(\boldsymbol{k})\neq \boldsymbol{0}$
\end{itemize}
\begin{eqnarray}\label{chippfinal2}
    \chi^{\mathrm{pp}}_{\boldsymbol{Q}}=\frac{1}{4T} \frac{|\nu|}{\mathrm{arctanh}|\nu|}\sum_{\boldsymbol{k}}|\boldsymbol{o}(\boldsymbol{k})|^2\left[2 n^2+\frac{1}{N} \tilde{\boldsymbol{b}}^R_{\boldsymbol{o}}(\boldsymbol{k}+\boldsymbol{Q}) \cdot \boldsymbol{b}(-\boldsymbol{k})\right]
\end{eqnarray}
where  $\tilde{\boldsymbol{b}}_{\boldsymbol{o}}^R(\boldsymbol{k}+\boldsymbol{Q})$ is the dressed Bloch vector, corrected by the order parameter:
\begin{eqnarray} \tilde{\boldsymbol{b}}^R_{\boldsymbol{o}}(\boldsymbol{k}+\boldsymbol{Q}) \equiv \boldsymbol{b}^R(\boldsymbol{k}+\boldsymbol{Q})-N(\hat{\boldsymbol{o}}(\boldsymbol{k}) \times \boldsymbol{b}^R(\boldsymbol{k}+\boldsymbol{Q}) \times \hat{\boldsymbol{o}}(\boldsymbol{k})).
\end{eqnarray}
The condition for obtaining the maximum of \eqnref{chippfinal2} is
\begin{eqnarray}\label{nestpp}
\tilde{\boldsymbol{b}}^R_{\boldsymbol{o}}(\boldsymbol{k}+\boldsymbol{Q}) \;\; \parallel\;\; \boldsymbol{b}(-\boldsymbol{k}),\;\;\;\;\;\;\;\;\;\; \forall \boldsymbol{k} \in \mathrm{BZ},
\end{eqnarray}
which indicates that the maximum susceptibility is achieved when all Bloch vectors $\boldsymbol{b}(\boldsymbol{k})$ across the BZ are parallel to the dressed Bloch vector $\tilde{\boldsymbol{b}}_{\boldsymbol{o}}^R(\boldsymbol{k}+\boldsymbol{Q})$.

\section{Effects of Interactions}\label{II}

In this section, we discuss the effect of interactions, which can enhance the bare susceptibility and induce actual long-range order. Furthermore, we provide additional derivation details on the connection between the quantum metric and the correlation length when the order parameter satisfies the nesting condition.

We start with a generic multi-orbital model,
\begin{eqnarray}\label{H0}
  \begin{aligned}
    H_0=\sum_{\bd{k}}  h_{\alpha \beta}(\bd{k}) c_{\alpha}^{\dagger}(\bd{k}) c_{\beta}(\bd{k}) =\sum_{\bd{k},m}\left[\epsilon_m(\bd{k})-\mu\right] c_{m}^{\dagger}(\bd{k}) c_{m}(\bd{k}),
\end{aligned}
\end{eqnarray}
where $\alpha, \beta$ denote the orbital and spin indices, and $m$ refers to the diagonal band. Assuming that the bands in the lower-energy branch are nearly degenerate so that we focus only on the physics occurring within the lower-energy sector --- meaning both the interaction scale and temperature scale are smaller than the energy gap separating the low-energy and high-energy sectors --- the action for the free Hamiltonian in \eqnref{H0} is:
\begin{eqnarray}\label{S0}
    S_0=-\sum_{\boldsymbol{k}, \boldsymbol{q}} \sum_{\omega_n, \nu_n} \sum_{m, m'\in L} c_m^{\dagger}(\boldsymbol{k}+\boldsymbol{q}, i\nu_n+i\omega_n)\left[G_0^{-1}(\boldsymbol{k}, i\omega_n) \delta_{\boldsymbol{q}, 0} \delta_{\nu_n, 0} \delta_{m, m'}\right] c_{m'}(\boldsymbol{k}, i\omega_n),
\end{eqnarray}
where the single particle Green's function is:
\begin{equation}
    G_0(\boldsymbol{k}, i\omega_n)=\frac{1}{i\omega_n - \epsilon_m(\bd{k})+\mu}.
\end{equation}
A four-fermion interaction term can be expressed as a product of a bilinear fermion operator $\hat{O}_{\boldsymbol{q}}$ (here we discuss only particle-hole order, but the particle-particle case is similar):
\begin{eqnarray}\label{int}
    S_V=\int_0^\beta d \tau \frac{1}{\mathcal{N}} \sum_{\boldsymbol{q}} \frac{W_{\boldsymbol{q}}}{2} \hat{O}_{-\boldsymbol{q}}^{\mathrm{ph}} \hat{O}_{\boldsymbol{q}}^{\mathrm{ph}},
\end{eqnarray}
where $\mathcal{N}$ denotes the total number of lattice sites and $W_{\boldsymbol{q}}$ is the strength of the interaction. 

\subsection{Response Function at the Random Phase Approximation (RPA) Level}

We now apply the Hubbard-Stratonovich 
by introducing the bosonic field $\phi$ and then shifting $\phi\rightarrow -i \phi +i {\hat O}^{\mathrm{ph}}$ :
\begin{eqnarray}\label{Svde}
    \begin{aligned}
    S_V &= \int_0^\beta d \tau \frac{1}{\mathcal{N}} \sum_{\boldsymbol{q}}\left(-\frac{W_{\boldsymbol{q}}}{2} \phi_{-\boldsymbol{q}} \phi_{\boldsymbol{q}}+\frac{W_{\boldsymbol{q}}}{2} \phi_{-\boldsymbol{q}} \hat{O}_{\boldsymbol{q}}^{\mathrm{ph}}+\frac{W_{\boldsymbol{q}}}{2} \hat{O}_{-\boldsymbol{q}}^{\mathrm{ph}} \phi_{\boldsymbol{q}}\right)\\
    &=\sum_{\boldsymbol{k}, \boldsymbol{q}} \sum_{\omega_n, \nu_n} \sum_{m, m' \in L}\frac{W_{\boldsymbol{q}}}{\mathcal{N}} c_m^{\dagger}(\boldsymbol{k}+\boldsymbol{q}, i\nu_n+i\omega_n) \phi_{-\boldsymbol{q}}(-\nu_n) \hat{\Gamma}_{mm'}(\boldsymbol{k}+\boldsymbol{q}, \boldsymbol{k}) c_{m'}(\boldsymbol{k}, i\omega_n)-\frac{1}{N} \sum_{\boldsymbol{q},\omega_n} \frac{W_{\boldsymbol{q}}}{2} \phi_{-\boldsymbol{q}}(-i\omega_n) \phi_{\boldsymbol{q}}(i\omega_n),
\end{aligned}
\end{eqnarray}
where the vertex is given by:
\begin{equation}
    \hat{\Gamma}_{m n}(\boldsymbol{k}+\boldsymbol{q}, \boldsymbol{k})=\sum_{\alpha \beta}U_{m \alpha}^{\dagger}(\boldsymbol{k}+\boldsymbol{q}) \mathcal{O}_{\alpha \beta}(\boldsymbol{k}) U_{\beta n}(\boldsymbol{k}).
\end{equation}
Then, the effective action for the bosonic field $\phi_{\boldsymbol{q}}$ at zero frequency is obtained by integrating out the fermionic field $c$ in $S_0+S_V$, resulting in
\begin{eqnarray}\label{Seff}
    \begin{aligned}
    S_{\mathrm{eff}}[\phi]=&-\frac{1}{\mathcal{N}} \sum_{\boldsymbol{q}} \frac{W_{\boldsymbol{q}}}{2} \phi_{-\boldsymbol{q}} \phi_{\boldsymbol{q}}+\frac{1}{2} \sum_{\omega} \sum_{\boldsymbol{k}, \boldsymbol{q}}\sum_{m,m'\in L} \frac{W^2_{\boldsymbol{q}}}{N^2} \phi_{-\boldsymbol{q}} \phi_{\boldsymbol{q}} \Gamma_{m m'}(\boldsymbol{k}+\boldsymbol{q}, \boldsymbol{k}) \Gamma_{m' m}(\boldsymbol{k}, \boldsymbol{k}+\boldsymbol{q}) G_0(i\omega_n, \boldsymbol{k}+\boldsymbol{q}) G_0(i\omega_n, \boldsymbol{k})\\
    &=\frac{1}{2N} \sum_{\boldsymbol{q}}\left(\chi_{\boldsymbol{q}}^\phi \right)^{-1}\phi_{-\boldsymbol{q}} \phi_{\boldsymbol{q}},
\end{aligned}
\end{eqnarray}
where
\begin{equation}\label{chiphiq}
    \chi_{\boldsymbol{q}}^\phi=\left \langle  \phi_{-\boldsymbol{q}}\phi_{\boldsymbol{q}}\right\rangle=\frac{1}{-W_{\boldsymbol{q}}-W_{\boldsymbol{q}}^2\chi_{\boldsymbol{q}}^{\mathrm{ph}}},
\end{equation}
and $\chi^{\mathrm{ph}}_{\boldsymbol{q}}$ in \eqnref{chiphiq} is:
\begin{equation}
    \chi_{\boldsymbol{q}}^{\mathrm{ph}} \equiv-\sum_{\boldsymbol{k}} \frac{1}{\mathcal{N}} \operatorname{Tr}\left[\mathcal O^{\dagger}(\boldsymbol{k}) P(\boldsymbol{k}+\boldsymbol{q}) \mathcal O(\boldsymbol{k}) P(\boldsymbol{k})\right] \sum_{\omega_n} G_0(i\omega_n, \boldsymbol{k}+\boldsymbol{q}) G_0(i\omega_n, \boldsymbol{k}).
\end{equation}
which should equal the susceptibility defined in \eqnref{chiph}. Therefore, from \eqnref{Seff}, the $\hat O-\hat O$ response function at the RPA level is given by:
\begin{equation}\label{RPA} \chi^{\mathrm{RPA}}_{\boldsymbol{q}}=\left\langle\hat{O}_{-\boldsymbol{q}}^{\mathrm{ph}} \hat{O}_{\boldsymbol{q}}^{\mathrm{ph}}\right\rangle =\frac{1}{W_{\boldsymbol{q}}} +\chi_{\boldsymbol{q}}^\phi = \frac{\chi_{\boldsymbol{q}}^{\mathrm{ph}}}{1+W_{\boldsymbol{q}} \chi_{\boldsymbol{q}}^{\mathrm{ph}}}.
\end{equation}
Note that when the interaction in \eqnref{int} is compatible, i.e. $W_{\boldsymbol{Q}}<0$ at the ordering vector $\boldsymbol{Q}$, then there is an instability at sufficiently low temperature at this wavevector, i.e. $\chi_{\boldsymbol{Q}}^{\mathrm{RPA}}$ diverges at some temperature. Specifically, the critical temperature is estimated by the condition:
\begin{equation}\label{Tcexp}
    1+W_{\boldsymbol{q}} \chi_{\boldsymbol{q}}^{\mathrm{ph}}(T_c)=0
\end{equation} Moreover, when the interaction $W_{\bm{q}}$ is roughly momentum-independent, then the first instability encountered upon lowering temperature is the nesting one at wavevector $\bm{Q}$. When these conditions are satisfied, we can expect the low temperature order to coincide with the one found by the high temperature analysis.

\subsection{Ginzburg-Landau Action and Correlation Length}

For simplicity, we ignore the momentum dependence of $W_{\boldsymbol{q}}$. This corresponds to the on-site interaction case. We consider the case of an ordering vector $\boldsymbol{Q}$ and shift the momentum $\boldsymbol{q}$ relative to $\boldsymbol{Q}$. The effective action in \eqnref{Seff} then becomes:
\begin{eqnarray}\label{seffphi1}
\begin{aligned}    
    S_{\mathrm{eff}}[\phi]&=-\frac{1}{2 \mathcal{N}} \sum_{\boldsymbol{q}} \phi_{-\boldsymbol{q}}\left[W+W^2 \sum_{\boldsymbol{k}} \frac{\operatorname{Tr}\left[\mathcal O^{\dagger}(\boldsymbol{k}) \mathcal O(\boldsymbol{k})\right]}{4NT}\left(\frac{N_L^2}{N}+\frac{1}{2} \tilde{\boldsymbol{b}}_{\boldsymbol{o}}(\boldsymbol{k}+\boldsymbol{Q}+\boldsymbol{q}) \cdot \boldsymbol{b}(\boldsymbol{k})\right)\right] \phi_{\boldsymbol{q}}\\
    =&-\frac{1}{2 \mathcal{N}} \sum_{-\boldsymbol{q}} \phi_{-\boldsymbol{q}}\left[W+W^2 \frac{|\mathcal O|^2}{4T}\left(\frac{N_L^2}{N}+\frac{1}{2} \sum_{\boldsymbol{k}} \tilde{\boldsymbol{b}}_{\boldsymbol{o}}(\boldsymbol{k}+\boldsymbol{Q}) \cdot \boldsymbol{b}(\boldsymbol{k})-\frac{1}{4} \boldsymbol{q}_\mu \boldsymbol{q}_\nu \sum_{\boldsymbol{k}} \partial_\mu \tilde{\boldsymbol{b}}_{\boldsymbol{o}}(\boldsymbol{k}+\boldsymbol{Q}) \cdot \partial_\nu \boldsymbol{b}(\boldsymbol{k})\right)\right] \phi_{\boldsymbol{q}},
    \end{aligned}
\end{eqnarray}
where $|\mathcal O|^2 \equiv \frac{1}{\mathcal{N}} \sum_{\boldsymbol{k}} \operatorname{Tr}\left[\mathcal O^{\dagger}(\boldsymbol{k}) \mathcal O(\boldsymbol{k})\right]$, and $\mathcal O(\boldsymbol{k})$ is assumed to be independent of momentum, which is true in most realistic cases. When the order satisfies the nesting condition given in \eqnref{phnest}, the relation $\tilde{\boldsymbol{b}}_{\boldsymbol{o}}(\boldsymbol{k}+\boldsymbol{Q}) \cdot \boldsymbol{b}(\boldsymbol{k})= 2\left(N_L-\frac{N_L^2}{N}\right)$ (which appears in \eqnref{seffphi1}) holds due to the expression of the magnitude of $\boldsymbol{b}$ given in \eqnref{magni}. 

Let us now show that the coefficient of the term quadratic in momentum in \eqnref{seffphi1} corresponds to the quantum metric. First, note that due to the property of the projection operator $P^2=P$, we have the relation:
\begin{equation}
    \boldsymbol{b}(\boldsymbol{k}) \star \boldsymbol{b}(\boldsymbol{k})=2\left(1-2 \frac{N_L}{N}\right) \boldsymbol{b}(\boldsymbol{k}).
\end{equation}
Then, taking the derivative on both sides leads to: 
\begin{equation}
    \partial_\nu \boldsymbol{b}(\boldsymbol{k}) \star \boldsymbol{b}(\boldsymbol{k})=(1-2\frac{N_L}{N}) \partial_\nu \boldsymbol{b}(\boldsymbol{k}),
\end{equation}
so that
\begin{equation}\label{ppb}
    \partial_\nu \boldsymbol{b}(\boldsymbol{k}) \star \partial_\mu \boldsymbol{b}(\boldsymbol{k}) \cdot \boldsymbol{b}(\boldsymbol{k})=(1-2 \frac{N_L}{N}) \partial_\nu \boldsymbol{b}(\boldsymbol{k}) \cdot \partial_\mu \boldsymbol{b}(\boldsymbol{k}).
\end{equation}
Now, the quantum geometric tensor of the low-energy sector can be expressed as:
\begin{eqnarray}\label{QGT}
\begin{aligned}
    T_{\mu \nu}(\boldsymbol{k})&=\operatorname{Tr}\left[\partial_\mu P(\boldsymbol{k})\left(\mathds{1}_N-P(\boldsymbol{k})\right)\partial_\nu P(\boldsymbol{k})\right]\\
    &=\frac{1}{2}  \partial_\nu \boldsymbol{b}(\boldsymbol{k}) \cdot \partial_\mu \boldsymbol{b}(\boldsymbol{k})-\frac{N_L}{2N}  \partial_\nu \boldsymbol{b}(\boldsymbol{k}) \cdot \partial_\mu \boldsymbol{b}(\boldsymbol{k})-\frac{1}{4} \left(\partial_\nu \boldsymbol{b}(\boldsymbol{k}) \star \partial_\mu \boldsymbol{b}(\boldsymbol{k})+i \partial_\nu \boldsymbol{b}(\boldsymbol{k}) \times \partial_\mu \boldsymbol{b}(\boldsymbol{k})\right) \cdot \boldsymbol{b}(\boldsymbol{k})\\
    &=g_{\mu \nu}(\boldsymbol{k})-\frac{i}{2} \Omega_{\mu \nu}(\boldsymbol{k}),
    \end{aligned}
\end{eqnarray}
where the real and minus twice the imaginary parts of the quantum geometric tensor correspond to the quantum metric and Berry curvature, respectively, and are given by:
\begin{eqnarray}\label{gOmega}
    \begin{aligned}
        g_{ \mu \nu}(\boldsymbol{k})\equiv \operatorname{Re} T_{\mu \nu}(\boldsymbol{k})&=\frac{1}{4} \partial_\mu \boldsymbol{b}(\boldsymbol{k}) \cdot \partial_\nu \boldsymbol{b}(\boldsymbol{k})\\
         \Omega_{ \mu \nu}(\boldsymbol{k})\equiv-2 \operatorname{Im} T_{\mu \nu}(\boldsymbol{k})&=-\frac{1}{2} \partial_\mu \boldsymbol{b}(\boldsymbol{k}) \times \partial_\nu \boldsymbol{b}(\boldsymbol{k})\cdot \boldsymbol{b}(\boldsymbol{k}) .
    \end{aligned}
\end{eqnarray}
Then, since $\tilde{\boldsymbol{b}}_{\boldsymbol{o}}(\boldsymbol{k}+\boldsymbol{Q})$ can be exactly identified as $\boldsymbol{b}(\boldsymbol{k})$ when the nesting condition in \eqnref{phnest} is fulfilled, it follows that:
\begin{equation}
    g_{ \mu \nu}(\boldsymbol{k})=\frac{1}{4} \partial_\mu \tilde{\boldsymbol{b}}_{\boldsymbol{o}}(\boldsymbol{k}+\boldsymbol{Q}) \cdot \partial_\nu \boldsymbol{b}(\boldsymbol{k}).
\end{equation}

Finally, \eqnref{seffphi1} leads to the Ginzburg-Landau action:
\begin{eqnarray}
    S_{\mathrm{eff}}[\phi]=\sum_{\boldsymbol{q}} \phi_{-\boldsymbol{q}}\left(\frac{a(T)}{2}+\frac{c(T)}{2} \boldsymbol{q}^2\right) \phi_{\boldsymbol{q}}+O\left(\phi^4\right),
\end{eqnarray}
where the mass term $a(T)$ is given by:
\begin{equation}
\begin{aligned}
        a(T)&=-W-W^2 \frac{N_L|\mathcal O|^2}{4T}\\
        &=\frac{W^2 N_L|\mathcal O|^2}{4}\left(\frac{1}{T_c}-\frac{1}{T}\right).
        \end{aligned}
\end{equation}
Here, $T_c$ is the critical temperature at which $a(T_c)=0$. Furthermore, the term $c(T)$, which controls the spatial fluctuations, is given by:
\begin{equation}
\begin{aligned}
        c(T)=\frac{W^2|\mathcal O|^2}{4T} \sqrt{\operatorname{det} \bar{g}},
        \end{aligned}
\end{equation}
where $\bar{g}$  is the average quantum metric over the entire BZ, defined as follows:
\begin{equation}\label{ag}
    \bar{g}_{\mu\nu} \equiv \sum_{\boldsymbol{k}} g_{\mu\nu}(\boldsymbol{k})/\mathcal{N}=\sqrt{\operatorname{det} \bar{g}_{\mu\nu}} \delta_{\mu\nu}.
\end{equation}
Therefore, the correlation length near the critical temperature is given by:
\begin{equation}\label{GLc}
\xi\equiv\sqrt{\frac{c(T)}{a(T)}}=\frac{(\operatorname{det} \bar{g})^{1 / 4}}{\sqrt{ N_L}} \left|1-\frac{T}{T_c}\right|^{-\frac{1}{2}}.
\end{equation}

\subsection{Lower Bound in Topologically Nontrivial Bands}\label{IIC}

In the following, we further demonstrate that the correlation length can have a lower bound in a topologically nontrivial band.

Since the quantum geometric tensor given in Eq.~\eqref{QGT} includes contributions from all bands within the low-energy sector, we first discuss the relationship between this composite quantum geometric tensor and the conventional quantum geometric tensor, in order to clarify the physical meaning of \( g_{\mu \nu}(\boldsymbol{k}) \). Using the relation in Eq.~\eqref{ppb} with \( N_L = 1 \),
then, the quantum geometric tensor for a given band \( m \) can be expressed as:
\begin{eqnarray}\label{QGT}
\begin{aligned}
    T_{\mu \nu}^{(m)}(\boldsymbol{k})&=\operatorname{Tr}\left[\partial_\mu \mathcal{P}_m(\boldsymbol{k})\left(\mathds{1}_N-\mathcal{P}_m(\boldsymbol{k})\right)\partial_\nu \mathcal{P}_m(\boldsymbol{k})\right]\\
    &=g^{(m)}_{\mu \nu}(\boldsymbol{k})-\frac{i}{2} \Omega^{(m)}_{\mu \nu}(\boldsymbol{k}),
    \end{aligned}
\end{eqnarray}
where the components of the quantum geometric tensor corresponding to the quantum metric and Berry curvature for the given band $m$ are given by:
\begin{eqnarray}\label{gG}
    \begin{aligned}
        g^{(m)}_{ \mu \nu}(\boldsymbol{k})\equiv \operatorname{Re} T^{(m)}_{\mu \nu}(\boldsymbol{k})&=\frac{1}{4} \partial_\mu \boldsymbol{v}_{m}(\boldsymbol{k}) \cdot \partial_\nu \boldsymbol{v}_{m}(\boldsymbol{k})\\
         \Omega^{(m)}_{ \mu \nu}(\boldsymbol{k})\equiv-2 \operatorname{Im} T^{(m)}_{\mu \nu}(\boldsymbol{k})&=-\frac{1}{2} \partial_\mu \boldsymbol{v}_{m}(\boldsymbol{k}) \times \partial_\nu \boldsymbol{v}_{m}(\boldsymbol{k})\cdot \boldsymbol{v}_{m}(\boldsymbol{k}) .
    \end{aligned}
\end{eqnarray}
Next, we consider the full Bloch vector \(\boldsymbol{b}\), which includes contributions from all bands in the low-energy sector, i.e., \(\boldsymbol{b}(\boldsymbol{k}) = \sum_{m \in L} \boldsymbol{v}_m(\boldsymbol{k})\), from which one obtains:
\begin{equation}\label{bb}
    \frac{1}{4} \partial_\mu \boldsymbol{b}(\boldsymbol{k}) \cdot \partial_\nu \boldsymbol{b}(\boldsymbol{k})=\frac{1}{4} \sum_{m \in L} \partial_\mu \boldsymbol{v}_m(\boldsymbol{k}) \cdot \partial_\nu \boldsymbol{v}_m(\boldsymbol{k})+\frac{1}{4} \sum_{m, n \in L}^{m \neq n} \partial_\mu \boldsymbol{v}_m(\boldsymbol{k}) \cdot \partial_\nu \boldsymbol{v}_m(\boldsymbol{k})
\end{equation}
It is apparent that the first term in Eq.~\eqref{bb} corresponds to the sum over the quantum metrics of the individual bands in the low-energy sector, with the expression given by Eq.~\eqref{gG}. Next, we discuss the physical meaning of the second term in Eq.~\eqref{bb}. Note that when \( m \neq n \), one obtains
\begin{equation}\label{T1}
    \operatorname{Tr}\left[\partial_\mu \mathcal{P}_m(\boldsymbol{k}) \partial_\nu \mathcal{P}_n(\boldsymbol{k})\right]=\frac{1}{4} \operatorname{Tr}\left[\partial_\mu \boldsymbol{v}_m(\boldsymbol{k}) \cdot \boldsymbol{\lambda}\; \partial_\nu \boldsymbol{v}_n(\boldsymbol{k}) \cdot \boldsymbol{\lambda}\right]=\frac{1}{2} \partial_\mu \boldsymbol{v}_m(\boldsymbol{k}) \partial_\nu \boldsymbol{v}_n(\boldsymbol{k}),
\end{equation}
Additionally, the left-hand side of Eq.~\eqref{T1} can be expressed as:
\begin{eqnarray}\label{Tr2}
    \begin{aligned}
&\operatorname{Re} \operatorname{Tr}\left[\partial_\mu \mathcal{P}_m(\boldsymbol{k}) \partial_\nu \mathcal{P}_n(\boldsymbol{k})\right]\\=&\operatorname{Re}\left\langle u_n(\boldsymbol{k}) \mid \partial_\mu u_m(\boldsymbol{k})\right\rangle\left\langle u_m(\boldsymbol{k}) \mid \partial_\nu u_n(\boldsymbol{k})\right\rangle+\operatorname{Re}\left\langle\partial_\nu u_n(\boldsymbol{k}) \mid u_m(\boldsymbol{k})\right\rangle\left\langle\partial_\mu u_m(\boldsymbol{k}) \mid u_n(\boldsymbol{k})\right\rangle\\
=&-2 \operatorname{Re}\left\langle\partial_\mu u_m(\boldsymbol{k}) \mid u_n(\boldsymbol{k})\right\rangle\left\langle u_n(\boldsymbol{k}) \mid \partial_\nu u_m(\boldsymbol{k})\right\rangle
    \end{aligned},
\end{eqnarray}
where $\left|u_m(\boldsymbol{k})\right\rangle$ denotes the states constructed from $U_{\alpha m}(\boldsymbol{k})$ in \eqnref{Uc}.
In \eqnref{Tr2}, we use the relation $\left\langle u_m(\boldsymbol{k}) \mid u_n(\boldsymbol{k})\right\rangle=0$ and $\left\langle\partial_\mu u_m(\boldsymbol{k}) \mid u_n(\boldsymbol{k})\right\rangle=-\left\langle u_m(\boldsymbol{k}) \mid \partial_\mu u_n(\boldsymbol{k})\right\rangle$. By combining \eqnref{T1} and \eqnref{Tr2}, one can define the quantity:
\begin{equation}\label{gmn2}
    g_{\mu v}^{(mn)}(\boldsymbol{k}) \equiv \frac{1}{4} \partial_\mu \boldsymbol{v}_m(\boldsymbol{k}) \cdot \partial_\nu \boldsymbol{v}_n(\boldsymbol{k})=-\operatorname{Re}\left[\left\langle\partial_\mu u_m(\boldsymbol{k}) \mid u_n(\boldsymbol{k})\right\rangle\left\langle u_n(\boldsymbol{k}) \mid \partial_\nu u_m(\boldsymbol{k})\right\rangle\right],
\end{equation}
which satisfies the relation:
\begin{equation}\label{gmndis}
    \left|\left\langle u_m(\boldsymbol{k}+d\boldsymbol{q}) \mid u_n(\boldsymbol{k})\right\rangle \right|^2 \approx - g_{\mu \nu}^{(m n)}(\boldsymbol{k}) dq_{\mu} dq_{\nu}.
\end{equation}
Therefore, \( g_{\mu \nu}^{(mn)}(\boldsymbol{k}) \) defined in Eq.~\eqref{gmn2} has the physical interpretation of measuring the distance between Bloch functions of different bands in momentum space~\cite{PhysRevB.109.214518}. This can be viewed as the inter-band component of the quantum metric. As a result, Eq.~\eqref{bb} represents the full quantum metric of the composite bands, consisting of both intra-band and inter-band contributions:
\begin{equation}
    g_{\mu \nu}(\boldsymbol{k}) =\sum_{m \in L}g_{\mu \nu}^{(m)}(\boldsymbol{k})+\sum_{m, n \in L}^{m \neq n}g_{\mu \nu}^{(mn)}(\boldsymbol{k}).
\end{equation}

According to the arithmetic–geometric mean inequality for a $2\times2$ positive semidefinite matrix, we have
\begin{eqnarray}
    \sum_{\boldsymbol{k}} \opr{Tr }g(\boldsymbol{k})\geq 2\sum_{\boldsymbol{k}} \sqrt{\opr{det} g(\boldsymbol{k})},
\end{eqnarray}
as well as a relation between the local quantum metric and the Berry curvature~\cite{PhysRevB.104.045103}:
\begin{equation}\label{gO}
    \sqrt{\opr{det} g(\boldsymbol{k}) }\geq \frac{\left|\Omega_{xy}(\boldsymbol{k})\right|}{2}.
\end{equation}
It is important to note that \( g(\boldsymbol{k}) \) here represents the full quantum metric of the composite bands, so the Berry curvature in Eq.~\eqref{gO} is also the total Berry curvature of the composite system as defined in Eq.~\eqref{gOmega}. For simplicity, we now assume that all bands in the subset are disconnected (so that the non-Abelian Berry curvature can be neglected; even if included, the final conclusion remains unchanged). Under this assumption, one obtains
\begin{eqnarray}
\begin{aligned}   
    \Omega_{\mu \nu}(\boldsymbol{k}) =&-2 \operatorname{Im} T_{\mu \nu}(\boldsymbol{k})=-2\sum_{mn\lambda\in L} \operatorname{Im} \operatorname{Tr}\left[\mathcal{P}_\lambda(\boldsymbol{k}) \partial_\mu \mathcal{P}_m(\boldsymbol{k}) \partial_\nu \mathcal{P}_n(\boldsymbol{k})\right]\\
    =&-2\sum_{mn\in L}\operatorname{Im}\left[\left\langle u_n(\boldsymbol{k}) \mid \partial_\mu u_m(\boldsymbol{k})\right\rangle\left\langle\partial_\nu u_m(\boldsymbol{k}) \mid u_n(\boldsymbol{k})\right\rangle+\left\langle u_n(\boldsymbol{k}) \mid \partial_\mu u_m(\boldsymbol{k})\right\rangle\left\langle u_m(\boldsymbol{k}) \mid \partial_\nu u_n(\boldsymbol{k})\right\rangle\right.\\
    &\left.+\left\langle\partial_\mu u_m(\boldsymbol{k}) \mid u_n(\boldsymbol{k})\right\rangle\left\langle\partial_\nu u_n(\boldsymbol{k}) \mid u_m(\boldsymbol{k})\right\rangle\right]-2\sum_{m\in L} \operatorname{Im}\left\langle\partial_\mu u_m(\boldsymbol{k}) \mid \partial_\nu u_m(\boldsymbol{k})\right\rangle\\
    =& \sum_{m\in L}\Omega_{\mu \nu}^{(m)}(\boldsymbol{k}),
    \end{aligned}
\end{eqnarray}
here we use the relation
\begin{equation}
    \sum_{mn\in L}\left\langle u_n(\boldsymbol{k}) \mid \partial_\mu u_m(\boldsymbol{k})\right\rangle\left\langle\partial_\nu u_m(\boldsymbol{k}) \mid u_n(\boldsymbol{k})\right\rangle \in \mathbb{R}.
\end{equation}
Therefore, the total Chern number of the composite bands can be considered as the sum of the Chern numbers of all bands within the subset. Consequently, the correlation length in Eq.~\eqref{GLc} satisfies
\begin{eqnarray}\label{xibon} \xi&=&\sqrt{\frac{\sum_{\boldsymbol{k}}\opr{Tr}g(\boldsymbol{k})}{2\mathcal{N} N_L}}\left|1-\frac{T}{T_c}\right|^{-\frac{1}{2}}\geq\sqrt{\frac{\sum_{\boldsymbol{k}}\left|\Omega_{xy}(\boldsymbol{k})\right|}{2\mathcal{N} N_L}}\left|1-\frac{T}{T_c}\right|^{-\frac{1}{2}}\label{ineq1}\\
    &\geq& \sqrt{\frac{|\sum_{m\in L}\mathcal{C}_m|}{4\pi N_L}}\left|1-\frac{T}{T_c}\right|^{-\frac{1}{2}},\label{ineq2}
\end{eqnarray}
where \(\mathcal{C}_m\) is the Chern number of band \(m\). The inequality in Eq.~\eqref{ineq2} indicates that the correlation length (or ``high-temperature stiffness'', as we called it in the main text) has a lower bound determined by the total Chern number of the low-energy sector. This implies that the correlation length of the order will be finite if the low-energy sector exhibits finite quantum Hall signals.Furthermore, the bound in Eq.~\eqref{ineq1} depends only on the {\em absolute value} of the Berry curvature and therefore provides a tighter bound than Eq.~\eqref{ineq2}. This means that even for topologically trivial bands—where the Berry curvature integrates to zero—a nonzero Berry curvature still sets a lower bound on the correlation length.

Finally, we would like to discuss in more detail the difference between a single flat band and a composite flat band with \(N_L \neq 1\). Note that when \(N_L = 1\), we have \(g_{\mu \nu}(\boldsymbol{k}) = g^{(m)}_{\mu \nu}(\boldsymbol{k})\) and \(\Omega_{\mu \nu}(\boldsymbol{k}) = \Omega^{(m)}_{\mu \nu}(\boldsymbol{k})\), so that \(\xi\) in Eq.~\eqref{xibon} is simply bounded by the Chern number of that single band, which is consistent with previous works~\cite{2015NatCo...6.8944P,PhysRevB.104.045103, PhysRevB.90.165139, PhysRevB.95.024515}.   However, for a composite band with \(N_L \neq 1\), the correlation length is determined by the total quantum metric \(g_{\mu \nu}(\boldsymbol{k})\), which includes both intraband and interband components, as given in Eq.~\eqref{gmn2}. According to Eq.~\eqref{gmndis}, it is apparent that the interband components \(g_{\mu \nu}^{(mn)}(\boldsymbol{k})\) are negative semi-definite, in contrast to the well-known positive semi-definite nature of the intraband metric \(g_{\mu \nu}^{(m)}(\boldsymbol{k})\).  
As a result, the total quantum metric is not simply the algebraic sum of the quantum metrics of the individual bands. Instead, interference between different bands leads to a suppression effect, indicating that the lower bound on the correlation length is not given by the sum of the absolute values of the Chern numbers but rather by the total Chern number itself. Within this scenario, tighter bounds can emerge if the low-energy composite bands can be partitioned into several elementary band representations~\cite{PhysRevB.109.214518}.

\subsection{Valid Energy Range for Nesting	}\label{Sec:valid}
In this section, we discuss the energy scale over which our nesting condition is applicable, and then provide a straightforward analysis of the behavior beyond this regime.

First, note that our analysis begins at finite temperature, above all critical temperatures (\(T > T_c\)), so that comparing the susceptibilities of different orders remains meaningful. Otherwise, at $T=0$, vanishing dispersion would cause all susceptibilities to diverge in the absence of an assumed order gap. Importantly, thermal excitation at finite temperature can smooth out dispersion features when the temperature scale exceeds both the band gap and the bandwidth of the low-energy subset. This ensures that our analysis and conclusions remain valid within this regime.

To be more specific, we revisit our starting point: the generic mean-field susceptibility presented in \eqnref{chiph}:
\begin{eqnarray}\label{chigen}
    \begin{aligned}    \chi_Q^{\mathrm{ph}}=\sum_{m, m^{\prime} \in L} \sum_{\boldsymbol{k}} \left[U_{m^{\prime} \alpha^{\prime}}^{\dagger}(\boldsymbol{k}) \mathcal{O}_{\alpha^{\prime} \beta^{\prime}}^{\dagger}(\boldsymbol{k}) U_{\beta^{\prime} m}(\boldsymbol{k}+\boldsymbol{Q}) U_{m \alpha}^{\dagger}(\boldsymbol{k}+\boldsymbol{Q}) \mathcal{O}_{\alpha \beta}(\boldsymbol{k}) U_{\beta m^{\prime}}(\boldsymbol{k})\right] \times g_{m,m'}(\boldsymbol{k},\boldsymbol{Q},T)
    \end{aligned}
\end{eqnarray}
There are two parts to this susceptibility: the first part corresponds to the contribution from the wavefunction (i.e., the band geometry discussed in our main text), while the second part arises from the dispersion, given by the expression:
\begin{equation}
     g_{m,m'}(\boldsymbol{k},\boldsymbol{Q},T)=\frac{n_F\left[\epsilon_m(\boldsymbol{k}+\boldsymbol{Q})\right]-n_F\left[\epsilon_{m^{\prime}}(\boldsymbol{k})\right]}{\epsilon_{m^{\prime}}(\boldsymbol{k})-\epsilon_m(\boldsymbol{k}+\boldsymbol{Q})}
\end{equation}
When the temperature scale under consideration is larger than both the bandwidth of the low-energy sector and the small energy gap between bands within this sector, one obtains \(\frac{1}{e^{\epsilon_m(\boldsymbol{k}) \beta}+1} \approx \frac{1}{2} - \frac{\epsilon_m(\boldsymbol{k}) \beta}{4}\) for any momentum \(\boldsymbol{k}\) and band index \(m\). As a result, the dispersion-dependent factors reduce to a constant that depends only on temperature:
\begin{equation}\label{scale}
    g_{m,m'}(\boldsymbol{k},\boldsymbol{Q},T)\approx \frac{1}{4T},
\end{equation}
which simply reflects the physical fact that any features of the dispersion become irrelevant due to thermal excitation. In this case, the generic susceptibility in Eq.~\eqref{chigen} depends only on the band geometry, corresponding to the situations discussed in our work, and all of our conclusions remain valid.

To clarify the scope of our conclusions more comprehensively and self-consistently, we next discuss the temperature scale used here. Since we require that the \emph{temperature \( T \) considered is higher than all possible \( T_c \)}, it is necessary to specify the scale of \( T_c \). Notably, when the condition in Eq.~\eqref{scale} is satisfied, the actual transition temperature \( T_c \) at the RPA level can be estimated using Eq.~\eqref{Tcexp} as follows (interaction assumed to be on-site \( U \) for simplicity)
:
\begin{equation}\label{Tc}
    T_c \sim \left( \sum_k \frac{1}{ 4\mathcal{N}} \operatorname{Tr}\left[\mathcal{O}^{\dagger}(k) \mathcal{O}(k) \right]\right)\frac{N_L}{N} U,
\end{equation}
which is determined by the magnitude of the interaction \( U \) and the degeneracy of the low-energy subset. Based on this, we can define a concrete parameter regime for our formalism: specifically, the total energy width \( w \) of the low-energy subset (including both the band gap and the dispersion) must be much smaller than the \( T_c \) estimated in Eq.~\eqref{Tc}, which itself depends on the interaction strength. In summary, the relevant energy scale is:
\begin{equation}\label{condi}
    w\ll U \ll \Delta,
\end{equation}
in which \(\Delta\) corresponds to the band gap between the low-energy and high-energy sectors. Under this condition, the conclusions and analysis in our main text remain valid.

In the following, we discuss the situation beyond the condition given in Eq.~\eqref{condi}. First, we can consider the case where the degeneracy within the low-energy subset is preserved, but the bandwidth \( w \) becomes comparable to or even larger than \( U \). In this regime, the band index in \( g_{m,m'}(\boldsymbol{k}, \boldsymbol{Q}, T) \) can be ignored, so that the dispersion and wavefunction contributions to \(\chi_Q^{\mathrm{ph}}\) can be completely separate, leading to
\begin{eqnarray}
\chi_{\boldsymbol{Q}}^{\mathrm{ph}}=\sum_{\boldsymbol{k}} g(\boldsymbol{k},\boldsymbol{Q},T) \frac{\operatorname{Tr}\left[\mathcal{O}^{\dagger}(\boldsymbol{k}) \mathcal{O}(\boldsymbol{k})\right]}{ N}\left(\frac{N_L^2}{N}+\frac{1}{2} \tilde{\boldsymbol{b}}_{\boldsymbol{o}}(\boldsymbol{k}+\boldsymbol{Q}) \cdot \boldsymbol{b}(\boldsymbol{k})\right),
\end{eqnarray}
which indicates that the dispersion effect here acts as a weighting factor in the Brillouin zone, determining the weight assigned to contributions from different momentum regions to the overall alignment of the \(\boldsymbol{b}\)-vector. When the bandwidth \( w \) is large enough to exceed the condition given in Eq.~\eqref{condi}, these weighting factors become significant, and the perfect nesting criterion presented in the main text is no longer applicable; instead, the analysis must account for the detailed structure of the dispersion. When the bandwidth is large, satisfying \( U \ll w \ll \Delta \), the weight factor \(\sum_{\boldsymbol{k}} g(\boldsymbol{k}, \boldsymbol{Q}, T)\) exhibits dominant peaks in the presence of a Fermi surface nesting structure. This indicates that, in this energy scale regime, the competition between different orders is governed by the conventional Fermi surface nesting scenario, as extensively studied in previous work.

Next, we discuss the case where a band gap within the low-energy subset is present, while the bands themselves remain flat. When this band gap is much smaller than the interaction strength \( U \), the condition given in Eq.~\eqref{condi} still holds, so the nesting criteria discussed in our work remain applicable. In the opposite limit, where the band gap is much larger than \( U \), the original partition of the low-energy subset needs to be revised: we must further project onto the truly active bands within the subset, while the other bands effectively become a new high-energy sector.  The most complex physics emerges when the band gap is comparable to \( U \). A typical and well-studied example is the flat-band superconductivity (as discussed in the main text, when there is time-reversal symmetry, this is perfectly satisfy the nesting condition), this parameter regime (where the band gap is approximately equal to the attractive interaction strength) leads to the well-known BCS–BEC crossover, corresponding to a strongly correlated regime~\cite{PhysRevB.106.014518,DENG2024169639}.

\section{Details of the Flat Band Model Example Discussed in the Main Text}\label{III}

In this section, we provide additional details on the properties of the model used as an example in the main text.

\subsection{Non-interacting Hamiltonian}

We start from a two-orbital spinful electronic model (with orbitals labeled $A$ and $B$), first introduced in Ref.~\onlinecite{Berg.Hofmann.2022}. The non-interacting Hamiltonian is given by 
\begin{equation}
    H_0= \sum_{\boldsymbol{k}} \psi_{\boldsymbol{k}}^{\dagger}h_{\boldsymbol{k}}\psi_{\boldsymbol{k}},
\end{equation}
where the basis is defined as:
\begin{equation} \psi_{\boldsymbol{k}}=\left[c_{A\uparrow}(\boldsymbol{k})\;\;c_{B\uparrow}(\boldsymbol{k})\;\;c_{A\downarrow}(\boldsymbol{k})\;\;c_{B\downarrow}(\boldsymbol{k})\right]^T. 
\end{equation}
We consider the Hamiltonian matrix:
\begin{eqnarray}\label{hk_sp}
h_{\boldsymbol{k}}=t\left(\begin{array}{cccc}
-\mu & -i e^{i\alpha_k^{\uparrow}} & 0 & 0 \\
i e^{-i \alpha_k^{\uparrow}} & -\mu & 0 & 0 \\
0 & 0 & -\mu & ie^{-i \alpha_k^{\downarrow }} \\
0 & 0 & -ie^{i \alpha_k^{\downarrow }} & -\mu,
\end{array}\right)
\end{eqnarray}
where 
\begin{equation}
    \alpha_k^\sigma=\eta_{\sigma}\left(\cos k_x+\cos k_y\right),
\end{equation}
and $\eta_\sigma$ controls the locality of the Wannier wave function for spin $\sigma$ \cite{souza2000,Vanderbilt_2018}. In the original model proposed in Ref.~\onlinecite{Berg.Hofmann.2022} time-reversal symmetry is preserved by setting $\eta_{\uparrow}=\eta_{\downarrow}$, but in this work, $\eta_{\sigma}$ can take different values for different spins to account for the possibility of breaking time-reversal symmetry. After diagonalization, there are a total of four bands ($N=4$) with energies:
\begin{equation}
    E_{ \pm,\sigma}=t( \pm 1-\mu),
\end{equation}
which represents two pairs of perfectly flat bands that are independent of the parameters $\eta_{\sigma}$. If the filling number $\nu < 2$, all electrons remain in the lower band sector (with $N_L=2$), and the energy gap between the two sectors is $2t$. Furthermore, the projection operator for the low-energy sector is given by:
\begin{eqnarray}\label{Pkfull}
    P(\boldsymbol{k})=\sum_\sigma \mathcal{P}_{-, \sigma}(\boldsymbol{k})=\frac{1}{2}\left(\begin{array}{cccc}
1 & i e^{i \alpha_k^{\uparrow}} & 0 & 0 \\
-i e^{-i \alpha_k^{\uparrow}} & 1 & 0 & 0 \\
0 & 0 & 1 & -i e^{-i \alpha_{k}^{\downarrow}} \\
0 & 0 & i e^{i \alpha_k^{\downarrow}} & 1
\end{array}\right)=\frac{1}{2} \mathds{1}_4+\frac{1}{2} \boldsymbol{b}(\boldsymbol{k}) \cdot \boldsymbol{\lambda},
\end{eqnarray}
where the Bloch vector is given by:
\begin{equation}\label{bfull}
 \boldsymbol{b}(\boldsymbol{k})=-\frac{1}{\sqrt{2}}\left(\sin \alpha_k^{\downarrow}+\sin \alpha_k^{\uparrow}\right) \hat{e}_{01}+\frac{1}{\sqrt{2}}\left(\cos \alpha_{k}^{\downarrow}-\cos \alpha_k^{\uparrow}\right) \hat{e}_{02}+\frac{1}{\sqrt{2}}\left(\sin \alpha_{k}^{\downarrow}-\sin \alpha_k^{\uparrow}\right) \hat{e}_{31}-\frac{1}{\sqrt{2}}\left(\cos \alpha_{k}^{\downarrow}+\cos \alpha_{k}^{\uparrow}\right) \hat{e}_{32}
\end{equation}
in the basis of  $\boldsymbol{\lambda}_{\alpha\beta}=\boldsymbol{\sigma}_\alpha \otimes \boldsymbol{\tau}_\beta /\sqrt{2} $, where $\sigma$ and $\tau$ denote the Pauli matrices representing spin and orbital degrees of freedom, respectively. The unit vector $\hat e_{\alpha\beta}$ in \eqnref{bfull} is the ``Bloch vector'' of $\boldsymbol{\lambda}_{\alpha\beta}$. Then, using the projection operator given in \eqnref{Pkfull}, the quantum geometric tensor defined in  \eqnref{QGT} can be expressed as:
\begin{eqnarray}
\begin{aligned}
T_{\mu \nu}(\boldsymbol{k})=&\operatorname{Tr}\left[\partial_\mu P(\boldsymbol{k})\left(\mathds{1}_4-P(\boldsymbol{k})\right) \partial_\nu P(\boldsymbol{k})\right]=\frac{1}{4} \partial_\mu \alpha_k^{\uparrow} \partial_\nu \alpha_k^{\uparrow}+\frac{1}{4} \partial_\mu \alpha_k^{\downarrow} \partial_\nu \alpha_k^{\downarrow}\\
=&\frac{1}{4} \eta_{\uparrow}^2 \sin \boldsymbol{k}_\mu \sin \boldsymbol{k}_\nu+\frac{1}{4} \eta_{\downarrow}^2 \sin \boldsymbol{k}_\mu \sin \boldsymbol{k}_\nu.
\end{aligned}
\end{eqnarray}
Thus, the average quantum metric from \eqnref{ag} is given by:
\begin{equation}\label{bg}
    \bar{g}=\frac{1}{8}\left(\eta_{\uparrow}^2+\eta_{\downarrow}^2\right)\mathds{1}_2,
\end{equation}
which is directly related to the parameters $\eta_\sigma$.

\begin{figure*}[t]
    \centering
    \includegraphics[width=0.9\linewidth]{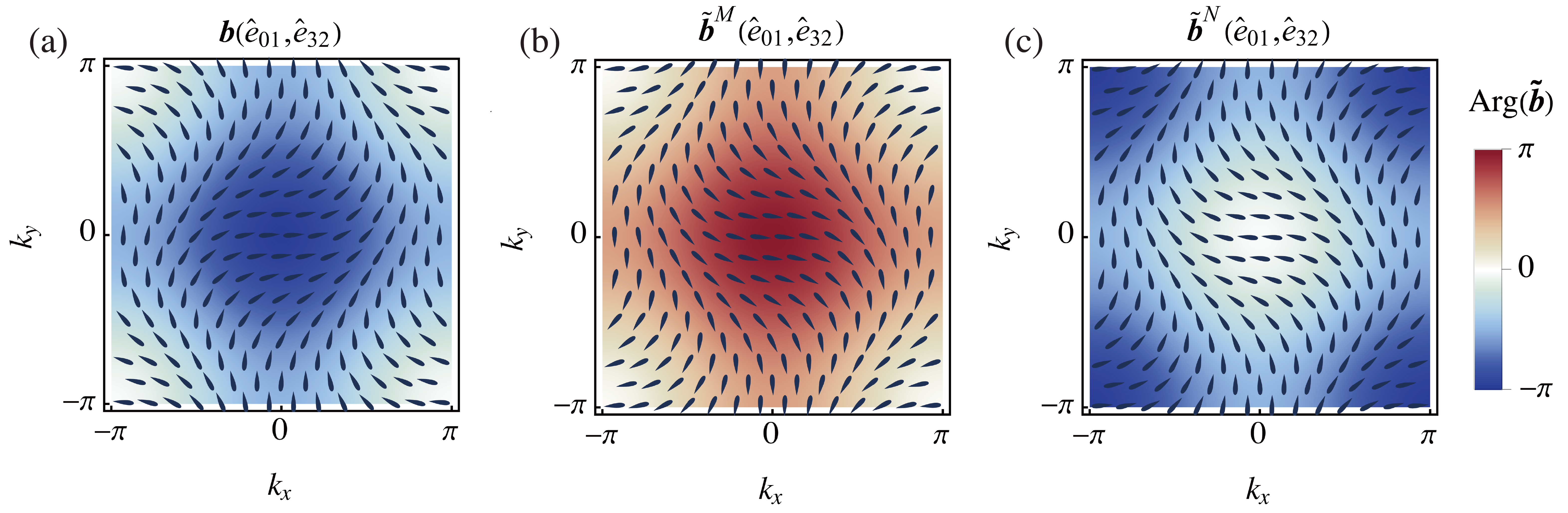}
    \caption{With the parameter $\eta_{\sigma}=0.75$: (a) Bloch vector $\boldsymbol{b}$. (b) Bloch vector $\tilde{\boldsymbol{b}}_{\boldsymbol{M}}$ for the $\boldsymbol{M}^x$ operator. (c) Bloch vector $\tilde{\boldsymbol{b}}_{\boldsymbol{N}}$ for the $\boldsymbol{N}^x$ operator. The drop arrows indicate the direction of the Bloch vector in the basis of $\hat{e}_{01}$ and $\hat{e}_{32}$, while the background color represents its angle, i.e., $\mathrm{Arg}(\boldsymbol{b})$.}
    \label{fig:app_spin}
\end{figure*}

\subsection{Repulsive Interaction: Spin Excitations}

First, we discuss the case of a local repulsive interaction:
\begin{eqnarray}\label{Hv}
    H_{V}=U \sum_i\left(n_{i A}^2+n_{i B}^2\right),
\end{eqnarray}
where $n_{iA}=\sum_\sigma c_{iA\sigma}^\dagger c_{iA\sigma}$ and $n_{iB}=\sum_\sigma c_{iB\sigma}^\dagger c_{iB\sigma}$ denote the electron densities in the $A$ and $B$ orbitals, respectively. Here, we assume that the interaction magnitude $U$ is smaller than the band gap $2t$, allowing us to project onto the low-energy sector. Then, by decomposing the interaction as follows:
\begin{eqnarray}\label{Hv2}
    H_{V}=-\frac{2}{3}U\sum_i \boldsymbol{M}_i^2 -\frac{2}{3}U\sum_i \boldsymbol{N}_i^2 + 2U\sum_{i}n_i,
\end{eqnarray}
it is evident that the channels for the total spin order, $\boldsymbol{M}_i=\boldsymbol{S}_{i, A}+\boldsymbol{S}_{i, B}=c^\dagger_{A\alpha}(\boldsymbol{k}) \boldsymbol{\sigma}_{\alpha \beta}c_{A\beta}(\boldsymbol{k})+c^\dagger_{B\alpha}(\boldsymbol{k}) \boldsymbol{\sigma}_{\alpha \beta}c_{B\beta}(\boldsymbol{k})$, and staggered spin order, $\boldsymbol{N}_i=\boldsymbol{S}_{i, A}-\boldsymbol{S}_{i, B}=c^\dagger_{A\alpha}(\boldsymbol{k}) \boldsymbol{\sigma}_{\alpha \beta}c_{A\beta}(\boldsymbol{k})-c^\dagger_{B\alpha}(\boldsymbol{k}) \boldsymbol{\sigma}_{\alpha \beta}c_{B\beta}(\boldsymbol{k})$, are both encoded in the interaction $H_V$. In the following, we analyze which spin order and which wave vector are most favorable. For simplicity, we preserve time-reversal symmetry by setting $\eta_\uparrow=\eta_\downarrow=\eta$. Then \eqnref{bfull} is:
\begin{equation}
    \boldsymbol{b}(\boldsymbol{k})=-\sqrt{2}\sin \alpha_k \hat{e}_{01}-\sqrt{2}\cos \alpha_{k}\hat{e}_{32},
\end{equation}
where $\alpha_k=\eta\left(\cos \boldsymbol{k}_x+\cos \boldsymbol{k}_y\right)$, and the matrix of the total spin order along the $x$-direction in the given basis  $\boldsymbol{\lambda}_{\alpha\beta}=\boldsymbol{\sigma}_\alpha \otimes \boldsymbol{\tau}_\beta /\sqrt{2} $ can be expressed as:
\begin{equation}
    \boldsymbol{M}_i^x=\boldsymbol{S}_{i, A}^x+\boldsymbol{S}_{i, B}^x=\frac{1}{2} \sigma_{10},
\end{equation}
whose corresponding unit vector is given by:
\begin{equation}
    \hat{\boldsymbol{o}}_{\boldsymbol{M}}=\hat{e}_{10}.
\end{equation}
Using the Lie algebra relation \eqnref{li}, we obtain
\begin{equation}
    \hat e_{10} \times \hat e_{01}=0, \;\;\;\;\;\;\hat e_{10} \times \hat e_{32}=-\frac{1}{\sqrt{2}} \hat e_{22},\;\;\;\;\;\;\hat e_{22} \times \hat e_{10}=-\frac{1}{\sqrt{2}} \hat e_{32},
\end{equation}
so the dressed Bloch vector of the $\boldsymbol{M}_i^x$ order, is given by:
\begin{equation}
    \tilde{\boldsymbol{b}}_{\boldsymbol{M}}(\boldsymbol{k}) \equiv \boldsymbol{b}(\boldsymbol{k})-4(\hat{\boldsymbol{o}}_{\boldsymbol{M}} \times \boldsymbol{b}(\boldsymbol{k}) \times \hat{\boldsymbol{o}}_{\boldsymbol{M}})=-\sqrt{2}\sin \alpha_{k} \hat{e}_{01}+\sqrt{2}\cos \alpha_{k}\hat{e}_{32},
\end{equation}
as shown in \figref{fig:app_spin}~(b). Compared to the Bloch vector $\boldsymbol{b}$ pattern in \figref{fig:app_spin}~(a), there are no translation vectors $\boldsymbol{Q}$ (including zero momentum) that make $\boldsymbol{b}(\boldsymbol{k})$ and $\tilde{\boldsymbol{b}}_{\boldsymbol{M}}(\boldsymbol{k}+\boldsymbol{Q})$ compatible. Furthermore, among all the non-perfect nesting momenta, it is evident that a zero-momentum translation leads to the most parallel alignment. Consequently, the ferromagnetic (FM) total spin order is relatively more favorable than all other possible total spin density wave order vectors.

Similarly, the matrix of the staggered spin order along the $x$-direction can be expressed as:
\begin{equation}
    \boldsymbol{N}_i^x=\boldsymbol{S}_{i, A}^x-\boldsymbol{S}_{i, B}^x=\frac{1}{2} \sigma_{13},
\end{equation}
whose corresponding unit vector is:
\begin{equation}
    \hat{\boldsymbol{o}}_{\boldsymbol{N}}=\hat{e}_{13}.
\end{equation}
Using the following relations:
\begin{equation}
    \hat e_{13} \times \hat e_{01}=\frac{1}{\sqrt{2}} \hat e_{12}, \;\;\;\;\;\;e_{13} \times \hat e_{32}=0,\;\;\;\;\;\;\hat e_{12} \times \hat e_{13}=\frac{1}{\sqrt{2}} \hat e_{01},
\end{equation}
we obtain the dressed Bloch vector of the staggered spin order:
\begin{equation}
    \tilde{\boldsymbol{b}}_{\boldsymbol{N}}(\boldsymbol{k}) \equiv \boldsymbol{b}(\boldsymbol{k})-4(\hat{\boldsymbol{o}}_{\boldsymbol{N}} \times \boldsymbol{b}(\boldsymbol{k}) \times \hat{\boldsymbol{o}}_{\boldsymbol{N}})=\sqrt{2}\sin \alpha_k \hat{e}_{01}-\sqrt{2}\cos \alpha_{k}\hat{e}_{32},
\end{equation}
as shown in \figref{fig:app_spin}~(c). After translating the pattern by the momentum $\boldsymbol{Q}=(\pi,\pi)$, the dressed Bloch vector $\tilde{\boldsymbol{b}}_{\boldsymbol{N}}$ in \figref{fig:app_spin}~(c) becomes perfectly compatible with the Bloch vector $\boldsymbol{b}$ pattern in \figref{fig:app_spin}~(a). This indicates that the AFM staggered order in the $x$-$y$ plane perfectly satisfies the nesting condition, making it the most favorable spin order.

Moreover, according to the average quantum metric given in \eqnref{bg}, the Ginzburg-Landau correlation length in \eqnref{GLc} is
\begin{equation}
\xi=\frac{|\eta|}{2\sqrt{2}} \left|1-\frac{T}{T_c}\right|^{-\frac{1}{2}}
\end{equation}
for the staggered spin order.

\subsection{Attractive Interaction}

We now turn to the case of a local attractive interaction, i.e., $U<0$ in \eqnref{Hv}. This typically leads to competition between charge density wave and superconductivity. These correspond to typical particle-hole and particle-particle orders, respectively, so we will analyze the two orders separately.

\subsubsection{Particle-Hole Excitations: Charge Order}
The charge order parameter is given by
\begin{equation}
    n_i=n_{i, A}+n_{i, B}= \mathds{1}_4,
\end{equation}
which corresponds to the case where $o_0(\boldsymbol{k})\neq 0$ but $\boldsymbol{o}(\boldsymbol{k}) =\boldsymbol{0}$ in \secref{IA}, resulting in $\hat{\boldsymbol{o}}(\boldsymbol{k})$ vanishing in this case. Consequently, $\tilde{\boldsymbol{b}}_{\boldsymbol{o}}(\boldsymbol{k})$ reduces to $\boldsymbol{b}(\boldsymbol{k})$ as given in \eqnref{bfull}. In the time-reversal symmetric case, this is shown in \figref{fig:app_spin}~(a), indicating that no finite translation momentum exists such that the $\boldsymbol{b}$-pattern after translation is perfectly compatible with the pattern before translation. This suggests that the charge density wave is not particularly favorable if other orders satisfy perfect nesting conditions.

\begin{figure*}[t]
    \centering
    \includegraphics[width=0.9\linewidth]{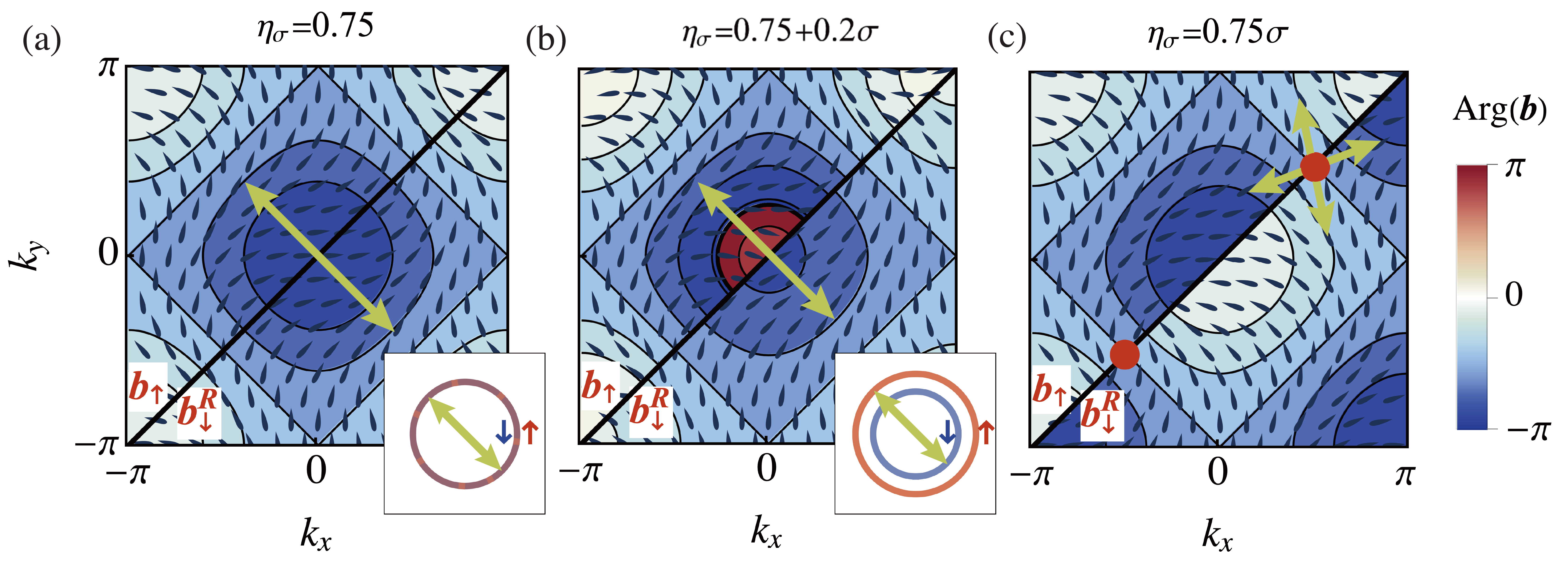}
    \caption{The Bloch vector $\boldsymbol{b}_{\uparrow}$ (top-left) and $\boldsymbol{b}^R_{\downarrow}$ (bottom-right) for the intra-orbital singlet pairing order, with the parameter (a) $\eta_{\sigma}=0.75$, (b) $\eta_{\sigma}=0.75+0.2\sigma$, and (c) $\eta_{\sigma}=0.75\sigma$. The drop arrows indicate the direction of the Bloch vector in the basis of $\hat{e}_{1}$ and $\hat{e}_{2}$, while the background color represents its angle, i.e., $\mathrm{Arg}(\boldsymbol{b})$. The yellow arrow serves as a visual guide for parallel Bloch vectors. The insets depict schematic illustrations of the pairing: (a) in a conventional metal with time-reversal symmetry and (b) with Zeeman splitting, where the red circle and blue circle denote the Fermi surface for $\uparrow$-spin and $\downarrow$-spin, respectively}
    \label{fig:supp}
\end{figure*}

\subsubsection{Particle-Particle Excitations: Pairing Order}

Next, we examine particle-particle excitations. In general, we need to compare the Bloch vector $\tilde{\boldsymbol{b}}^R_{\boldsymbol{o}}(\boldsymbol{k}+\boldsymbol{Q})$ and $\boldsymbol{b}(-\boldsymbol{k})$, as shown in \eqnref{nestpp}. However, for simplicity, given spin $U(1)$ rotational symmetry, different spin sectors decouple, allowing the projection matrix to be decomposed into distinct spin sectors as   
\begin{eqnarray}
P(\boldsymbol{k})=P_\uparrow(\boldsymbol{k})\oplus P_\downarrow(\boldsymbol{k}),
\end{eqnarray}
where 
\begin{eqnarray}
    P_{\sigma}(\boldsymbol{k})=\frac{N_L}{N} \mathds{1}_{N/2}+\frac{1}{2} \boldsymbol{b}_{\sigma}(\boldsymbol{k}) \cdot \boldsymbol{\lambda}
\end{eqnarray}
involves only the states with spin $\sigma$. The intra-orbital singlet pairing order parameter is given by
\begin{equation}
    \Delta_{\boldsymbol{Q}}=\sum_{\boldsymbol{k}} \left[c_{A \uparrow}(\boldsymbol{k}+\boldsymbol{Q}) c_{A \downarrow}(-\boldsymbol{k})+c_{B \uparrow}(\boldsymbol{k}+\boldsymbol{Q}) c_{B \downarrow}(-\boldsymbol{k})\right].
\end{equation}
Then, following the procedure from \secref{IB}, the susceptibility of $\Delta_{\boldsymbol{Q}}$ is
\begin{eqnarray}
    \chi^\Delta (\boldsymbol{Q})\sim \sum_{\boldsymbol{k}}\boldsymbol{b}_{\uparrow}^R(\boldsymbol{k}+\boldsymbol{Q}) \cdot \boldsymbol{b}_{\downarrow}(-\boldsymbol{k}),
\end{eqnarray}
indicating that the nesting condition for singlet pairing with $U(1)$ spin symmetry can be rewritten as:
\begin{eqnarray} \boldsymbol{b}^R_{\uparrow}(\boldsymbol{k}+\boldsymbol{Q}) \;\; \parallel\;\; \boldsymbol{b}_{\downarrow}(-\boldsymbol{k}),\;\;\;\;\;\;\;\;\;\; \forall \boldsymbol{k} \in \mathrm{BZ},
\end{eqnarray}
meaning the Bloch vector $\boldsymbol{b}_{\uparrow}^R$ at momentum $\boldsymbol{k}+\boldsymbol{Q}$ is compatible (parallel) with $\boldsymbol{b}_{\downarrow}$ at momentum $-\boldsymbol{k}$. In the basis of  $\boldsymbol{\lambda}_{\alpha}=\boldsymbol{\tau}_\alpha$, where $\tau$ denotes the Pauli matrix representing orbital degrees of freedom, the Bloch vector is:
\begin{eqnarray}\label{bSC}
    \begin{aligned}
& \boldsymbol{b}_\sigma(\boldsymbol{k})=- \sin \alpha_k^\sigma \hat{e}_1-\sigma  \cos \alpha_k^\sigma \hat{e}_2, \\
& \boldsymbol{b}_\sigma^R(\boldsymbol{k})=- \sin \alpha_k^\sigma \hat{e}_1+\sigma  \cos \alpha_k^\sigma \hat{e}_2
\end{aligned}
\end{eqnarray}
where the $\alpha$ index in the unit vector $\hat e_{\alpha}$ in \eqnref{bSC} corresponds to that in $\boldsymbol{\lambda}_{\alpha}$. 

First, when time-reversal symmetry is preserved, i.e., $\eta_{\uparrow}=\eta_{\downarrow}=0.75$, the $\boldsymbol{b}_{\downarrow}^R(\boldsymbol{k})$ and $\boldsymbol{b}_{\uparrow}(\boldsymbol{k})$ patterns are shown in \figref{fig:supp}~(a), and satisfy the perfect nesting condition. Specifically, each $\downarrow$-state with momentum $\boldsymbol{k}$ is connected to a compatible $\uparrow$-state with momentum $-\boldsymbol{k}$ (labeled by the yellow arrow), which is consistent with the pairing nesting scenario in conventional metals with a Fermi surface in the presence of time-reversal symmetry, as shown in the inset of \figref{fig:supp}~(a).

Second, upon breaking time-reversal symmetry by introducing distinct $\eta_{\sigma}$ for different $\sigma$-spins, the perfect nesting condition is broken, leading to the shift of the $\boldsymbol{b}_{\downarrow}^R(\boldsymbol{k})$ and $\boldsymbol{b}_{\uparrow}(\boldsymbol{k})$ patterns, as shown in \figref{fig:supp}~(b). In analogy to a conventional metal in a magnetic field, where the Fermi surface shifts for opposite spins due to Zeeman splitting  (shown in the inset of \figref{fig:supp}~(b)), so that an FFLO state can be induced, the spin shift of the Bloch vector pattern can also potentially introduce an FFLO state with a finite center-of-mass momentum for Cooper pairs. However, it is important to note that perfect nesting is not naturally satisfied in this case, like in the conventional FFLO case, where perfect pairing nesting also does not exist. Therefore, the FFLO state is not always the most favorable state and requires fine-tuning of the parameters. One suitable parameter is $\eta_\sigma=1.25+2.5\sigma$, such that the $(\pi,\pi)$ CDW is weak at the mean-field level, but the SC susceptibility exhibits peaks at $(\pm\pi/2,\pm\pi/2)$, as shown in Fig.~3 (c) in the main text. 

An interesting pairing case that satisfies the perfect nesting condition is $\eta_{\uparrow} = -\eta_{\downarrow}$, for which the Bloch vector pattern is shown in \figref{fig:supp}~(c), with nesting momentum $\boldsymbol{Q}=(\pi,\pi)$.

\section{Supplementary DQMC Simulation Data}

\subsection{Simulation Details}

The DQMC susceptibilities are obtained using $5 \times 10^4$ warm-up sweeps per Markov chain, and, in total on the order of $10^6$ measurement sweeps.
The imaginary-time Trotter discretization step for susceptibilities is chosen to be $d\tau=0.05/t$, which is sufficiently small to control Trotter errors, satisfying the heuristic criterion $U\delta \tau^2 \leq W^{-1}$, with $W$ the bandwidth of the tight-binding model. For our model with perfectly flat bands ($W=0$), this choice of $d\tau$ is conservative, remaining sufficiently small even if $W$ is taken as the hopping energy scale $t$ or the band gap scale $2t$.
To achieve a target filling $\nu$, we first perform DQMC calculations to determine the dependence of $\langle \nu \rangle$ on the chemical potential $\mu$, and then obtain the best $\mu$ by interpolating $\langle \nu \rangle$ versus $\mu$. For the tuning process, we use $d\tau=0.02/t$.
All simulations are performed on an $8\times 8$ lattice with periodic boundary conditions, with the interaction strength fixed at $|U|/t=1$ for both attractive and repulsive interactions.
Error bars in the DQMC results denote $\pm 1$ standard error of the mean, estimated by jackknife resampling. They are smaller than the marker size and therefore not visible.

\subsection{Repulsive Interaction}

\begin{figure*}[t]
    \centering
    \includegraphics[width=0.65\linewidth]{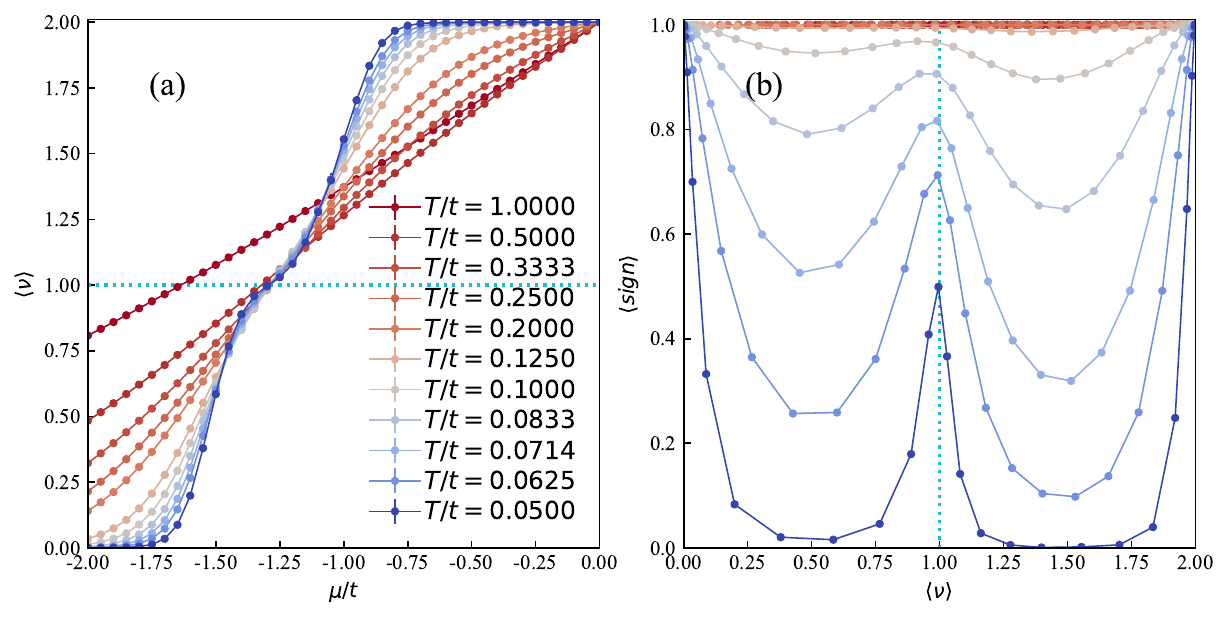}
    \caption{(a) Filling number $\langle \nu \rangle$ as a function of chemical potential $\mu$ at various temperatures. (b) Average fermion sign $\langle \text{sign} \rangle$ as a function of $\langle \nu \rangle$. Parameters: $\eta_\sigma = 0.75$, repulsive interaction $U/t = 1$, on an $8 \times 8$ square lattice, identical to those used in Fig.~2 of the main text.}
    \label{fig:den_rep}
\end{figure*}

In Fig.~\ref{fig:den_rep}~(a), we show the filling number $\nu$ as a function of the chemical potential $\mu$ at various temperatures. This is used for the selection of $\mu$ corresponding to the target $\nu$ during the tuning procedure. The average fermion sign $\langle \text{sign} \rangle$ is shown in Fig.~\ref{fig:den_rep}~(b). With decreasing temperature, as spin order sets in, the slope of the $\nu$ versus $\mu$ curves at $\nu=1$ decreases, indicating a tendency towards a charge gap. With increased incompressibility at $\nu=1$, correspondingly, the fermion sign improves noticeably.

\begin{figure*}[t]
    \centering
    \includegraphics[width=0.8\linewidth]{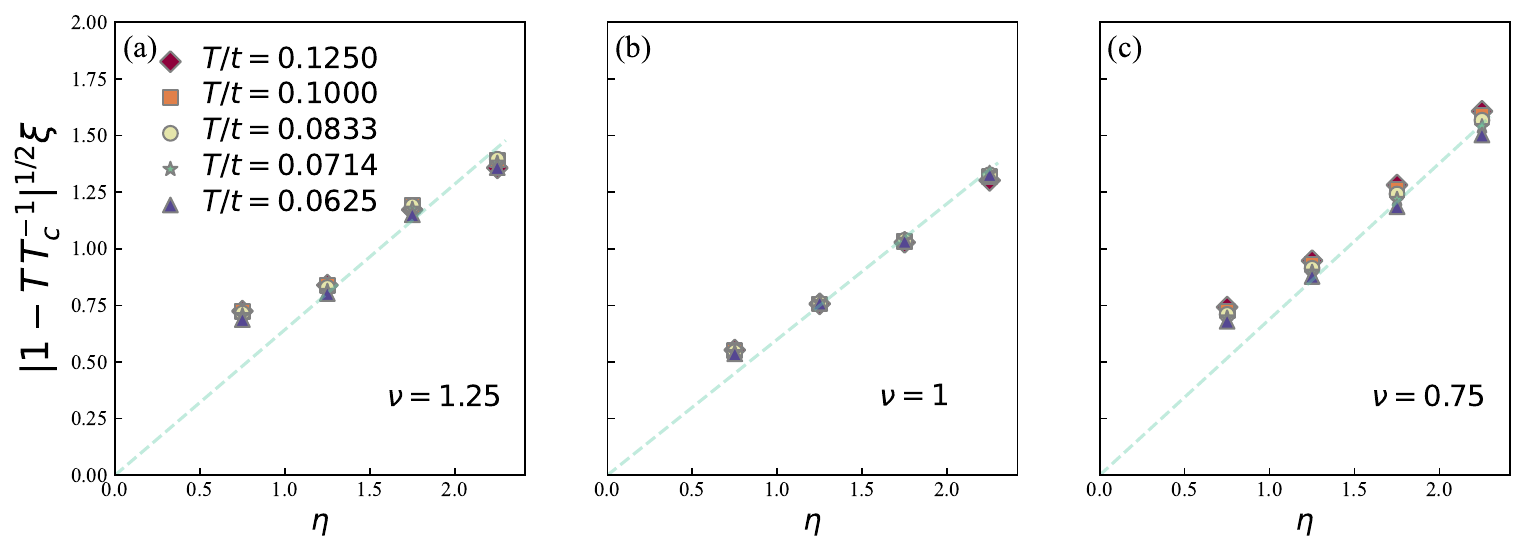}
    \caption{Correlation length $\xi$ of the $\mathbf{N}^x$ order at $\mathbf{Q}=(\pi,\pi)$, normalized by the temperature scaling factor $|1-T/T_c|^{1/2}$, for various fillings $\nu$. 
    The correlation length $\xi$ is extracted by fitting the $\mathbf{Q}$-dependent $\boldsymbol{N}^x$ spin susceptibility data along the $x$-direction to the Lorentzian form $\chi(\Delta \mathbf{Q})=C/(1+\xi^2 \Delta |\mathbf{Q}|^2)$, using the point $\mathbf{Q}=(\pi,\pi)$ and its two closest neighboring points along the $x$-direction. Here, $\Delta \mathbf{Q}=\mathbf{Q}-(\pi,\pi)$, and $C$ is a fitting parameter.
    $T_c$ is determined by linear extrapolation of the inverse $\boldsymbol{N}^x$ spin susceptibility at $\mathbf{Q}=(\pi,\pi)$ from temperatures $T/t=0.0714$ and $0.0625$. Dashed lines are guides to the eye, obtained from linear fits to the lowest-temperature ($T/t=0.0625$) data points. All data shown are obtained on $8 \times 8$ lattices.}
    \label{fig:corr_len}
\end{figure*}

In Fig.~\ref{fig:corr_len}, the quantity $|1-T/T_c|^{1/2}\xi$ shows minimal temperature dependence, consistent with the scaling discussed and predicted in the main text. Importantly, we observe a monotonic increase of the correlation length with $\eta$, approximately following a linear trend. This confirms that  the correlation length is proportional to the quantum metric.

\subsection{Attractive Interaction}

\begin{figure*}[t]
    \centering
    \includegraphics[width=\linewidth]{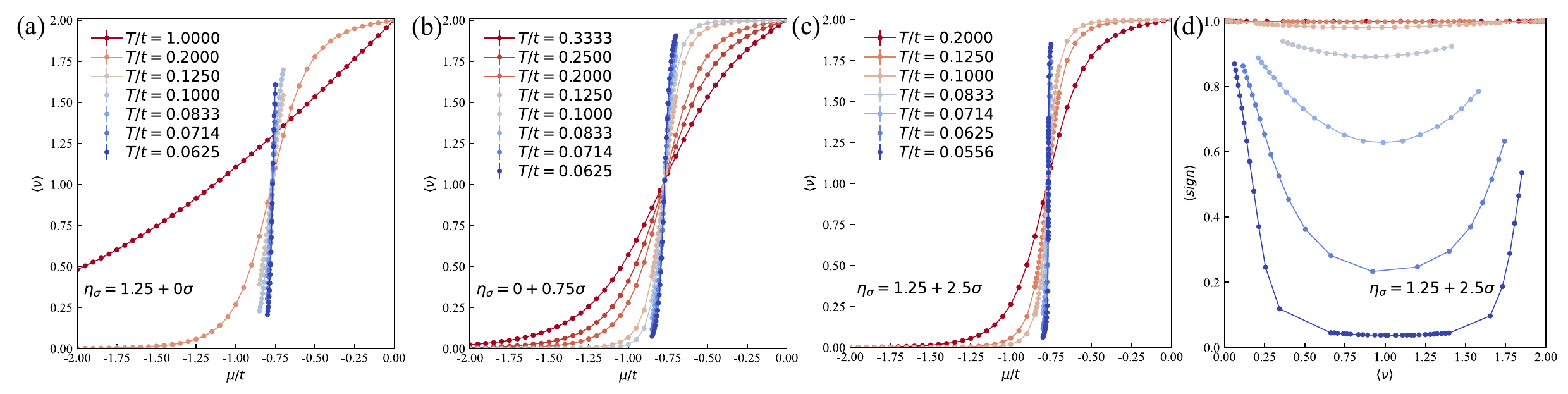}
    \caption{(a)–(c) Filling number $\langle \nu \rangle$ as a function of chemical potential $\mu$ at various temperatures for three different parameter sets. Among them, only the case with $\eta_\sigma = 1.25 + 2.5\sigma$ in (c) exhibits a fermion sign deviating from $1$, which is shown in (d). The fermion sign in (d) shares the same legend as (c). Parameters: Attractive interaction $U/t = -1$, and the lattice size is $8 \times 8$.}
    \label{fig:den_att}
\end{figure*}

In Fig.~\ref{fig:den_att}, the $\langle \nu \rangle$ versus $\mu$ curves at low temperatures indicate that the band remains highly flat, with its position modified due to interactions. Unlike in the repulsive case in Fig.~\ref{fig:den_rep}, no charge gap appears at $\nu=1$.

\begin{figure*}[t]
    \centering
    \includegraphics[width=0.7\linewidth]{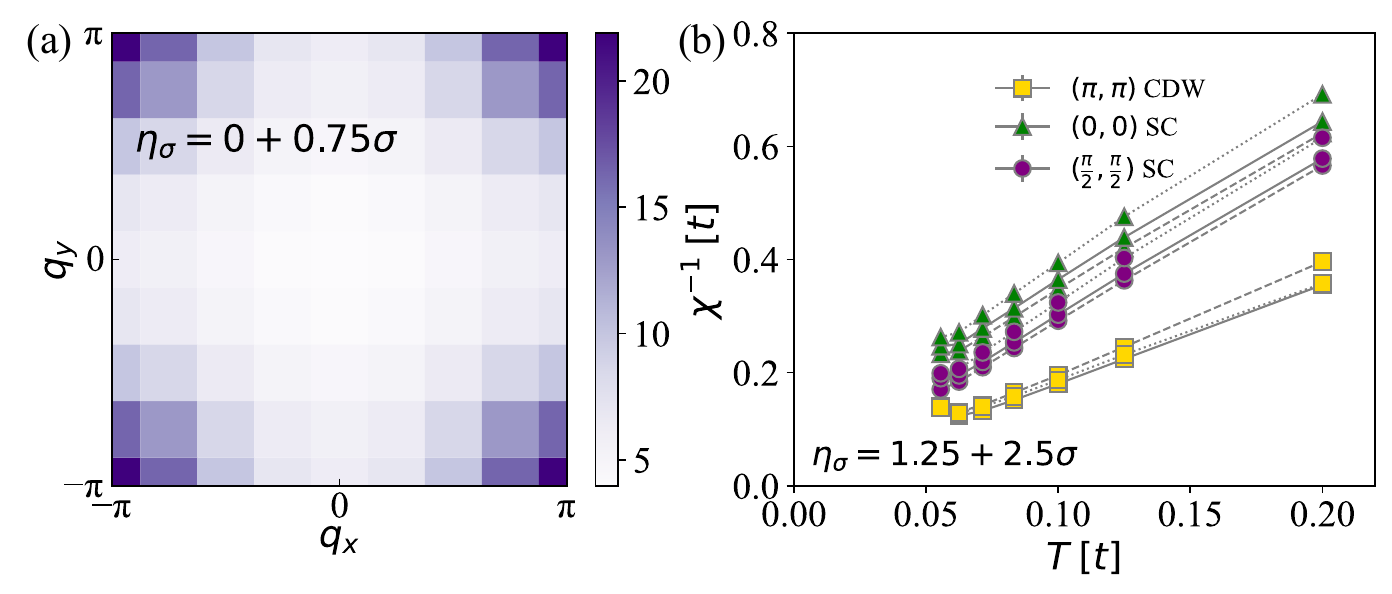}
    \caption{
   Supplementary DQMC data for attractive interaction ($U/t=-1$) cases discussed in the main text.
(a) Momentum distribution of the SC susceptibility from DQMC with $\eta_\sigma = 0.75\sigma$, for $\nu=1.25$ at $T/t=0.0625$ as a representative example.
(b) Temperature dependence of the inverse CDW and SC susceptibilities. Dashed, solid, and dotted lines correspond to $\nu=0.75$, $1$, and $1.25$, respectively.
       }
    \label{fig:supp_q_scaling}
\end{figure*}

Fig.~\ref{fig:supp_q_scaling} provides supplementary DQMC data for cases with attractive interactions.
Fig.~\ref{fig:supp_q_scaling}~(a) shows a peak at momenta $\mathbf{Q}=(\pi,\pi)$, consistent with Fig.~3~(b) in the main text, where the $(\pi,\pi)$ SC order dominates over other orders and exhibits a nonzero $T_c$.
Fig.~\ref{fig:supp_q_scaling}~(b) suggests that we cannot definitively determine whether the $(\pi/2,\pi/2)$ SC can dominate over CD and whether SC can reach a nonzero $T_c$ and become the true ground state.

\begin{figure*}[t]
    \centering
    \includegraphics[width=0.7\linewidth]{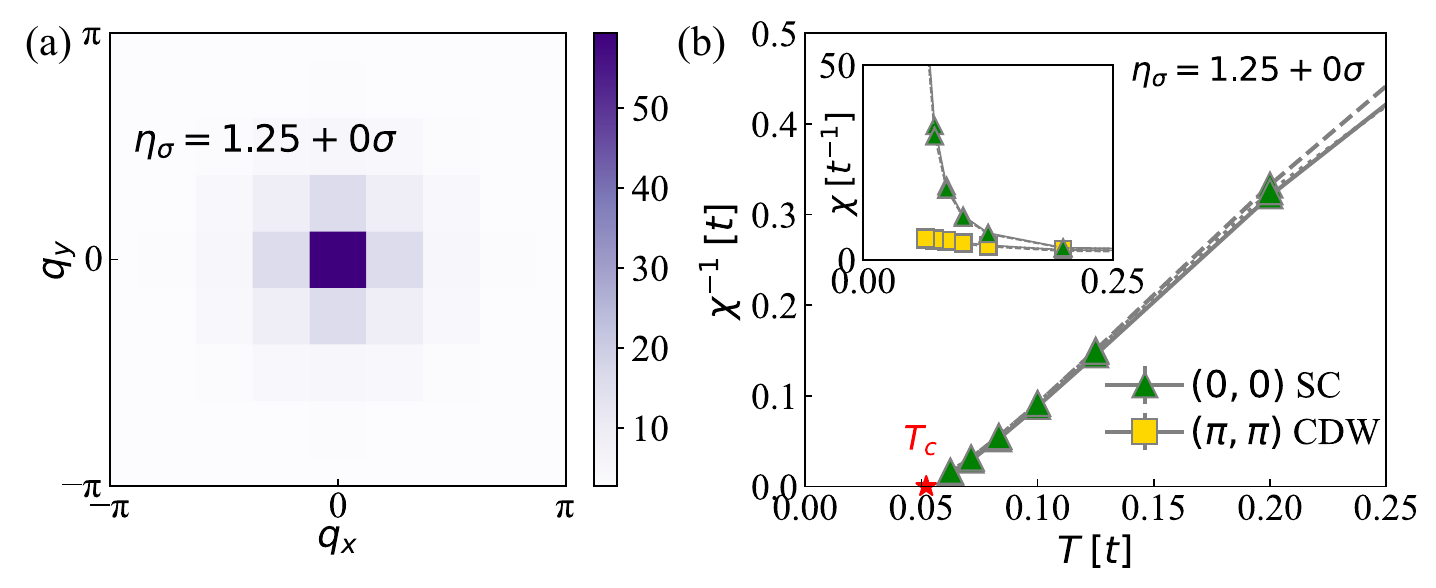}
    \caption{Analysis for the case without time-reversal symmetry breaking for $\eta_\sigma = 1.25$.
(a) Momentum distribution of the SC susceptibility, for $\nu=1.25$ at $T/t=0.0625$ as a representative example.
(b) Temperature dependence of the inverse susceptibility for uniform SC order. The inset compares the diverging susceptibility of the uniform SC order with the weak susceptibility of the $(\pi,\pi)$ charge order. The star marks the extrapolated critical temperature $T_c$ for $\nu=1$.
Dashed, solid, and dotted lines correspond to $\nu=0.75$, $1$, and $1.25$, respectively.
    }
    \label{fig:notbreaking_supp}
\end{figure*}

To compare with the attractive cases discussed in the main text, where time-reversal symmetry is broken, Fig.~\ref{fig:notbreaking_supp} presents the results of a situation where time-reversal symmetry is preserved under the same interaction, $U/t = -1$. In Fig.~\ref{fig:notbreaking_supp}~(a), the SC susceptibility exhibits a peak at momentum $\mathbf{Q} = (0,0)$. Correspondingly, Fig.~\ref{fig:notbreaking_supp}~(b) shows a nonzero $T_c$ for the uniform SC, where the competing $\mathbf{Q} = (\pi,\pi)$ CDW is weak.

\section{DQMC Results for Finite Bandwidth Effects}
In the following, we present DQMC numerical results to investigate the effects of finite bandwidth, as analytically discussed in \appref{Sec:valid}. We begin with the model introduced in Eq.\eqref{hk} of the main text. To disentangle the roles of band geometry (i.e., nesting extent) and band dispersion in determining the leading instability, we design the system such that the band geometry still satisfies the condition $\tilde{\boldsymbol{b}}_{\boldsymbol{o}}(\boldsymbol{k}+\boldsymbol{Q}) \;\|\; \boldsymbol{b}(\boldsymbol{k})$. 

With the basis $\psi_k = \left[c_{A\uparrow}(\boldsymbol{k})\;\;c_{B\uparrow}(\boldsymbol{k})\;\;c_{A\downarrow}(\boldsymbol{k})\;\;c_{B\downarrow}(\boldsymbol{k})\right]^T$, the Hamiltonian is given by:
\begin{eqnarray}\label{hksu}
\mathcal{H}_k = 
h_{\boldsymbol{k}}+ \epsilon_k \sigma_{00}=\left(\begin{array}{cccc}
-\mu  +\epsilon_k  & -it e^{i\alpha_k^{\uparrow}} & 0 & 0 \\
it e^{-i \alpha_k^{\uparrow}} & -\mu+\epsilon_k & 0 & 0 \\
0 & 0 & -\mu +\epsilon_k & it e^{-i \alpha_k^{\downarrow }} \\
0 & 0 & -it e^{i \alpha_k^{\downarrow }} & -\mu  +\epsilon_k
\end{array}\right)
\end{eqnarray}
with the resulting band dispersion:
\begin{equation}
E_{\pm, \sigma}(\boldsymbol{k}) = \pm t - \mu + \epsilon_k
\end{equation}
Here, $h_{\boldsymbol{k}}$ corresponds to the flat-band Hamiltonian given in Eq.~\eqref{hk} of the main text, while $\epsilon_k$ introduces dispersion via intraorbital hopping. Since $\epsilon_k$ is proportional to the identity matrix, it leaves the eigenvectors $U_{\alpha m}(\boldsymbol{k})$ unchanged. This allows us to isolate and analyze the effects of dispersion without modifying the band geometry.

\begin{figure*}[t]
    \centering
    \includegraphics[width=1\linewidth]{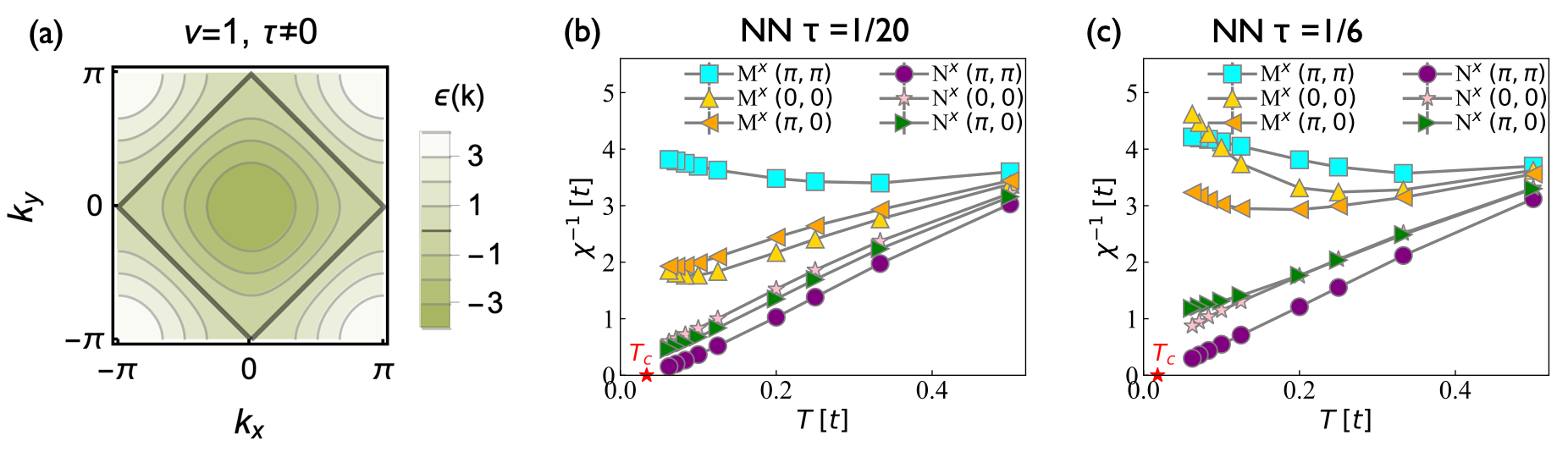}
    \caption{(a) Dispersion relation on the square lattice with nearest-neighbor (NN) hopping $\tau$. The thick gray line indicates the Fermi surface at half-filling ($\nu=1$). (b)-(c) DQMC results for the temperature dependence of the inverse spin susceptibilities $\boldsymbol{M}^x$ and $\boldsymbol{N}^x$ at momenta $\mathbf{Q} = (0,0)$, $(\pi,0)$, and $(\pi,\pi)$ at filling $\nu = 1$. Intraorbital NN hopping $\tau = 1/20 $ and $1/6$ for (b) and (c), respectively. Stars mark the extrapolated critical temperature $T_c$ for the $N_x$ order at $(\pi,\pi)$.
    }
    \label{fig:NN}
\end{figure*}

\begin{figure*}[t]
    \centering
    \includegraphics[width=1\linewidth]{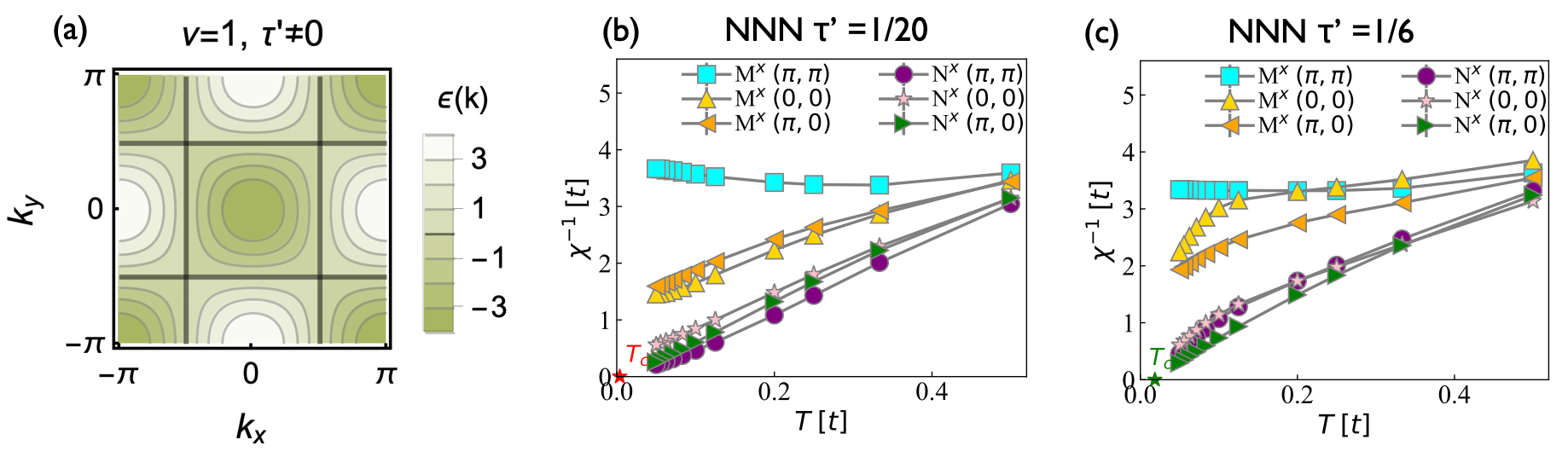}
    \caption{(a) Dispersion relation on the square lattice with NNN hopping $\tau'$. The thick gray line indicates the Fermi surface at half-filling ($\nu=1$). (b)-(c) DQMC results for the temperature dependence of the inverse spin susceptibilities $\boldsymbol{M}^x$ and $\boldsymbol{N}^x$ at momenta $\mathbf{Q} = (0,0)$, $(\pi,0)$, and $(\pi,\pi)$ at filling $\nu = 1$. Intraorbital NNN hopping $\tau' = 1/20 $ and $1/6$ for (b) and (c), respectively. Stars mark the extrapolated critical temperature $T_c$ for the $N_x$ order at $(\pi,\pi)$ for (b) and at $(\pi,0)$ for (c).
    }
    \label{fig:NNN}
\end{figure*}

We consider two distinct types of dispersion:
\begin{itemize}
    \item \textbf{NN Intraorbital Hopping:} By including only NN intraorbital hopping term \(\tau\) in the \(\nu=1\) case, the band dispersion within the low-energy subset is given by $\epsilon_k=-2 \tau\left(\cos k_x+\cos k_y\right)$ with the corresponding band structure and Fermi surface at the half-filling ($\nu=1$) illustrated in \figref{fig:NN}(a).  We examine two representative values: \(\tau = 1/20 t\), where the resulting bandwidth is much smaller than the interaction strength ($w \ll U$), and \(\tau = 1/6 t\), where the bandwidth becomes comparable to the interaction strength ($w \simeq U \ll \Delta$). The temperature dependence of susceptibilities for various ordering channels is shown in Fig.~\ref{fig:NN} (b)-(c). In both regimes, the leading instability is consistently found to be the AFM $\boldsymbol{N}$ order with ordering vector $(\pi,\pi)$, identical to the case of a perfectly flat band. This robustness arises from the Fermi surface nesting induced by the NN hopping $\tau$, which yields a nesting vector of $(\pi,\pi)$, as shown in Fig.~\ref{fig:NN}(a), which matches precisely with the nesting vector obtained from the nesting condition defined in our framework.
    \item \textbf{NNN Intraorbital Hopping:} Including only the NNN intraorbital hopping term \(\tau'\) in the \(\nu=1\) case, the band dispersion in the low-energy subset is given by $\epsilon_k=-2 \tau'\left(\cos (k_x+k_y)+\cos (k_x-k_y)\right)$, with the resulting band structure and Fermi surface at the half-filling ($\nu=1$) shown in Fig.~\ref{fig:NNN}(a). For \(\tau' = 1/20t\), corresponding to $w \ll U$, the dominant order remains the $(\pi,\pi)$ AFM $\boldsymbol{N}$, as shown in Fig.~\ref{fig:NNN}(b), consistent with our claim that the nesting condition derived in this work remains valid as long as the energy scale constraint in Eq.~\eqref{condi} is satisfied. On the other hand, for \(\tau' = 1/6t\), corresponding to $w \simeq U$, fluctuations of the $N$ order at $(\pi,0)$ are progressively enhanced and tend to replace the $(\pi,\pi)$ order,  as shown in Fig.~\ref{fig:NNN}(c). This is because the Fermi surface nesting vector induced by the NNN hopping $\tau'$ shifts to $(\pi,0)$, distinct from the flat-band limit. Fermi surface features become progressively more prominent as the bandwidth approaches the magnitude of interaction $U$, signaling a crossover toward dispersion-dominated instabilities.
\end{itemize}

\end{document}